\documentclass[aps,prd,nofootinbib]{revtex4}
\usepackage[caption=false]{subfig}

\usepackage[T1]{fontenc}
\usepackage[utf8]{inputenc}
\usepackage{lmodern}
\usepackage{amsmath}
\usepackage{amsfonts}
\usepackage{booktabs}
\usepackage{multirow}
\usepackage{tabularx}
\usepackage{graphicx}
\usepackage{todonotes}
\usepackage[english]{babel}
\usepackage{placeins}

\usepackage{hyperref}

\graphicspath{ {figures/} {SpectralDimPics/}}
\newcommand{\pd}[2]{\frac{\partial #1}{\partial #2}}

\newcommand{\mc}[1]{\mathcal{#1}}

\newcommand{\av}[1]{\left\langle #1 \right\rangle}
\newcommand{\Tr}[1]{\mathrm{Tr}\left( #1\right)}
\DeclareMathOperator\tr{Tr} 

\begin{document}

\title{Scaling behaviour in random non-commutative geometries}
\date{\today}
\author{Lisa Glaser}
\affiliation{Radboud University Nijmegen}

\begin{abstract}
Random non-commutative geometries are a novel approach to taking a non-perturbative path integral over geometries.
They were introduced in \cite{barrett_monte_2015}, where a first examination was performed.
During this examination we found that some geometries show indications of a phase transition.
In this article we explore this phase transition further for geometries of type $(1,1)$, $(2,0)$, and $(1,3)$.
We determine the pseudo-critical points of these geometries and explore how some of the observables scale with the system size.
We also undertake first steps towards understanding the critical behaviour through correlations and in determining critical exponents of the system.
\end{abstract}
\maketitle
\section{Introduction}
The spectral approach is an intriguing reformulation of geometry.
A manifold can be described as a spectral triple $(\mc{A},\mc{H},D)$, in which the algebra $\mc{A}$ encodes the functions of the coordinates which act on the Hilbert space $\mc{H}$ and the Dirac operator $D$ contains metric and differential information~\cite{connes_noncommutative_1994,van_suijlekom_noncommutative_2015}.
For finite spaces, e.g. a set of points, the algebra consists of diagonal matrices with the function value at point $i$ in the $i$-th position.
To describe continuous spaces through the spectral approach we can use the Gelfand duality, which prescribes a one-to-one correspondence between compact Hausdorff topological spaces and commutative $C^*$ algebras~\cite{van_suijlekom_noncommutative_2015}.
Generalising this definition to allow for non-commutative algebras $\mc{A}$ leads to general infinite dimensional non-commutative geometries.

While these infinite dimensional non-commutative geometries are very interesting from a mathematical perspective, the simpler case of fuzzy geometries is much more tractable for physical applications.
In a fuzzy space, the algebra and the Dirac operator become finite dimensional matrices, acting on a Hilbert space which is a product space of a Clifford module and a matrix space.
The best known example of a fuzzy space is the fuzzy sphere, with the Grosse-Presnajder Dirac operator
\begin{align}
D = \gamma + \sum_{i<j} \sigma_{i} \sigma_{j} \otimes [ L_{ij}, \;\cdot \; ]
\end{align}
with $\sigma_i$ the Pauli matrices and $L_{ij}$ the generators of the $SU(2)$ Lie algebra~\cite{madore_fuzzy_1992,grosse_dirac_1995,barrett_matrix_2015}.
Other well understood examples are the fuzzy torus~\cite{Connes:1997cr,Schreivogl:2013qza} or the fuzzy $CP^2$~\cite{alexanian_fuzzy_2002}.
In \cite{barrett_matrix_2015} a prescription for general spectral triples that should correspond to fuzzy geometries is given, however constructiong fuzzy spaces for less symmetric cases has proven to be very hard.

Representing these geometries through their spectral data also serves as a discretisation, and the encoding as matrices is ideal for computer simulations \cite{oconnor_monte_2006}.
In \cite{barrett_monte_2015} we used fuzzy spaces as a discretisation of space that would make the path integral over geometries amendable to Monte Carlo simulations.

The aim of this research is to use non-commutative geometry to quantise space-time, with the non-commutative structure serving as a regulator of the path integral over geometries.

Past work has found non-commutative spaces as emerging states of random matrix-models~\cite{delgadillo-blando_geometry_2008, oconnor_critical_2013}.
They find that in a three matrix model with the most general single trace action the ground state, which dominates the low temperature phase, is the fuzzy sphere with fluctuations around it described by a Yang-Mills field.
In their work the random matrices take on the values of the angular momentum generators of the sphere, and the Dirac operator is not considered.
In contrast our model concentrates mainly on the Dirac operator, which naturally leads to a multi trace action.

One open question in considering non-commutative geometry as a quantum description of space-time is how the Lorentzian structure should be included, since no satisfactory description of finite Lorentzian spectral triples exists yet.
Another problem, towards which we are currently working, is how to identify non-commutative geometries as approximated by classical manifolds, and which action should be used to weight these geometries in the path integral.

A particular strength of our approach is that, once we understand how to resolve these issues, coupling the standard model of particle physics to the quantised geometries is well understood and prescribed, as it is a simple generalisation of the almost-commutative description of the standard model given in \cite{connes_noncommutative_2006,barrett_lorentzian_2007}.
In the almost-commutative standard model space-time remains continuous, but the particle content is described through non-commutative extra-dimensions.
A particularly nice feature of this approach is that after writing down the fermionic content of the standard model the correct bosonic content arises inevitably through the structure of the theory.

The simulations in \cite{barrett_monte_2015} were of an exploratory character.
Looking at the eigenvalue distribution of the Dirac operator we saw some clear indications for a phase transition, and tantalising hints at two dimensional behaviour of the spectrum around it.
We will follow up on the hints of two dimensional behaviour in a forthcoming publication~\cite{barrett_druce_glaser}, while we will here further explore the phase transition.
Unfortunately the amount of data gathered before was not enough to make a definite case or determine the order of the transitions.
To better understand the phase transition and the thermodynamic behaviour associated with it we have started new, more extensive simulations.
While we tried to explore as many spaces with low $(p,q)$ as possible in \cite{barrett_monte_2015}, for this in depth exploration we decided to focus on three cases.
The first case are geometries of type $(1,3)$, as these are of the same class as the fuzzy sphere \cite{barrett_matrix_2015} which gives us a well understood geometry to compare them against.
The other two cases are the $2$ dimensional cases with phase transition $(1,1)$ and $(2,0)$.
In our first explorations these two cases presented as remarkably similar, and our hope is that a better understanding of the phase transition might hint at a reason for this.

In this paper we further explore the phase transition for the $(1,1),(2,0),(1,3)$ geometry.
In section \ref{sec:intro} we review the results from~\cite{barrett_monte_2015} and introduce our methods.
In section \ref{sec:phases} we explore the phase transitions further, and in section \ref{sec:scaling} we explore how the transition points scale with the volume.
We end with a conclusion in \ref{sec:conclusion}.

\section{\label{sec:intro}Random geometries}
While we refer to~\cite{barrett_monte_2015} for the full details of our implementation let us quickly recap the most important parts.
The fuzzy spaces used are real, finite spectral triples.
In these the triple of $(\mc{A},\mc{H},D)$ is enlarged to also include a real structure $J$ and a chirality $\Gamma$.
The fuzzy spaces can then be defined with a product Hilbert space $\mc{H}=V\otimes M(n,\mathbb{C})$, in which $V$ is a $(p,q)$-Clifford module, that is a Clifford module spanned by $p$ hermitian and $q$ anti-hermitian $\gamma$ matrices.
The algebra acting on this is $\mc{A}=M(n,\mathbb{C})$ and for $a \in \mc{A}, v\otimes m \in \mc{H}$ it acts through matrix multiplication as $v \otimes am$.
The real structure in this situation introduces a right action, so $J^{-1} a J$ acts on $v \otimes m $ as $v \otimes m a$.
In this framework the Dirac operator can be expressed as a linear combination of all possible products of $\gamma$ matrices with commutators and anti-commutators of anti-hermitian/ hermitian matrices, as described in~\cite{barrett_matrix_2015}.

In particular this means that the Dirac operator can be entirely parametrised through these matrices, hence they form the space of geometries.
In the cases we explore here we have,
\begin{align}
D^{(1,1)}&=\gamma^1 \otimes \{ H,\cdot\} + \gamma^2 \otimes [L,\cdot] \\
D^{(2,0)}&=\gamma^1 \otimes \{H_1,\cdot \} + \gamma^2 \otimes \{H_2,\cdot\} \\
D^{(1,3)}&=\sum_{j<k=1}^{3} \gamma^{0} \gamma^{j} \gamma^{k} \otimes [ L_{jk}, \cdot] + \gamma^{1}\gamma^{2}\gamma^{3} \otimes \{H_{123},\cdot\}+ \gamma^{0}\otimes \{ L_{0},\cdot\} + \sum_{i=1}^3 \gamma^{i} \otimes [L_i,\cdot]
\end{align}
where the $\gamma$ matrices are those of the respective Clifford modules. For $(1,1)$ $\gamma^1$ is the hermitian matrix, with $\gamma^2$ being anti hermitian, for $(2,0)$ both matrices are hermitian and for $(1,3)$ all except $\gamma^0$ are anti-hermitian.
The matrices $L_i$ are anti hermitian and all matrices $H_i$ are hermitian.
A spectral triple is characterised by the matrix size of $H_i,L_i$, which we denote as $N$, the type of the Clifford-algebra, which we denote as $(p,q)$, and the matrices $H_i,L_i$.

In the random geometries as defined in \cite{barrett_monte_2015} the Dirac operator, and hence the matrices $H_i,L_i$ are part of the ensemble defined as
\begin{align}
  Z(\beta, g_2,g_4) = \int \mathcal{D}[D] e^{- \beta S(D,g_2,g_4)}
\end{align}
with the action
\begin{align}\label{eq:action}
S(D,g_2,g_4) = g_2 \Tr{D^2} + g_4 \Tr{D^4} \;.
\end{align}
A possible motivation for this action is that it contains the $D$ dependent terms that arise up to this order in the heat kernel expansion~\cite{chamseddine_spectral_1997}.
Of course this motivation is stretched a bit, since the expansion is only valid for in some sense `small' $D$ and our path integral is non-perturbative and would effectively integrate over all possible $D$, however for particular choices of the values of $g_2,g_4$ the action itself suppresses all `large' $D$ and thus ensures its own viability.
This action also has the advantage that we can use the specific structure of the Dirac operator to rewrite it in a computationally efficient form, by reducing it to traces of the $H_i,L_i$~\cite{barrett_monte_2015}.
For example the action for the $(1,1)$ case becomes
\begin{align*}
  S(D,g_2,g_4)=& g_2 \bigg\{ 4 \;n (\tr{H^2}-\tr{L^2})+ 4\; (\tr{H})^2+4\;(\tr{L})^2\bigg\} \\
  		&+ g_4 \bigg\{ 4 n  \bigg(\tr{H^4}+\tr{L^4} -4 \tr{H^2L^2} +2 \tr{HLHL}\bigg) \\
  		&+ 16  \bigg(\tr{H}\left(\tr{H^3}-\tr{L^2H}\right)
  		+\tr{L}\left(- \tr{L^3}+\tr{H^2L} \right) +(\tr{HL})^2\bigg)\\
      +& 12  \bigg((\tr{H^2})^2+(\tr{L^2})^2\bigg) - 8 \tr{H^2}\tr{L^2} \bigg\} \;.
\end{align*}
This form is preferable for implementation on the computer, since the matrices $H,L$ are $N \times N$ matrices, while the Dirac operator in this case is a $ 2 N^2 \times 2 N^2$ matrix.

Our choice of action is markedly different from the Connes-Chamseddine spectral action \cite{chamseddine_spectral_1997}.
The Connes-Chamseddine spectral action is defined as
\begin{align}
  S(D,\Lambda) = \Tr{\chi\left(\frac{D}{\Lambda}\right)}
\end{align}
with $\chi(x)$ a regularisation function that goes to $0$ for $x>1$, and $\Lambda$ a cut-off scale.
For the almost-commutative standard model, this action recovers the correct standard model action, coupled to the Einstein Hilbert action for gravity.
The problem with this action for our framework is that it does not have a well localised minimum, and thus can not be explored using Monte Carlo simulations.
Exploring the continuum limit of our action, and trying to find if it can also recover the Einstein-Hilbert action is one interesting direction for future research.

In \cite{barrett_monte_2015} we found the phase transition of the $(1,1)$ geometry to lie around $g_2\sim -2$ and for $(2,0)$ around $g_2\sim -3$.
To determine this location we used the splitting point of the distribution of eigenvalues of the Dirac operator, and the autocorrelation time of the term $\Tr{D^2}$.
The type $(1,3)$ geometry was not examined there.
This data was very preliminary, since we only scanned the $g_2$ range with a step size of $0.5$.
To identify the phase transition we took $2000$ measurements, one after every $4\cdot N$ Monte Carlo steps.
In the present paper we try to pin down the location of the phase transition more precisely and to explore the scaling of the geometries, both at the phase transition and away from it.
Of particular importance to analyse the phase transition is the variance of observables, which as a second moment of the distribution requires us to sample the distribution much more extensively than a first moment like the average.
To do this we took $100\,000$ measurements for type $(1,3)$ and $97\,500$ measurements for type $(1,1)$ and $(2,0)$ each.
The $100\,000$ measurements for type $(1,3)$ consisted of $20$ Markov chains of length $5000$ each, these chains all started from the same thermalised configuration but developed independently afterwards.
For type $(1,1)$ and $(2,0)$ we started $5$ Markov chains of length $ 19\,500$ each of which started from a random Dirac operator and went through a thermalisation phase before starting the $19\,500$ steps\footnote{The chains were originally $20\,000$ steps long, however we found that the gods of thermalisation required a further sacrifice of 500 steps per chain.}.
Hence the data for types $(1,1)$ and $(2,0)$ is of better quality, since the $5$ chains are completely independent.
This was not possible for type $(1,3)$ since the thermalisation process for this case took considerably longer.
We also extended the length of our sweeps, now only measuring after every $4 \cdot N^2$ Monte Carlo steps, to reduce the overall correlation in our sample and thus improve our data  while only moderately increasing runtime.

\begin{figure}
\subfloat[][\label{fig:11dataS}$\av{S}$ for type $(1,1)$]{\includegraphics[width=0.49\textwidth]{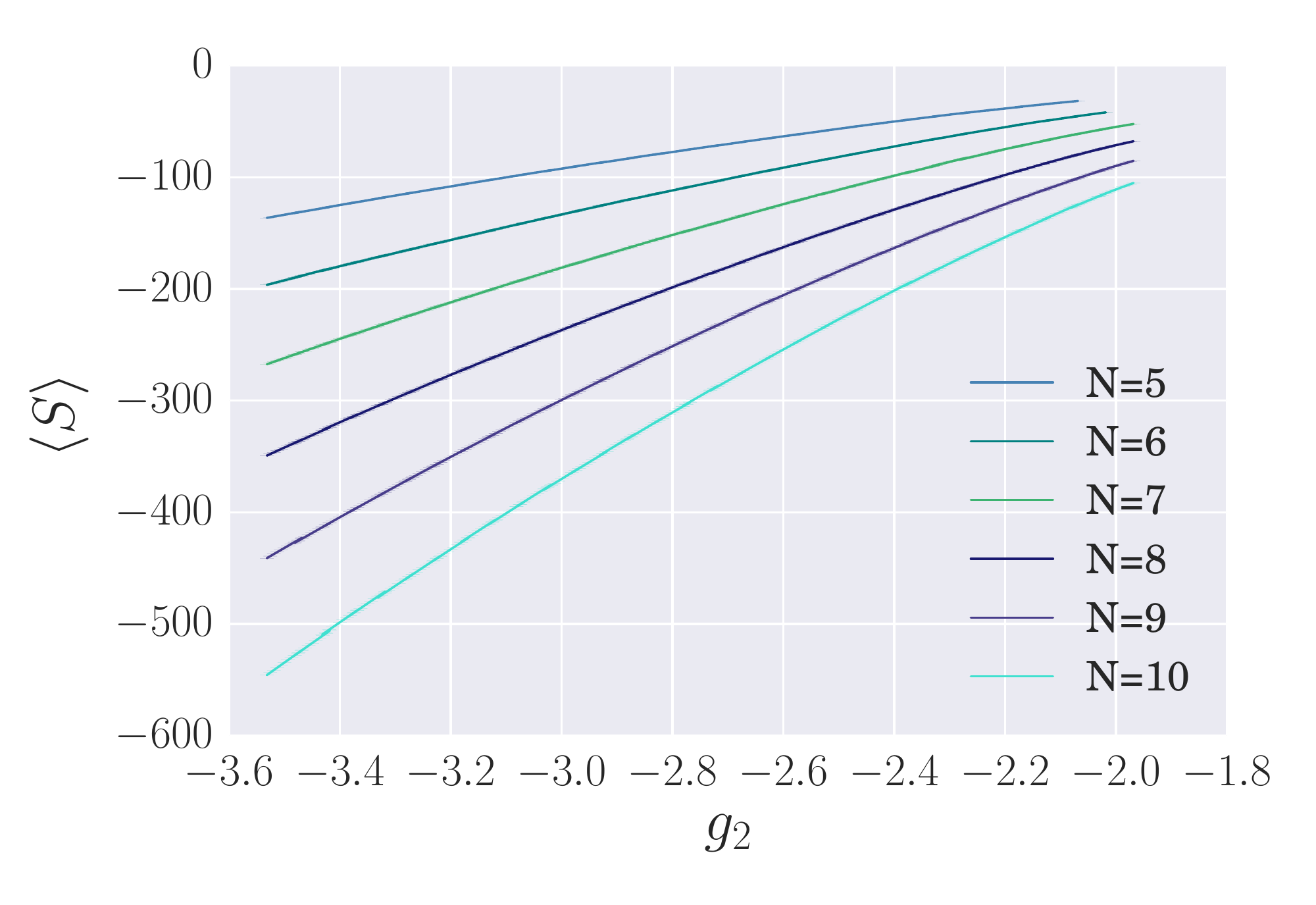}}
\subfloat[][\label{fig:11dataSVar}$\mathrm{Var}(S)$ for type $(1,1)$]{\includegraphics[width=0.49\textwidth]{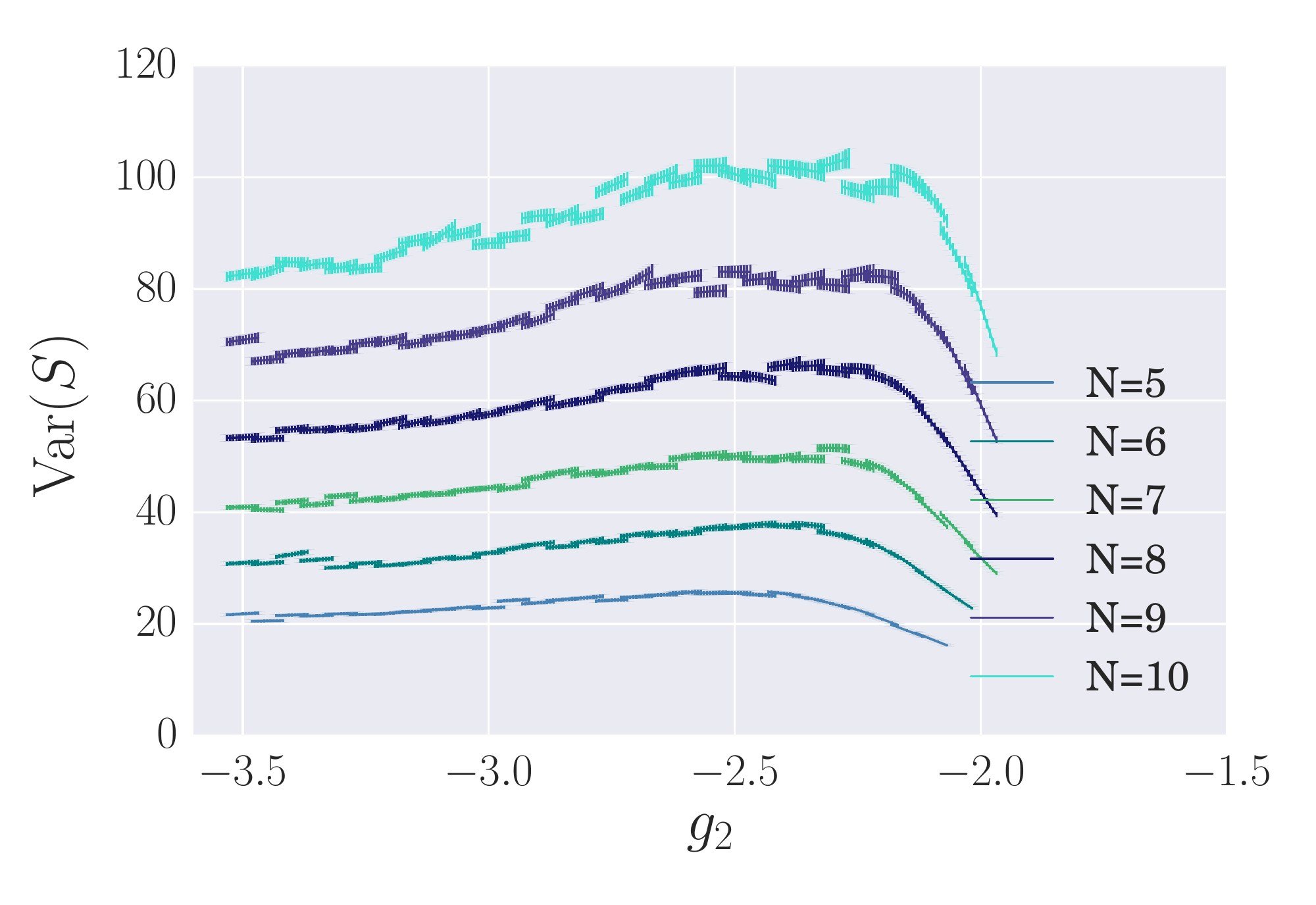}}
\caption{Average action and Variance of the average action plotted against $g_2$ for type $(1,1)$.}
\end{figure}

\begin{figure}
\subfloat[][\label{fig:13dataS}$\av{S}$ for type  $(1,3)$]{\includegraphics[width=0.49\textwidth]{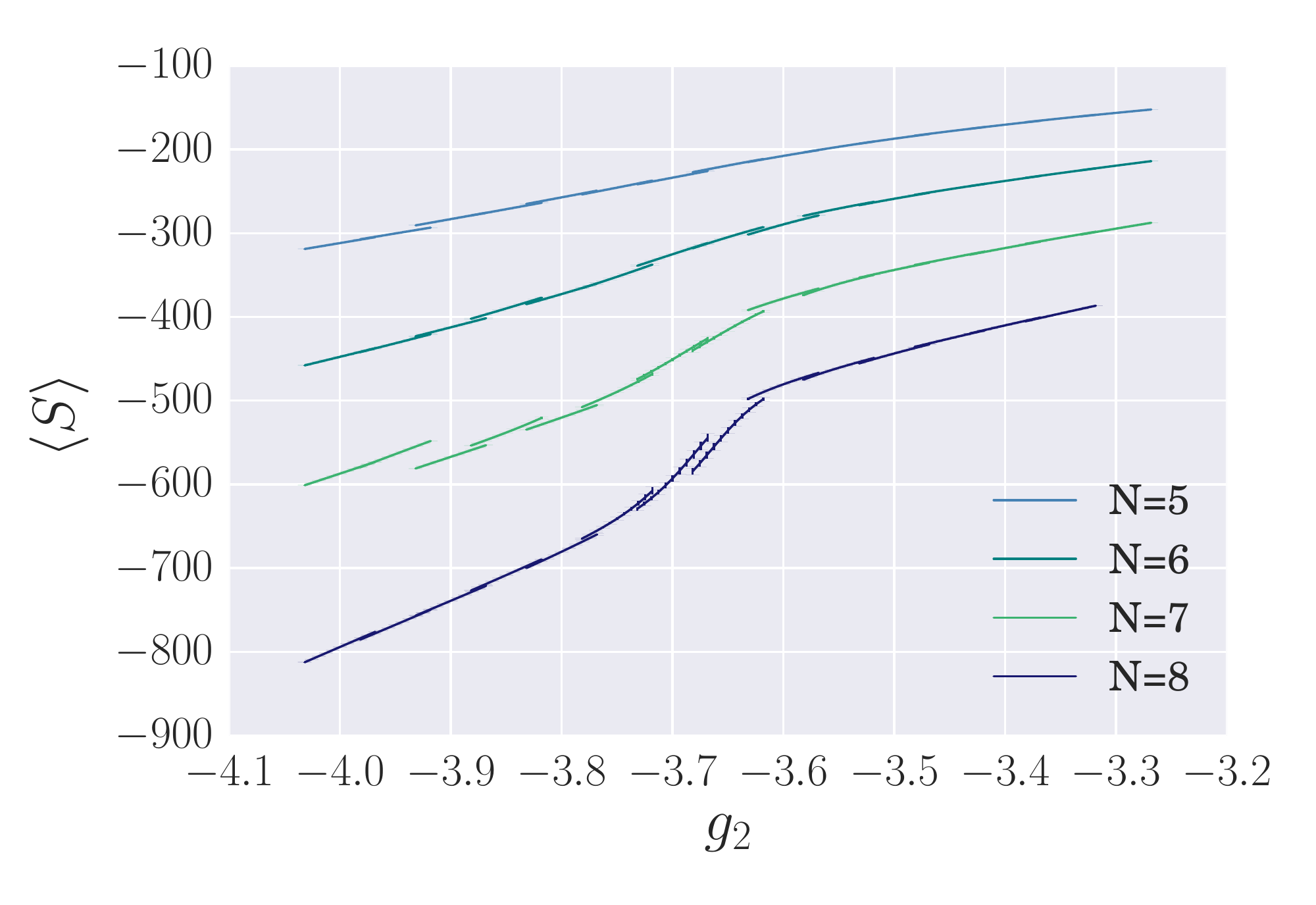}}
\subfloat[][\label{fig:13dataSVar}$\mathrm{Var}(S)$ for type $(1,3)$]{\includegraphics[width=0.49\textwidth]{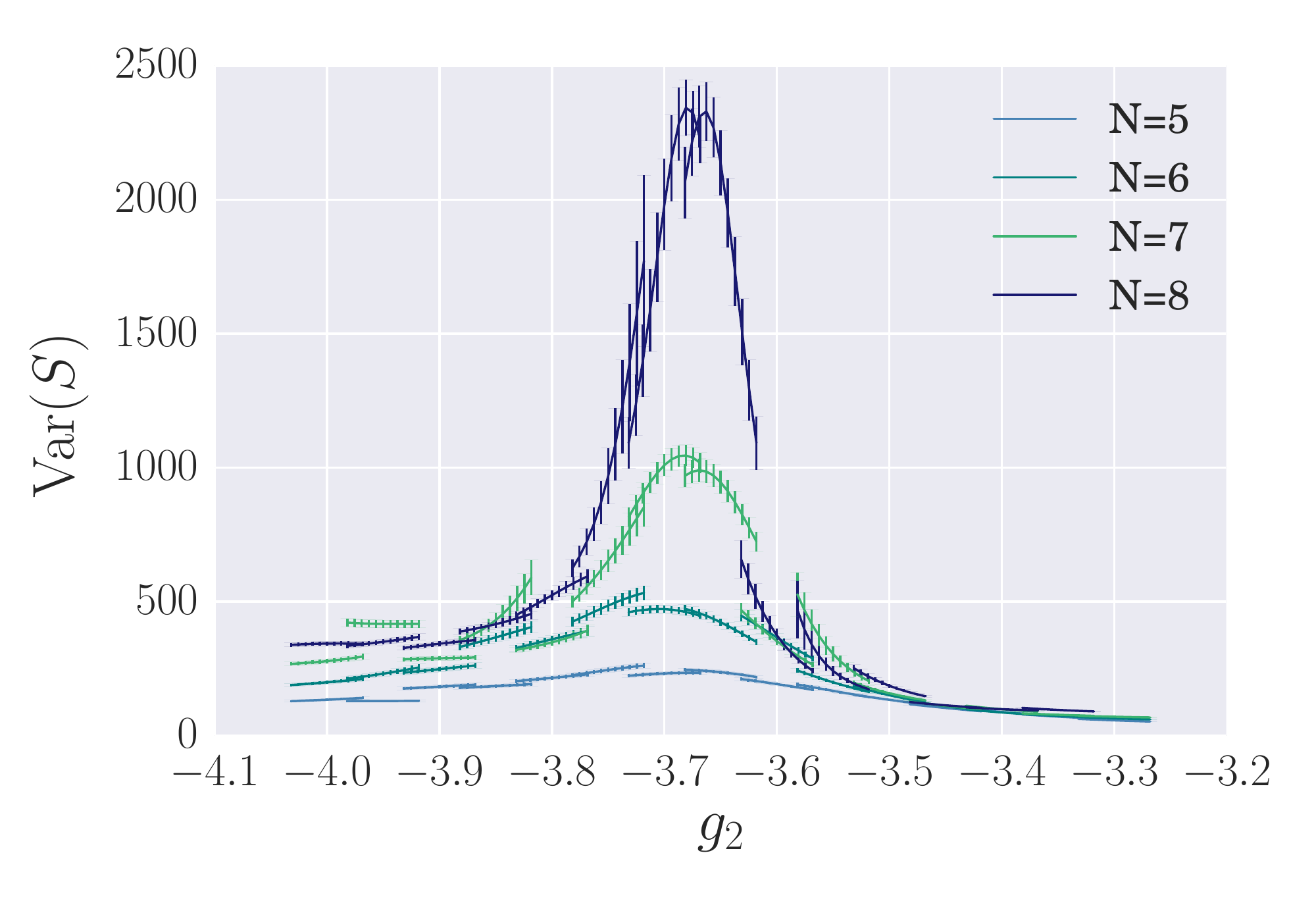}}
\caption{\label{fig:13data} Average action and Variance of the average action  plotted against $g_2$ for type $(1,3)$.}
\end{figure}

To locate the phase transition precisely we chose a range of $g_2$ adjusted to the critical region for each of the geometries, and vary the value with a step size of $0.05$.
For type $(1,1)$ we scan the range $g_2=-3.5,\dots,-2.0$, for type $(2,0)$ $g_2=-3.5,\dots,-2.5$, and for type $(1,3)$ $g_2=-4.0,\dots,-3.3$. All data pertaining to the simulations is summarised in Table \ref{tab:simdata}.
To allow an even more precise location of the phase transition we use a reweighting of the data to interpolate points, as described in \cite[Chapter~8.1]{newman_monte_1999}.
From each measured value of $g_2$ we interpolate $10$ more values, $5$ to either side of the original.
We have arranged these points so that the interpolated ranges for neighbouring points slightly overlap, to serve as a consistency check.
As we can see in, for example, Figure \ref{fig:11dataS} this method works very well for the average observables, while for the variances ( Figure \ref{fig:11dataSVar}) the interpolated regions do not necessarily overlap within their error bars.
This is indicative that the errorbars underestimate the error, by missing systematic sampling biases, and becomes even clearer for type $(1,3)$ (see Figure \ref{fig:13dataS}) in which these gaps are visible even for the average observables.
While this situation is not ideal, the interpolated regions are consistent with each other if we assume a systematic error of the same order of magnitude as the statistical error.
This error could be reduced through more data, however undertaking these additional simulations was beyond the scope of the current project.


To explore the scaling of the theory we conducted simulations at the values $N=5$ to $10$ for types $(1,1)$ and $(2,0)$ and for $N=5$ to $8$ for type $(1,3)$.
While these numbers do not sound impressive, it is important to remember that the Dirac operator is a $n_D=\gamma \cdot N^2$ dimensional matrix, where $\gamma$ is the dimension of the Clifford algebra, $\gamma=2$ for types $(1,1)$ and $(2,0)$ and $\gamma=4$ for type $(1,3)$, hence the matrix models we are observing scale from $n_D=50$ to $n_D=256$.
The original plan was to also simulate type $(1,3)$ at $N=9,10$, however this was reconsidered after not being able to generate thermalised initial configurations within a few months of runtime.
\begin{table}
\caption{\label{tab:simdata}Data about the Monte Carlo simulations}
\begin{tabularx}{\textwidth}{l X X X X X }
\toprule
type $\quad$ & range $g_2$ & step size $g_2 $ & number of samples & number of chains & length of chains\\
\midrule
$(1,1)$ & $-3.5,\dots,-2.0$ & $0.05$ & $97\,500$  & $5$ &$19\,500$\\
$(2,0)$ & $-3.5,\dots,-2.5$ & $0.05$ & $97\,500$  & $5$ &$19\,500$\\
$(1,3)$ & $-4.0,\dots,-3.3$ & $0.05$ & $100\,000$  & $20$ &$5000$\\
\bottomrule
\end{tabularx}
\end{table}

\section{\label{sec:phases}Exploring the phase transition}
\begin{figure}
\includegraphics[width=0.3\textwidth]{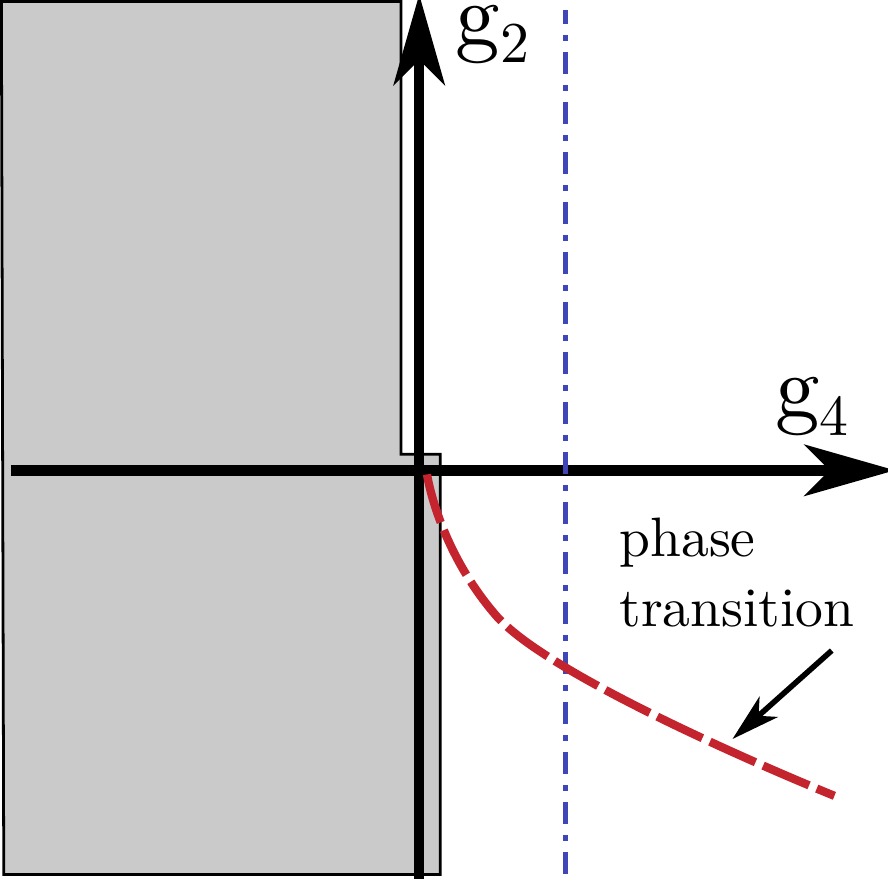}
\caption{\label{fig:phases} Qualitative image of the phase diagram of the theory. In the grey region the integral does not exist, the blue line indicates a fixed $g_4$ value, like we chose to explore in our simulations, and the red line is the phase transition line.}
\end{figure}

We use Monte Carlo simulations to explore the thermodynamic partition function of the space of geometries
\begin{align}\label{eq:partition}
Z(\beta, g_2,g_4) = \int \mathcal{D}[D] e^{- \beta S(D,g_2,g_4)}
\end{align}
and our action for the Dirac operator,
\begin{align}\label{eq:action}
S(D,g_2,g_4) = g_2 \Tr{D^2} + g_4 \Tr{D^4} \;,
\end{align}
leads to a $2$ dimensional phase diagram as sketched in Figure \ref{fig:phases}.
For $g_4<0$ or $g_4=0, g_2 <0$ the action \eqref{eq:action} is unbounded from below, and the integral does not exist, this region is marked in grey in Figure \ref{fig:phases}.

Our exploration is along the line of $\beta g_4=1$ (the blue, dash dotted line in the plot), however as argued in \cite{barrett_monte_2015} the Lebesgue measure leads to a scaling symmetry of the action and the Dirac operator and hence the critical point scales as well.
For any $g_2, g_4$ we can rescale $D$ with $g_4^{1/4}$ and $g_2 \to g_2/g_4^{1/2}$ without changing the physical system, which leads to the dashed red phase transition line.
The scaling symmetry implies that the phase transition line should end at $g_2,g_4=0$, however this point is outside the region we can explore.
The symmetry also means that the true phase diagram of our theory is $1$ dimensional, however the $2$ dimensional representation is helpful to compare to other theories of quantum gravity, in which such a symmetry of the measure does not exist, and the phase diagram studied remains $2$ dimensional.

To explore the phase transition we can look at the following observables
\begin{align}
\av{S} &= -\pd{\log Z}{\beta} & \mathrm{Var} (S) &= \pd{^2\log Z}{\beta ^2} \\
\beta \av{\Tr{D^2}} &= -\pd{\log Z}{g_2} & \beta^2 \mathrm{Var} (\Tr{D^2}) &= \pd{^2 \log Z}{g_2^2} \;. \label{eq:TrD2}
\end{align}
The inverse temperature $\beta$ is convenient for the theoretical analysis of the system, however with $g_2,g_4$ as independent couplings it is redundant, and for our analysis we set it to $1$.
Because of the scaling symmetry we can also fix $g_4 =1$ for our simulations and only vary $g_2$.
The conjugate variable to $g_2$ is $\av{\Tr{D^2}}$ which we would expect it to show the clearest signs of the phase transition.

\subsection{Location of the phase transition}
\begin{figure}
\subfloat[][$(1,1)$]{\includegraphics[width=0.33\textwidth]{type11/combined_VarS_type_11.pdf}}
\subfloat[][$(2,0)$]{\includegraphics[width=0.33\textwidth]{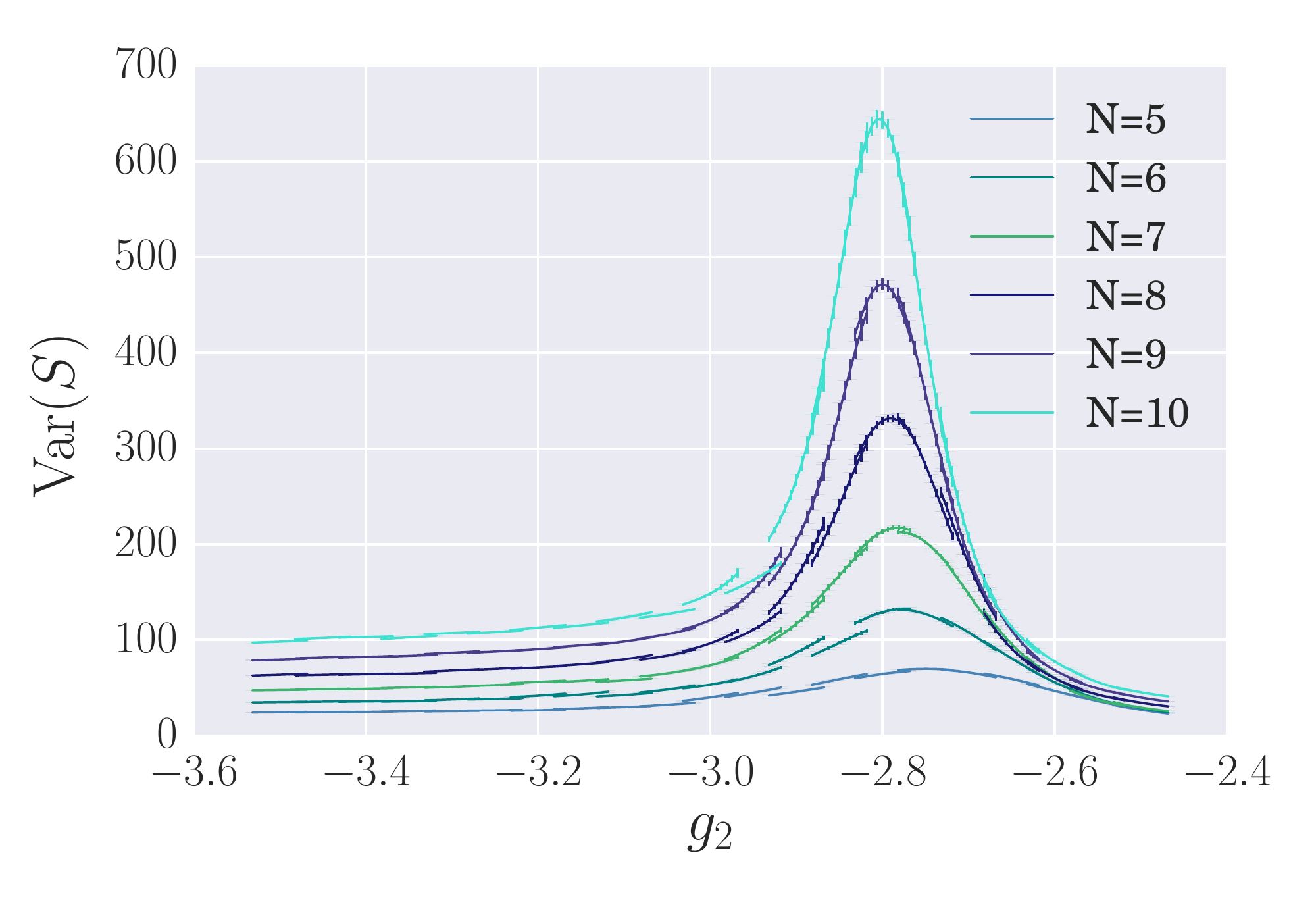}}
\subfloat[][$(1,3)$]{\includegraphics[width=0.33\textwidth]{type13/combined_VarS.pdf}}
\caption{\label{fig:VarS}$\mathrm{Var}(S)$ peaks around the pseudo-critical point.}
\end{figure}

\begin{figure}
\subfloat[][\label{fig:gctype11}$(1,1)$]{\includegraphics[width=0.5\textwidth]{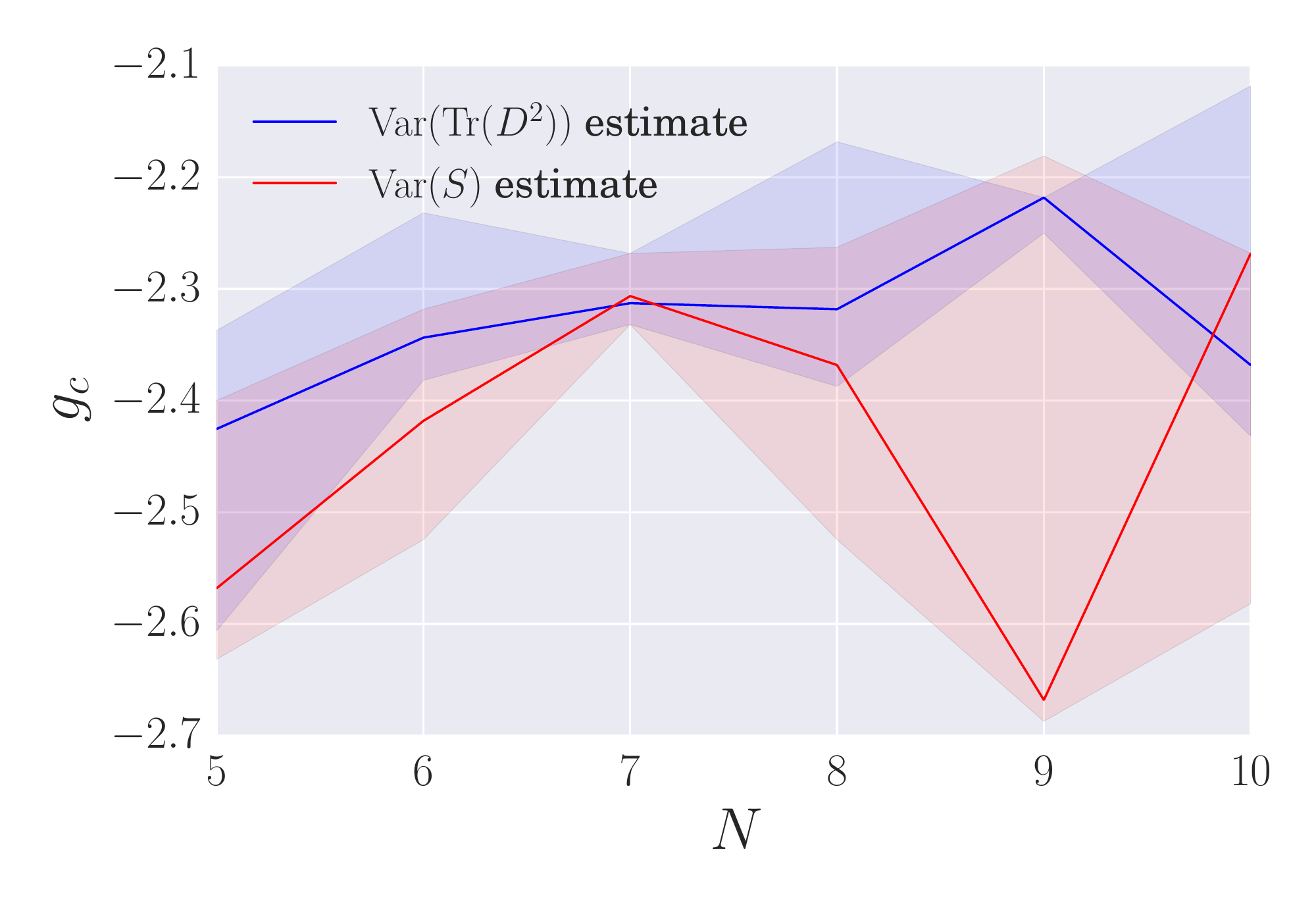}}
\subfloat[][\label{fig:gctype20}$(2,0)$]{\includegraphics[width=0.5\textwidth]{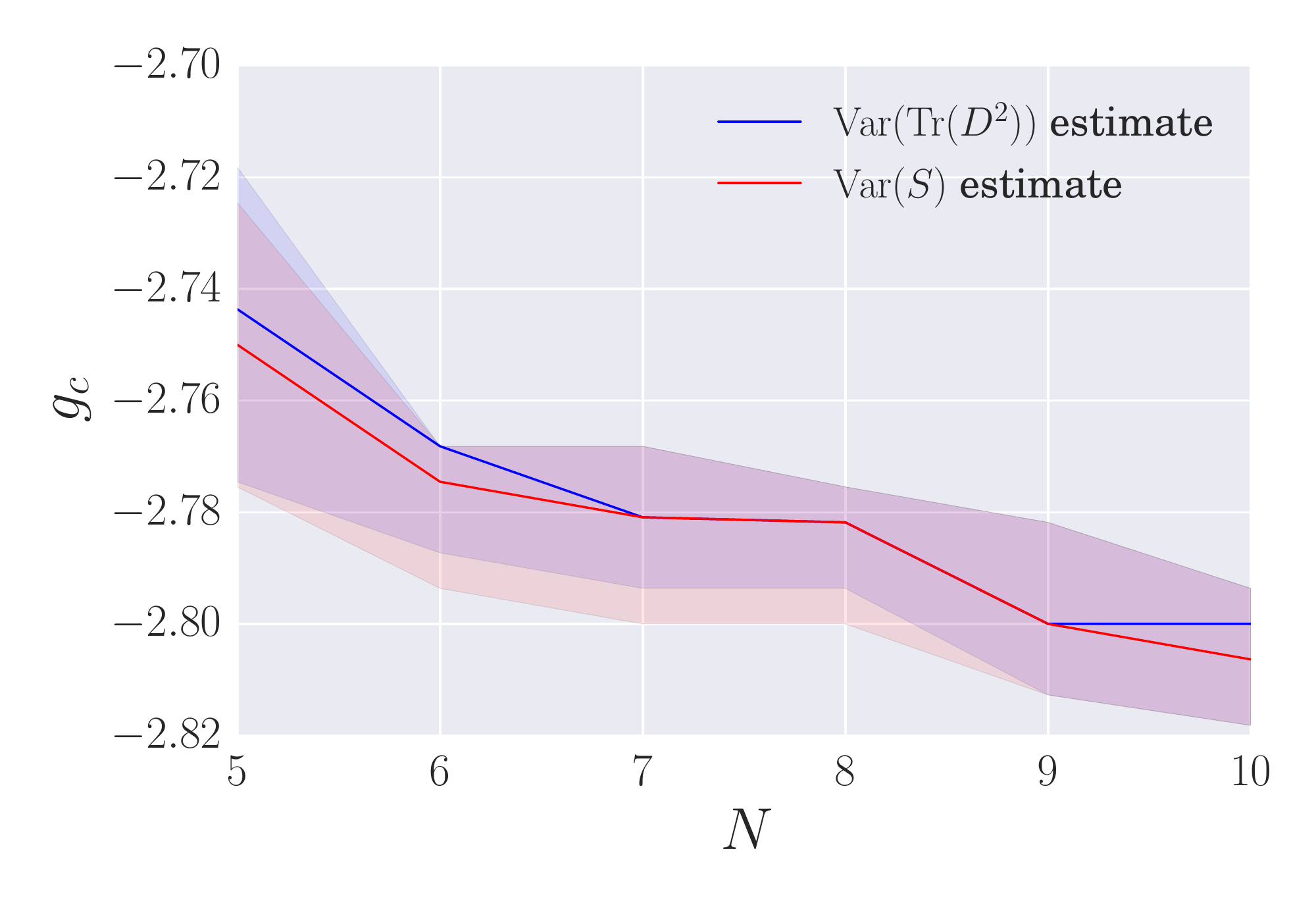}}

\subfloat[][\label{fig:gctype13}$(1,3)$]{\includegraphics[width=0.5\textwidth]{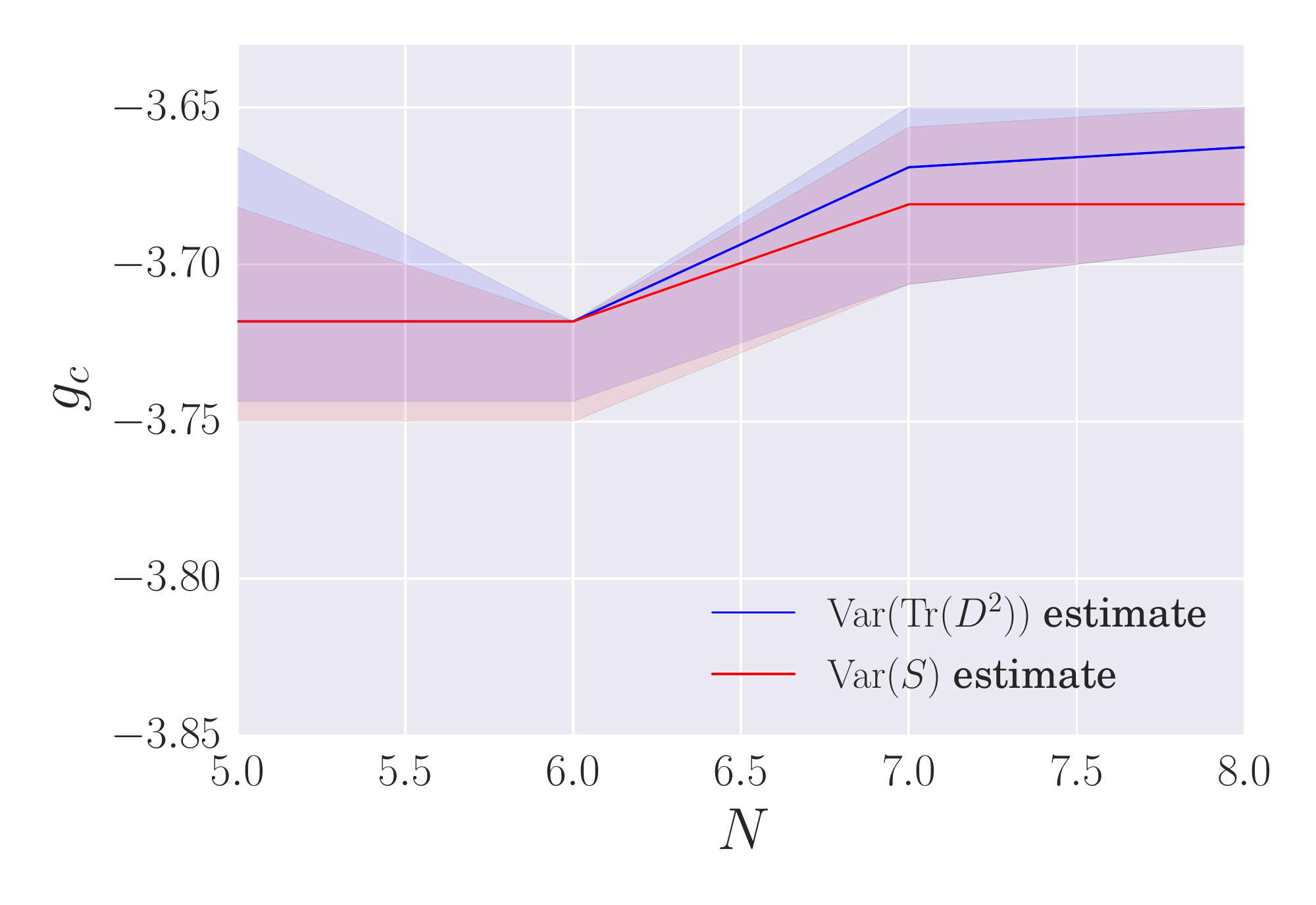}}
\caption{\label{fig:gcplot}Critical values of $g_2$ determined from $S$ and $\Tr{D^2}$ in comparison, the shaded regions indicate the range of uncertainty on the determined values.}
\end{figure}

Our first task is to determine the location of the phase transition more precisely.
Since we are working at finite system size the value of $g_2$ we are looking at is a pseudo-critical point, and not a phase transition in the strict sense.
To determine this pseudo-critical point we use the variance of both the action and the $\Tr{D^2}$ term.
For a second or higher order phase transition these should diverge at the critical point, in the infinite system size limit.
At finite system size they should still show a peak at the pseudo-critical point $g_c$, as visible in Figure \ref{fig:VarS}, and this peak should move closer towards the true phase transition point as the system size increases.
To determine the pseudo-critical value of $g_2$ and the error on it we find the maximal values of the variance of $\Tr{D^2}$ and $S$ for each given $N$.
We then determine the uncertainty region, shown shaded in Figure \ref{fig:gcplot} by determining the maximal and minimal values of $g_2$ for which the measured value of the variance and this maximal value overlap within their errorbars.
Since we know that these errorbars underestimate the real error this uncertainty region does not directly correspond to the error, which is why we decided to list the pseudo-critical points without errors, they are collected in Table \ref{tab:criticalnum}.
\begin{table}
\caption{\label{tab:criticalnum}Table of critical points for all three types}
\begin{tabularx}{\textwidth}{X X X X X X X X}
\toprule
type & & $5$& $6$& $7$ &$8$ &$9$&$10$ \\
\midrule
\multirow{2}{*}{$(1,1)$}& $\Tr{D^2}$&$-2.425$&$-2.344$&$-2.313$&$-2.318$&$-2.218$&$-2.368$\\
& $S$ &$-2.568$&$-2.418$&$-2.306$&$-2.368$&$-2.668$&$-2.268$\\
\multirow{2}{*}{$(2,0)$}& $\Tr{D^2}$&$-2.744$&$-2.768$&$-2.781$&$-2.782$&$-2.800$&$-2.800$\\
& $S$ &$-2.750$&$-2.775$&$-2.781$&$-2.782$&$-2.800$&$-2.806$\\
\multirow{2}{*}{$(1,3)$} & $\Tr{D^2}$&$-3.718$&$-3.718$&$-3.669$&$-3.663$& - & -\\
& $S$ &$-3.718$&$-3.718$&$-3.681$&$-3.681$& -&-\\
\bottomrule
\end{tabularx}
\end{table}
The difference between the pseudo-critical coupling determined from these two observables is minimal, as also illustrated in Figure \ref{fig:gcplot}.
In principle it is possible for different observables to have different pseudo-critical points, however we find that the pseudo-critical points for the action and $\Tr{D^2}$ agree within the uncertainty region, which allows us to average over both to improve our estimate.
While the exact location of the pseudo-critical point is usually $N$ dependent we are not able to determine this from our data set.
One can detect some drift of $g_c$ for the type $(2,0)$ geometry in Figure \ref{fig:gctype20}, however it is slight, and the other two cases fluctuate too much to make such a claim.
We hence decided to average the pseudo-critical values of $g_2$ for $N$ to determine a critical value $g_c$ for each of the geometries, and find $g^{(1,1)}_{c}=-2.382\pm  0.289$, $g^{(2,0)}_c =-2.781\pm  0.289$, and $g^{(1,3)}_c =-3.696\pm  0.354$.

\subsection{Order of the phase transition}
Since the Dirac operator is fundamentally a complicated observable on a complicated matrix model (the $(1,1)$ and $(2,0)$ cases are two-matrix models with up to quartic couplings and the $(1,3)$ case is an eight-matrix model with up to quartic couplings), we expect the phase transition to be higher order, like the phase transition in ordinary matrix models~\cite{kazakov_exact_1986}.

It is worthwhile to try and confirm this expectation, for this we histogrammed the action and $\Tr{D^2}$ normalised by $N^2$, which is proportional to the size of the Dirac operator and hence the number of terms that contributes to a trace.
This normalisation is necessary to plot the observables `per degree of freedom', which should be comparable at different sizes of the system.
To avoid spurious correlations we determined the autocorrelation time $\tau_{a}$ for each $N,g_2$ pair and only included the action (or $\Tr{D^2}$ respectively) every $\tau_{a}$ steps in the histogram.

For a first order phase transition at the pseudo-critical point the observables should have two separate peaks, arising from the system jumping between the different phases.
With increasing system size these peaks would grow further apart and the jumps would become rarer.
For a second order phase transition the observables should be peaked around a central volume, or if they form two peaks these should become closer as the size of the system increases.

\begin{figure}
\subfloat[][\label{fig:HistCritS11}$(1,1)$ $g_2=-2.4$]{\includegraphics[width=0.49\textwidth]{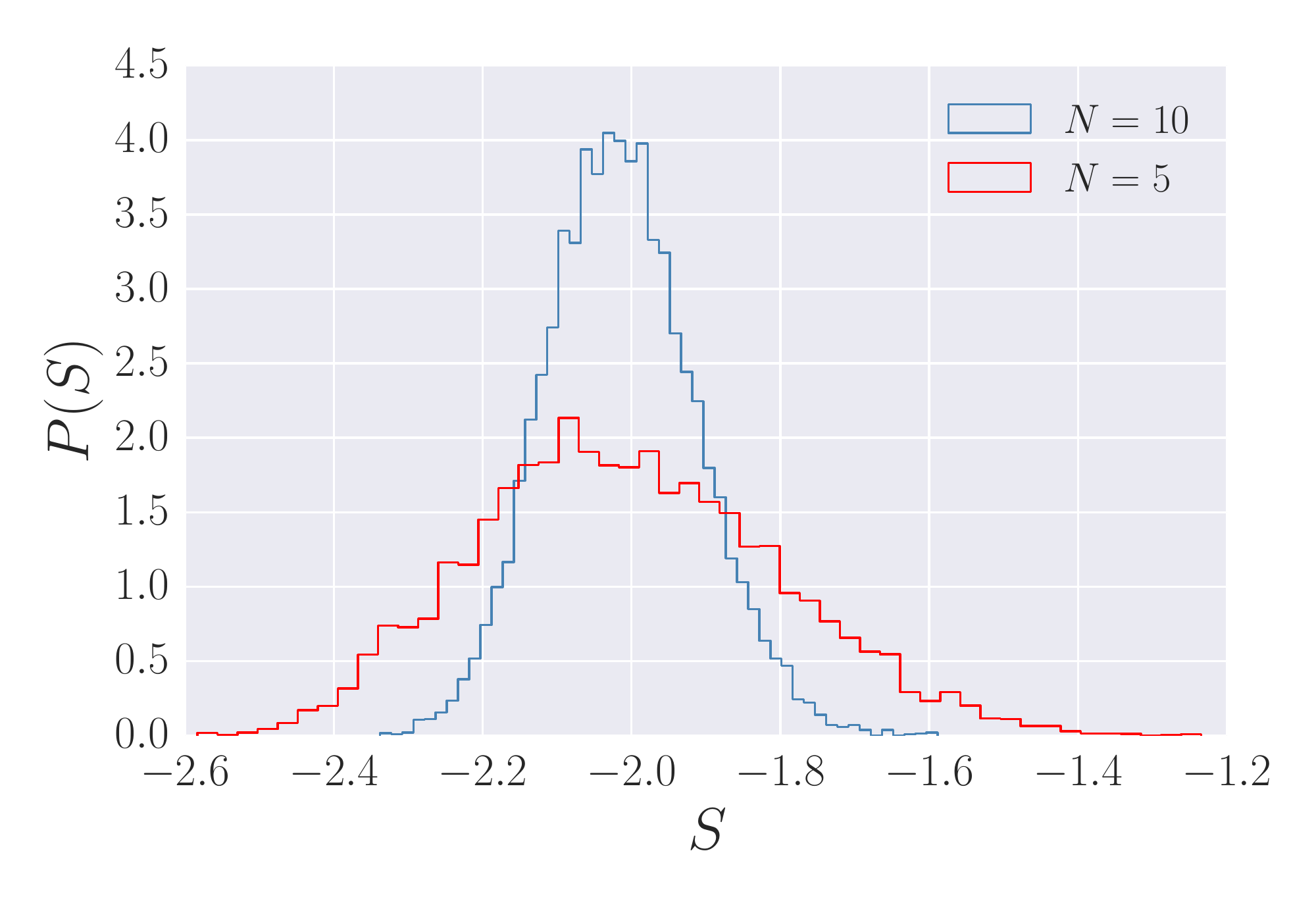}}
\subfloat[][\label{fig:HistCritT11}$(1,1)$ $g_2=-2.4$]{\includegraphics[width=0.49\textwidth]{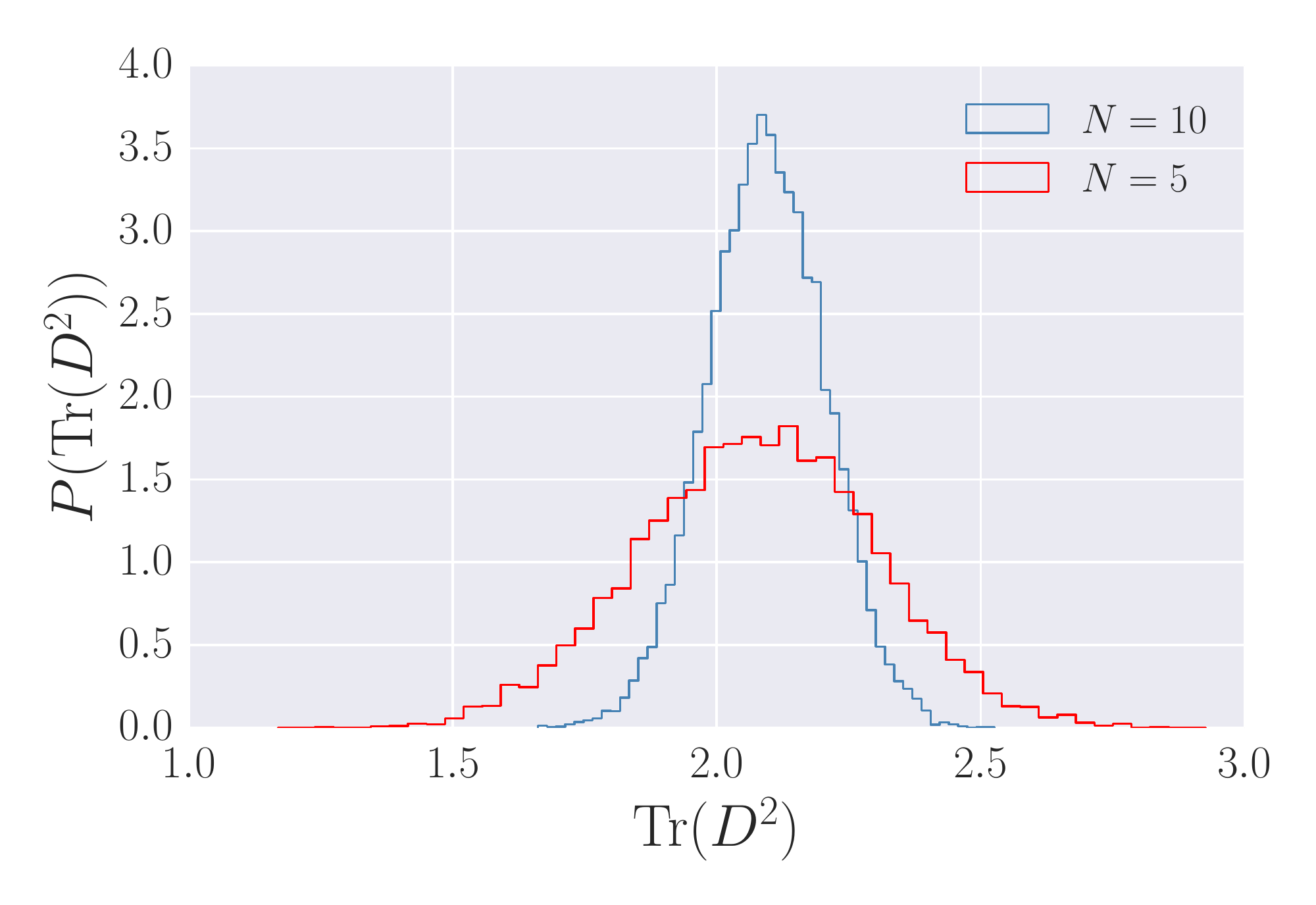}}

\subfloat[][\label{fig:HistCritS20}$(2,0)$ $g_2=-2.8$]{\includegraphics[width=0.49\textwidth]{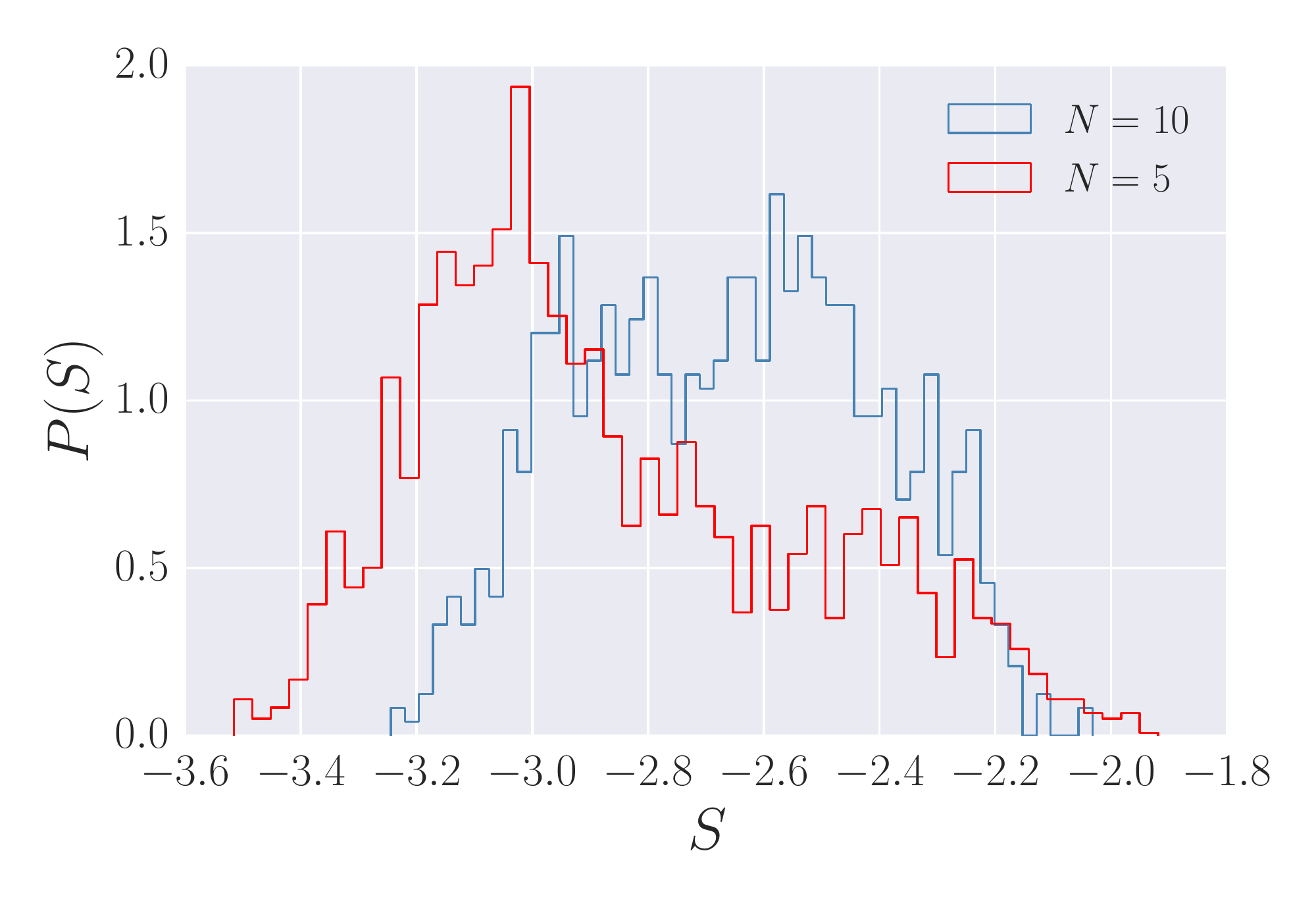}}
\subfloat[][\label{fig:HistCritT20}$(2,0)$ $g_2=-2.8$]{\includegraphics[width=0.49\textwidth]{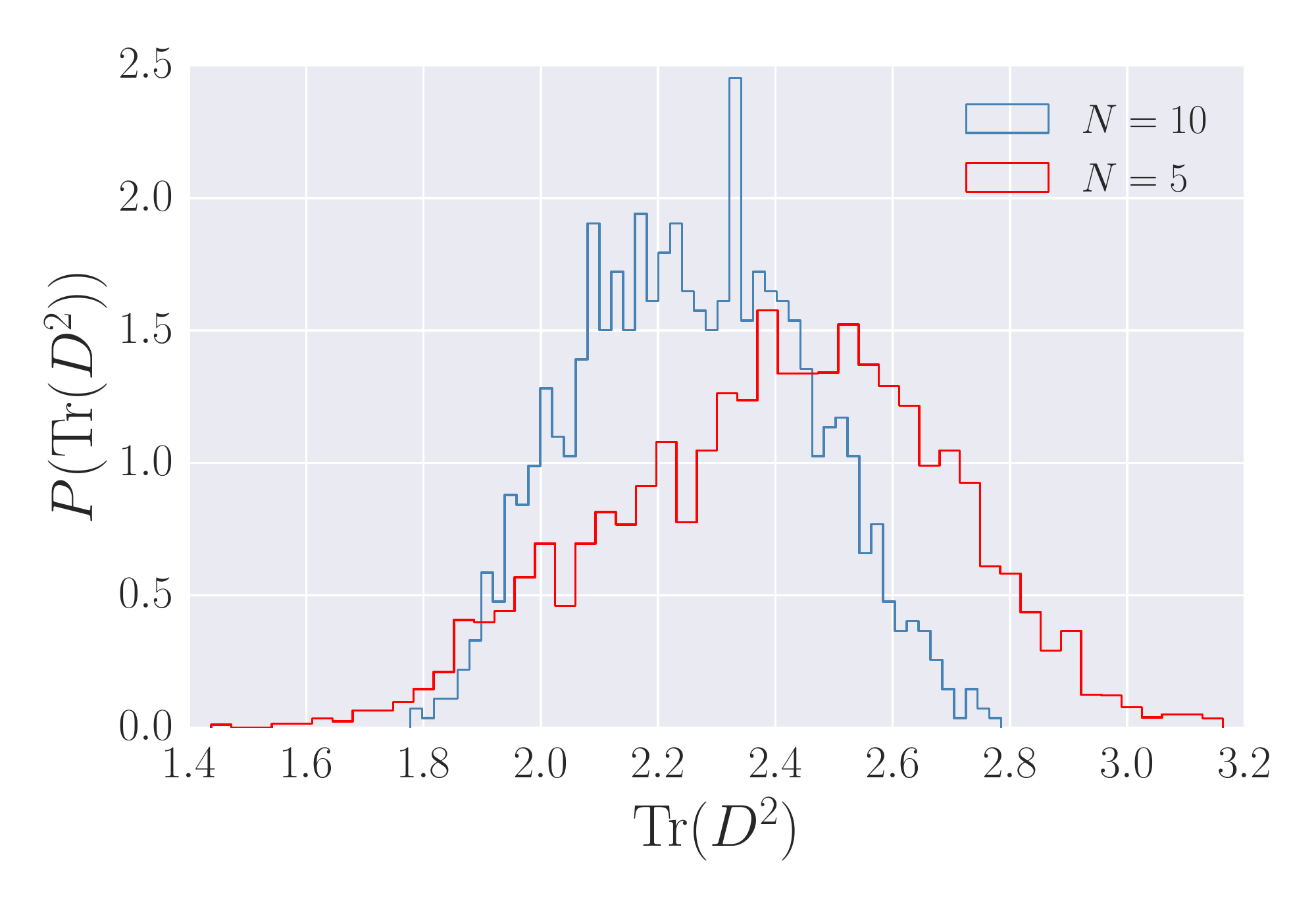}}

\subfloat[][\label{fig:HistCritS13}$(1,3)$ $g_2=-3.7$]{\includegraphics[width=0.49\textwidth]{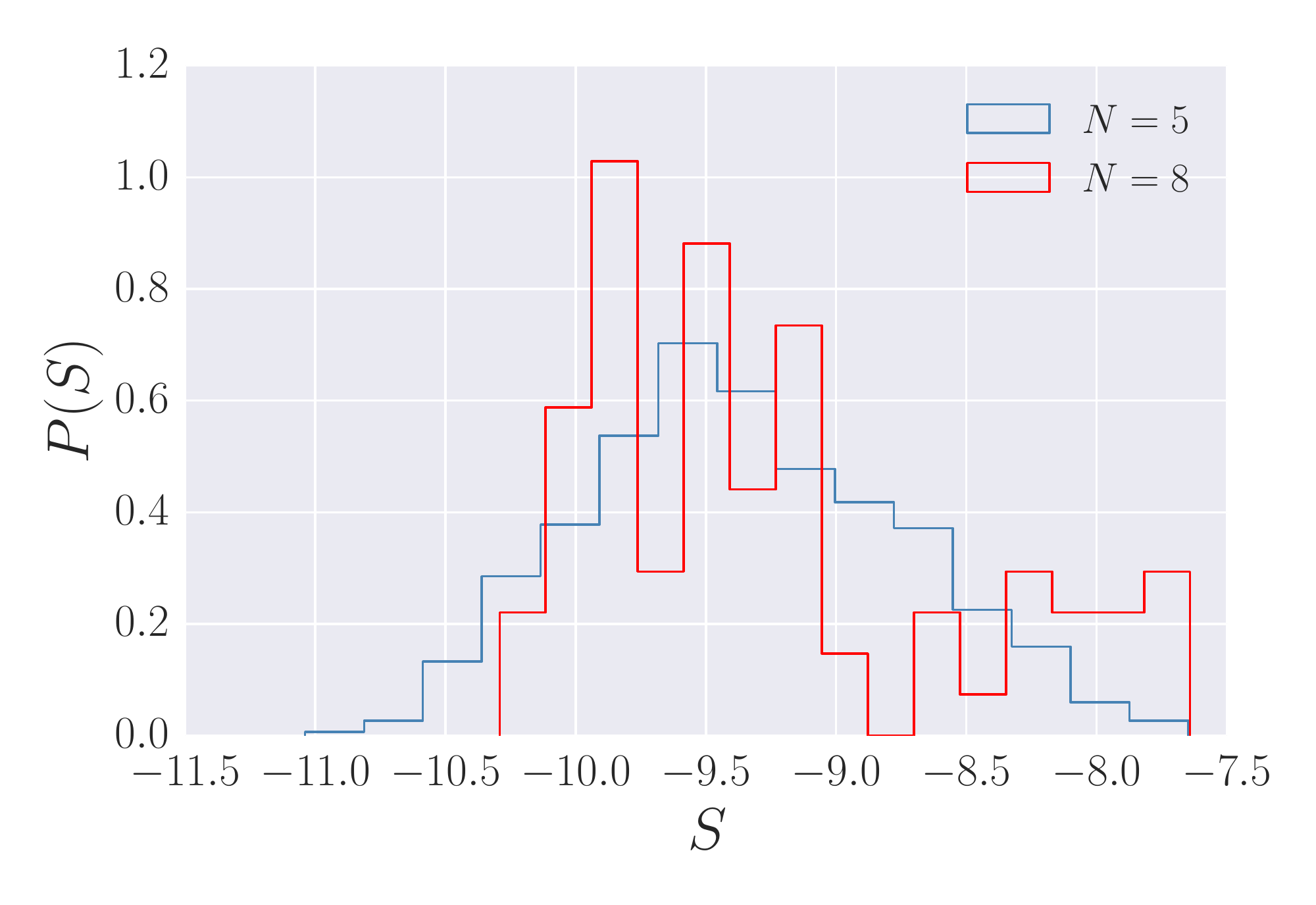}}
\subfloat[][\label{fig:HistCritT13}$(1,3)$ $g_2=-3.7$]{\includegraphics[width=0.49\textwidth]{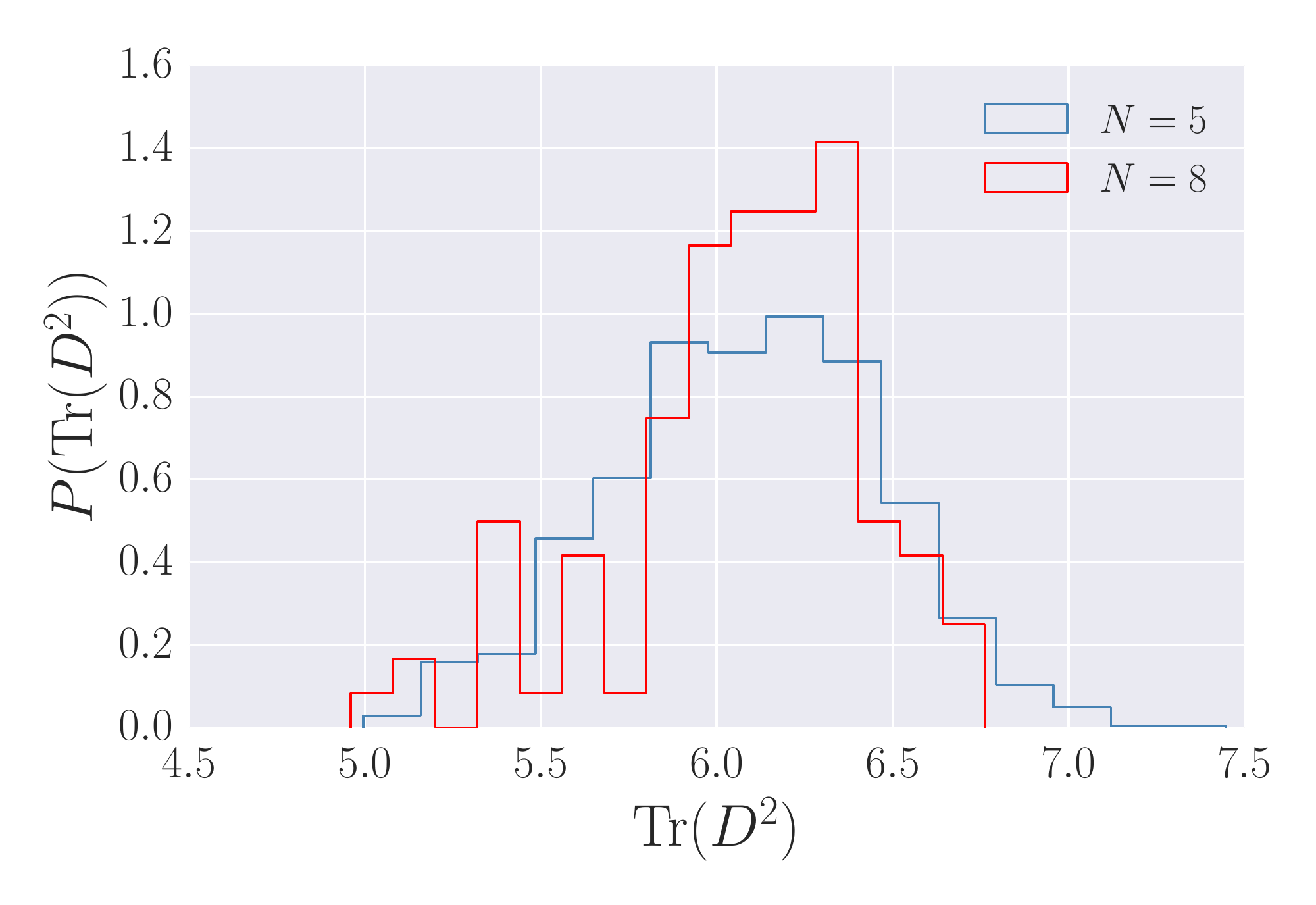}}

\caption{\label{fig:HistCrit} Trying to use the histogram of the probability density of the action or the $\Tr{D^2}$ term to determine the order of the phase transition. For type $(1,1)$ and $(2,0)$ it points towards a $2$nd order transition while it does not work for the data we have available for type $(1,3)$.}
\end{figure}

For type $(1,1)$ (Figures \ref{fig:HistCritS11} and \ref{fig:HistCritT11}) the indication is clearly a $2$nd, or possibly higher order transition.
The data is clearly peaked around a central value and for larger $N$ this peak becomes sharper.

For type $(2,0)$ (Figures \ref{fig:HistCritS20} and \ref{fig:HistCritT20}) we are still reasonably confident that the transition is $2$nd order.
While the distributions for $N=5$ show a slight double peak structure, for $N=10$ this disappears and the spread of the distribution becomes smaller.

The peak structure of type $(1,3)$ is very hard to determine at the phase transition due to the critical slowing down of the simulations.
While we have $200$ or more uncorrelated samples away from the phase transition for $N=8$ at the phase transition we only have $67$ uncorrelated samples.
This makes the histogram to determine the order of the phase transition of limited use, as is shown in Figure \ref{fig:HistCritT13}.
It is then clear that we need to improve our algorithms or use more computing time to be able to study this aspect of the $(1,3)$ geometry further.

\section{\label{sec:scaling}Scaling of the theory}
After having examined the pseudo-critical points and the order of the phase transition we next explore the scaling of the theory.

We expect that the action and the variance grow with the matrix size $N$, a scaling for which we would need to correct when trying to compare observables between simulations at different sizes or when taking a $N\to \infty$ limit.

\begin{figure}
\subfloat[][\label{fig:20dataTrD2scale}$(2,0)$]{\includegraphics[width=0.49\textwidth]{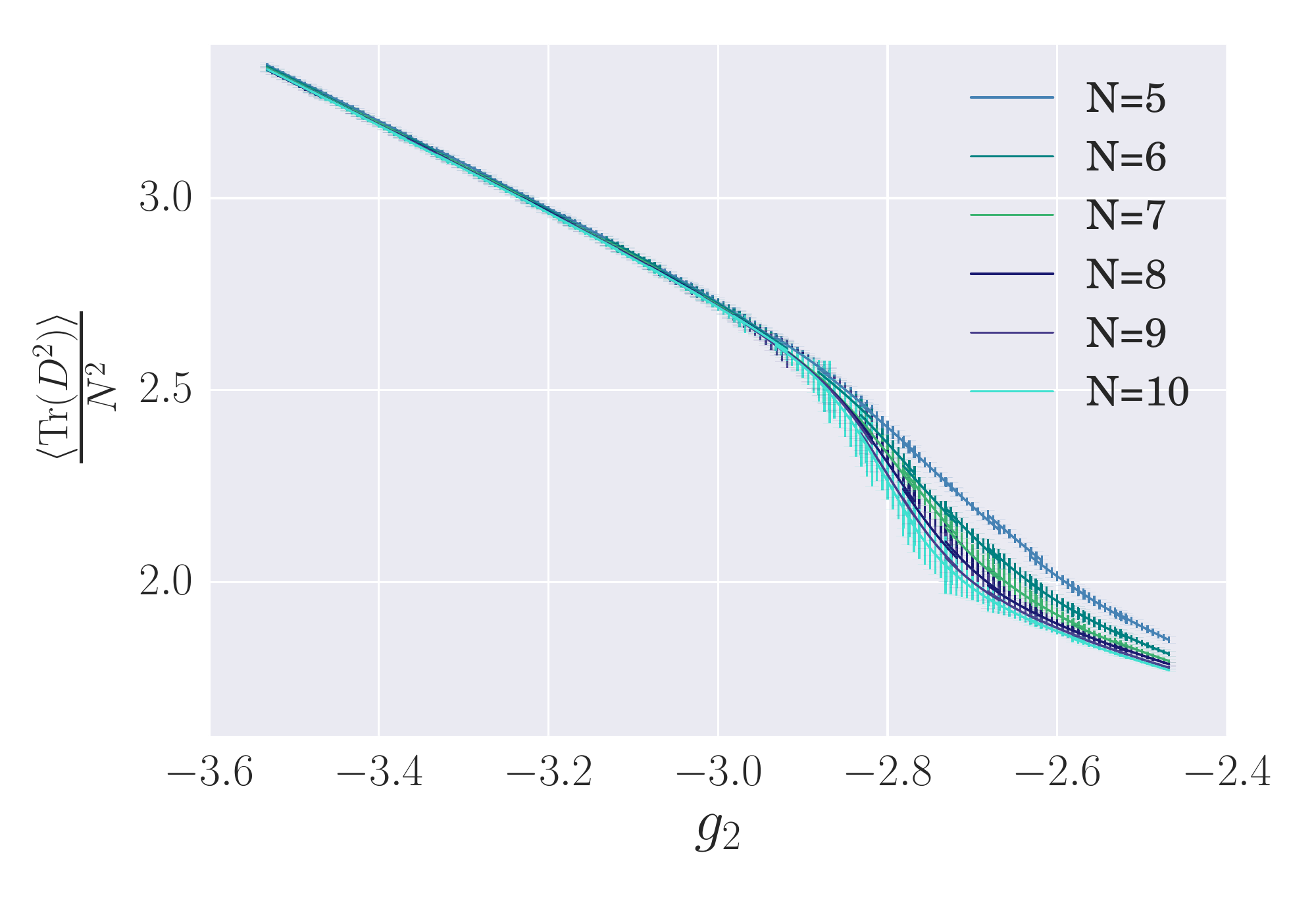}}
\subfloat[][\label{fig:13dataTrD2scale}$(1,3)$]{\includegraphics[width=0.49\textwidth]{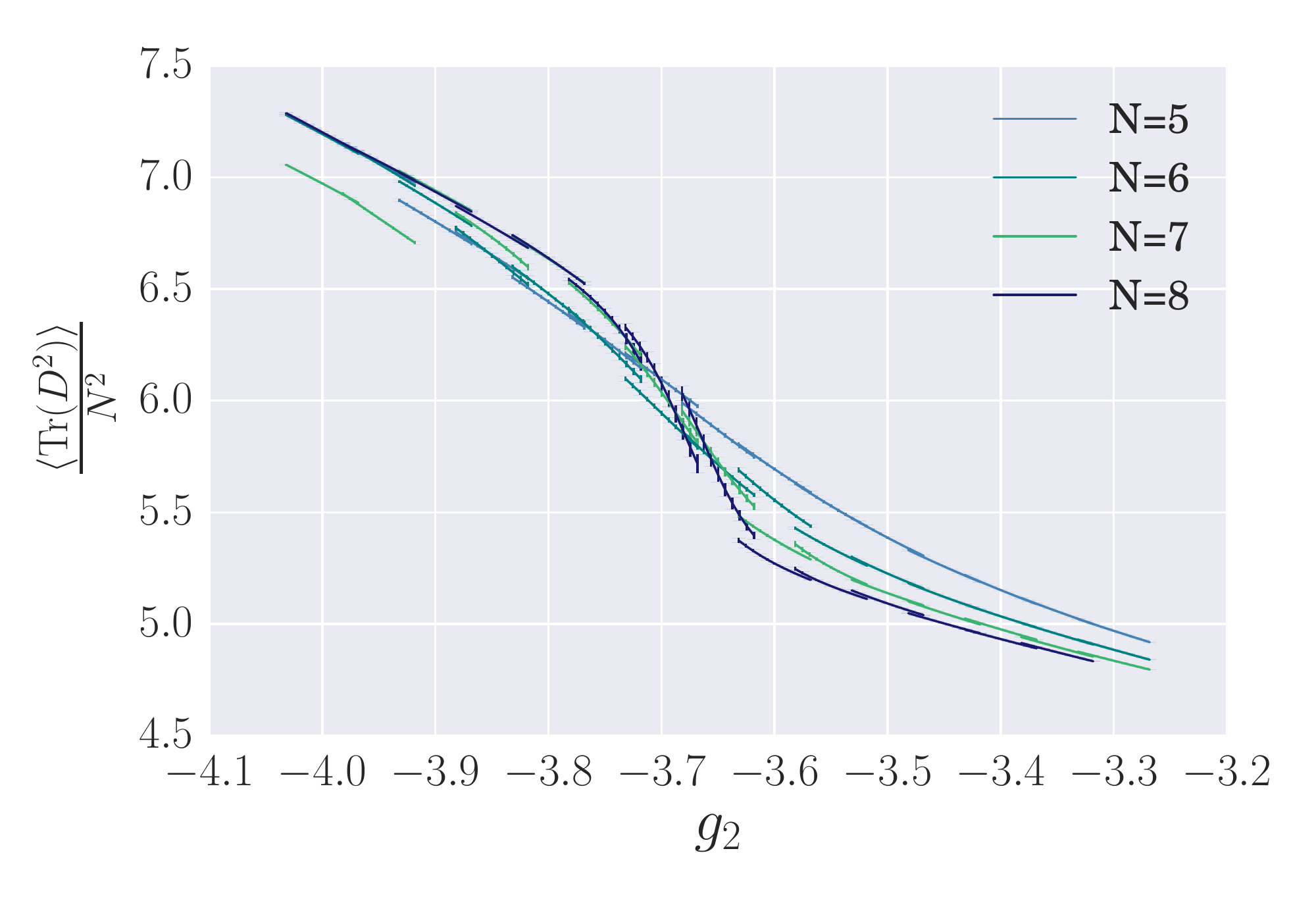}}

\subfloat[][\label{fig:11dataTrD2scale}$(1,1)$]{\includegraphics[width=0.6\textwidth]{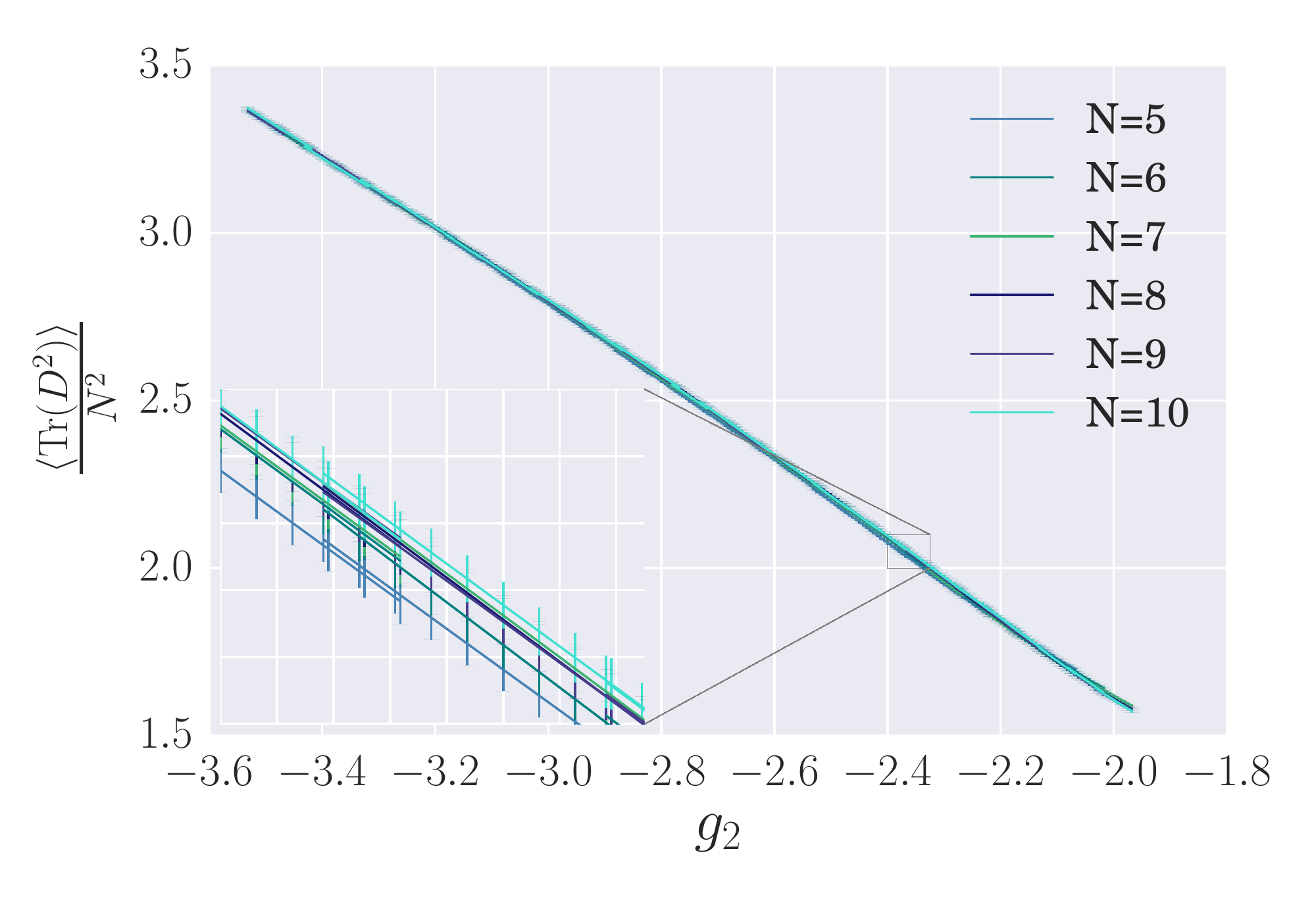}}
\caption{\label{fig:TrD2Scale}Rescaled plot of $\av{\Tr{D^2}}$ for type $(1,1)$, rescaled with $N^2$. For type $(1,1)$ a zoomed in inlay is provided, since the collapse is so good as to make the different lines indistinguishable otherwise.}
\end{figure}

We can predict the behaviour we expect for the scaling of the $\Tr{D^2}$ term and its variance from simple arguments\footnote{A very similar argument can be made for the action, however we do not reproduce it here.}
\begin{align}
\av{\mathrm{Tr}(D^2)}&=\av{ \sum_{i=0}^{n_d} \lambda_i^2} = \sum_{i=0}^{\sim n_d} \av{\lambda_i^2} \sim N^2 \;.
\end{align}
This scaling behaviour is consistent for all three types of geometries we have examined and is shown in Figure \ref{fig:TrD2Scale}.

\begin{figure}
\subfloat[][\label{fig:13dataTrD2scale}$(1,3)$]{\includegraphics[width=0.49\textwidth]{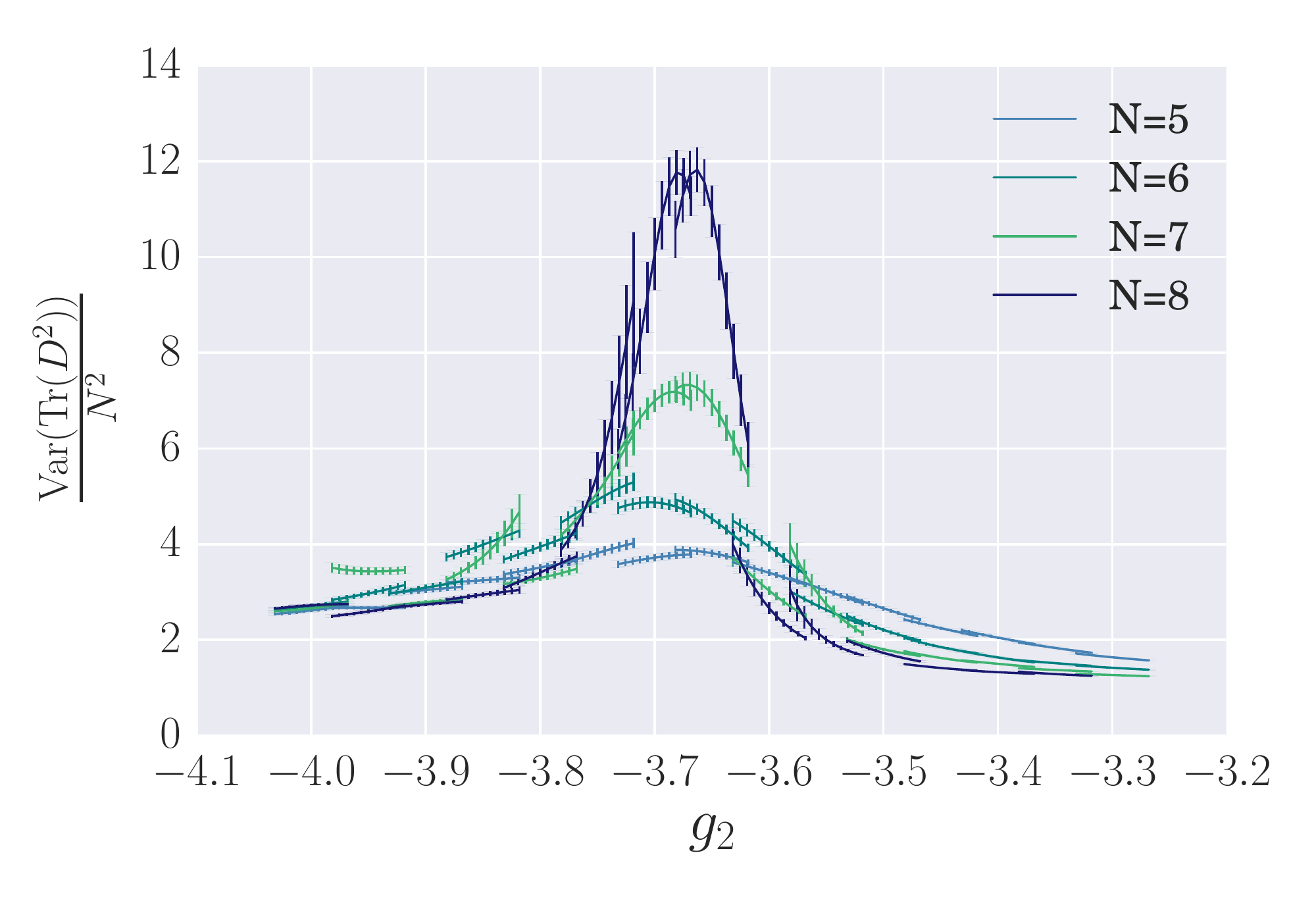}}
\subfloat[][\label{fig:11dataTrD2scale}$(1,1)$]{\includegraphics[width=0.49\textwidth]{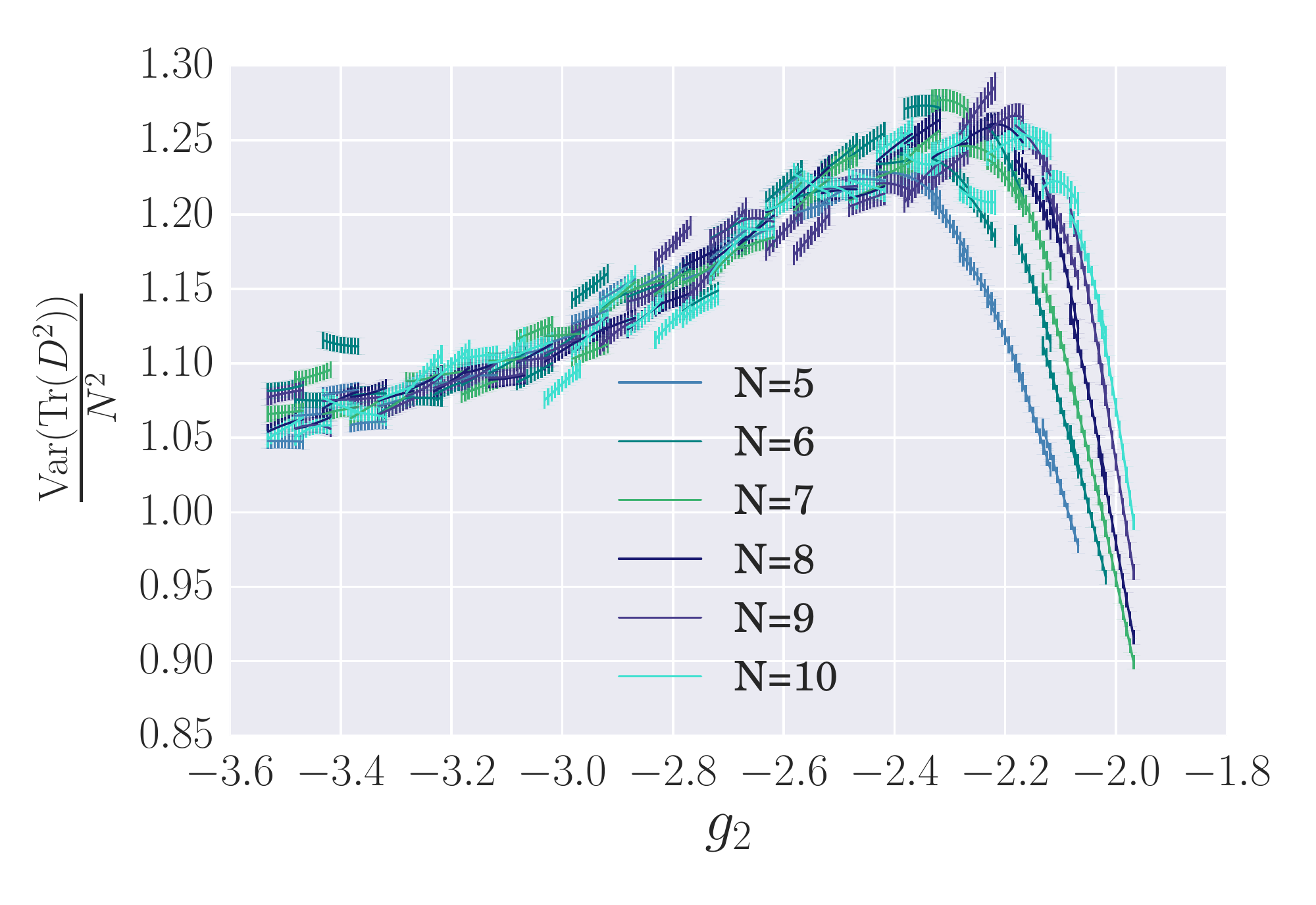}}

\subfloat[][\label{fig:20dataTrD2scale}$(2,0)$]{\includegraphics[width=0.6\textwidth]{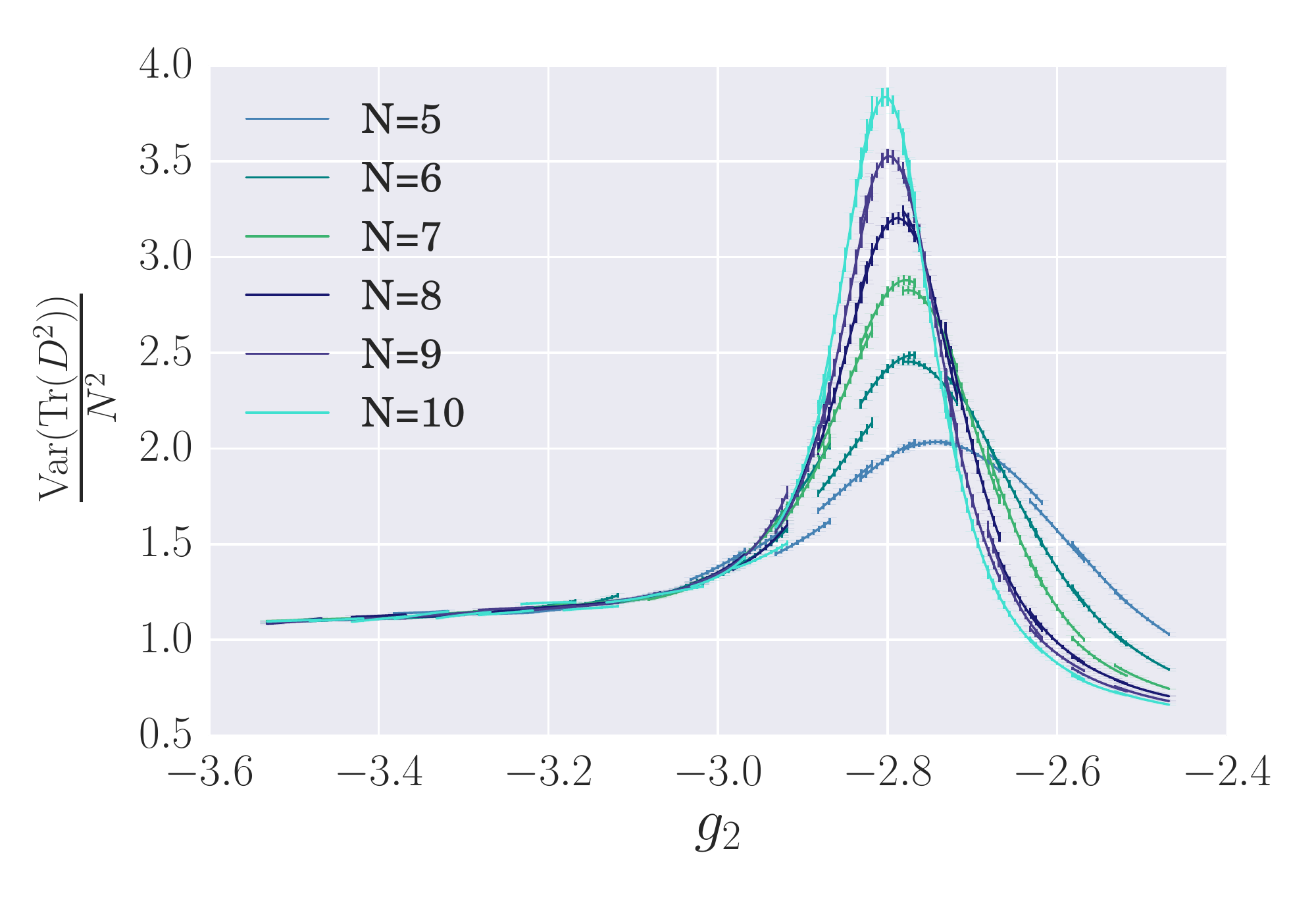}}
\caption{\label{fig:dataVarTrD2scale}Rescaled plot of $\mathrm{Var}{\Tr{D^2}}$ for type $(1,1)$, rescaled with $N^2$.}
\end{figure}

We can then try a similar analysis for the Variance, and find for the $\mathrm{Tr}(D^2)$ term
\begin{align}
\mathrm{Var}(\mathrm{Tr}(D^2))&=\av{\mathrm{Tr}(D^2)^2} - \av{\mathrm{Tr}(D^2)}^2\\
&= \sum_i^{n_d}\sum_j^{n_d} \av{ \lambda_i^2 \lambda_j^2 } - \av{\lambda_i^2} \av{\lambda_j^2} \\
&=\sum_i^{n_d}\sum_j^{n_d} \mathrm{Cov}(\lambda_i^2,\lambda_j^2)\;.
\end{align}
The sums over $i,j$ with $n_d \sim N^2$ terms each would naively lead one to expect a scaling $\sim N^4$.
However we find that, away from the phase transition, the scaling of the variances does follow the $N^2$ law as shown in Figure \ref{fig:dataVarTrD2scale}.

\begin{figure}
\subfloat[][$N=5$ $g_2=-2.00$]{\includegraphics[width=0.32\textwidth]{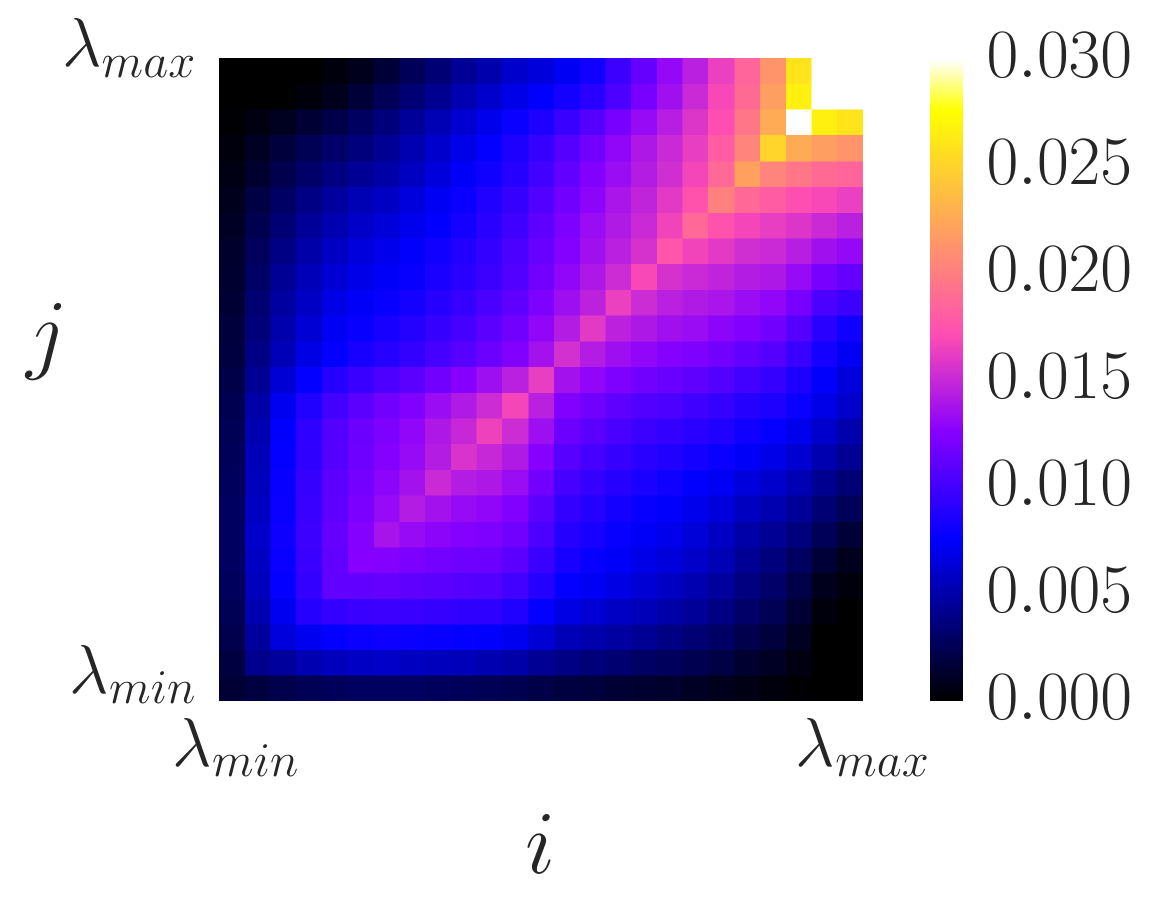}}
\subfloat[][\label{fig:cov2211_gc5}$N=5$ $g_2=-2.40$]{\includegraphics[width=0.32\textwidth]{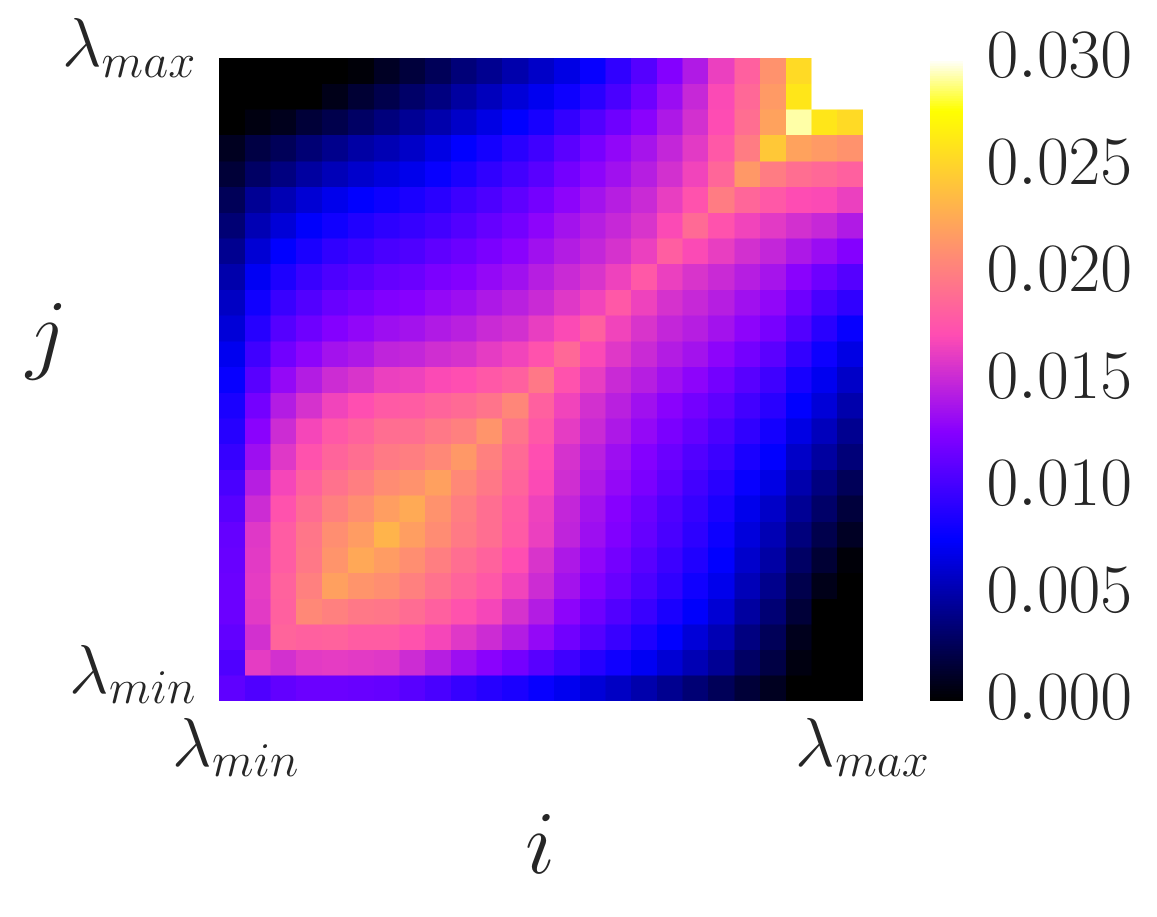}}
\subfloat[][$N=5$ $g_2=-3.50$]{\includegraphics[width=0.32\textwidth]{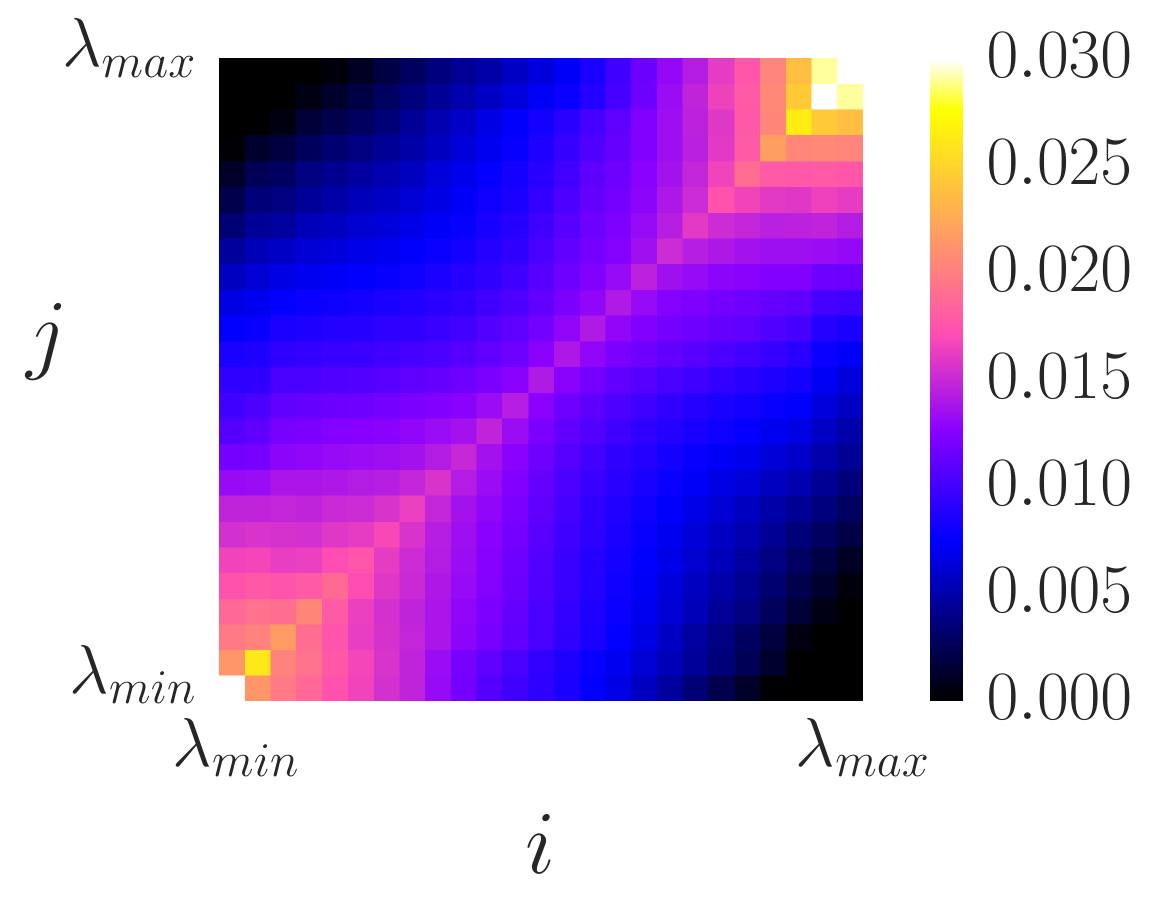}}

\subfloat[][$N=5$ $g_2=-2.00$]{\includegraphics[width=0.32\textwidth]{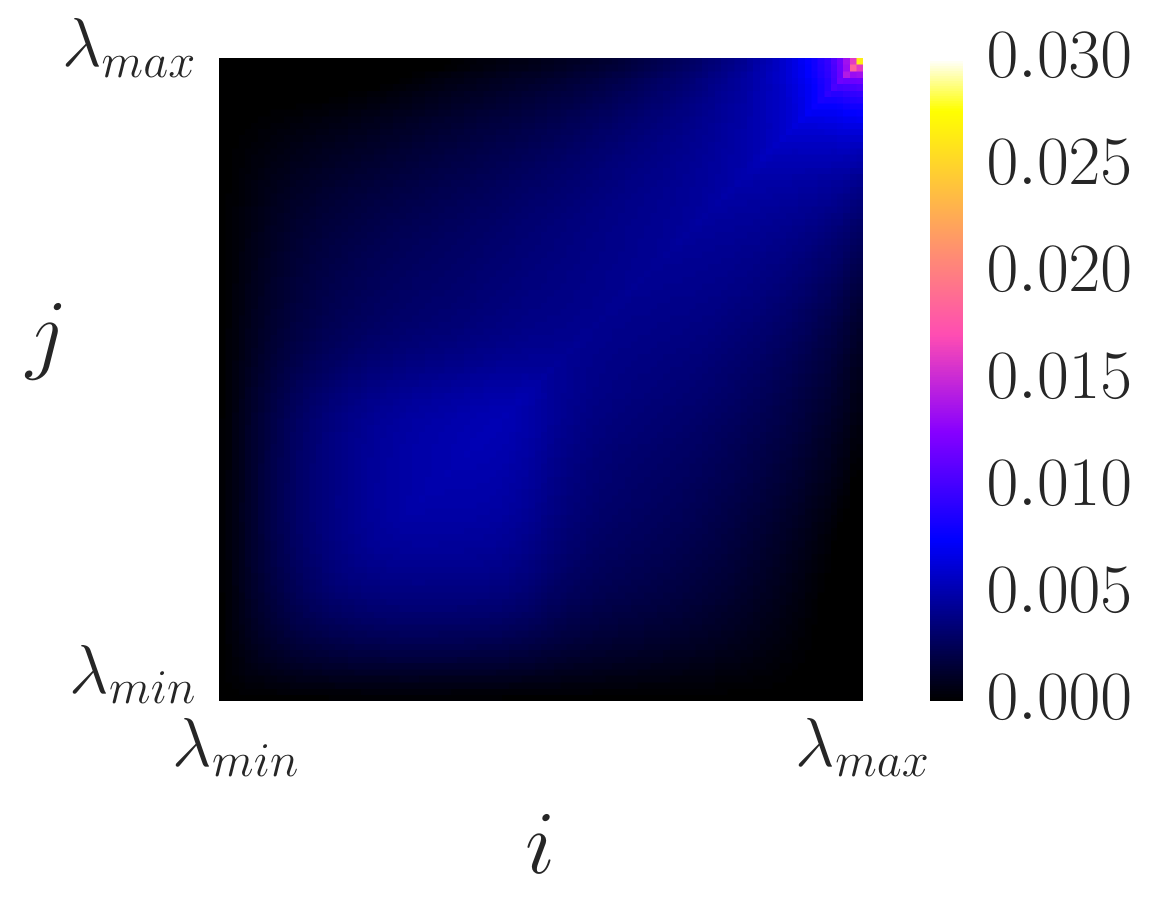}}
\subfloat[][\label{fig:cov2211_gc10}$N=5$ $g_2=-2.40$]{\includegraphics[width=0.32\textwidth]{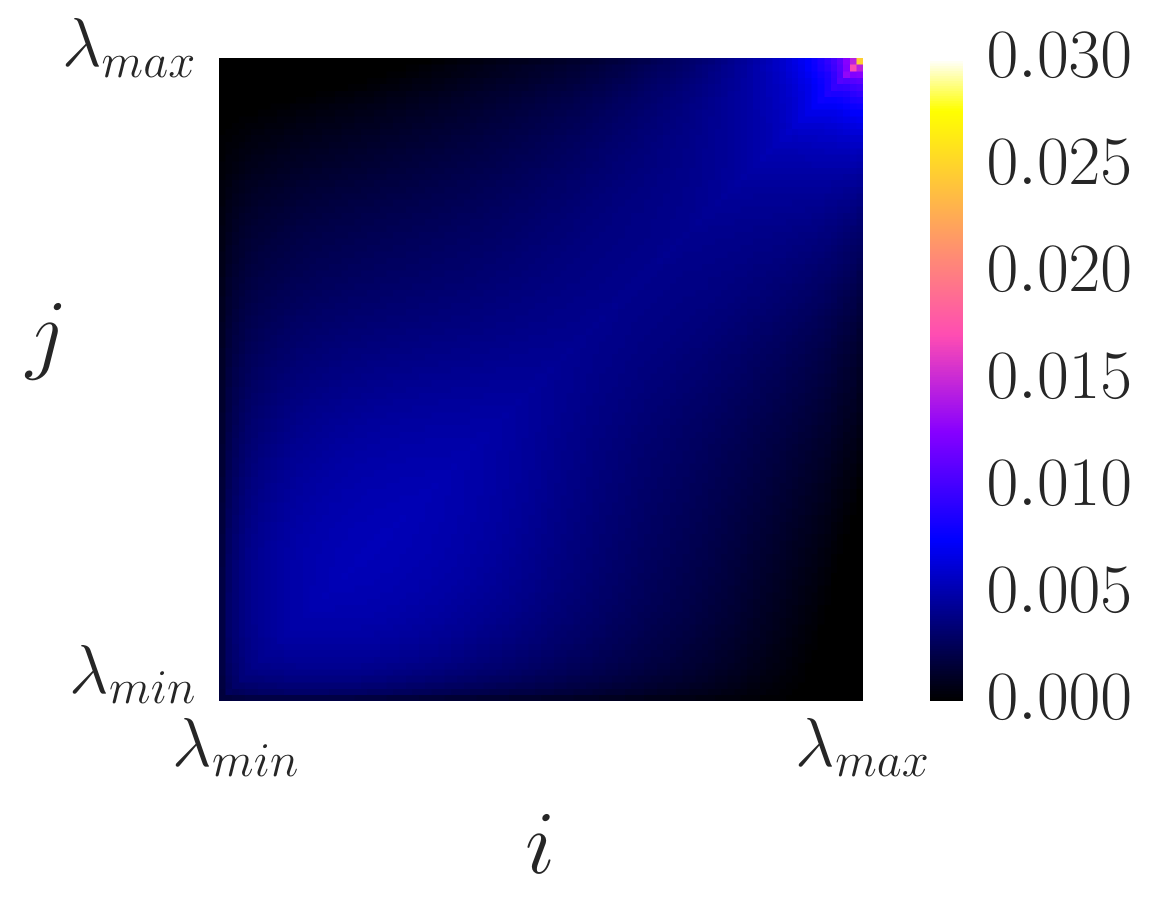}}
\subfloat[][$N=5$ $g_2=-3.50$]{\includegraphics[width=0.32\textwidth]{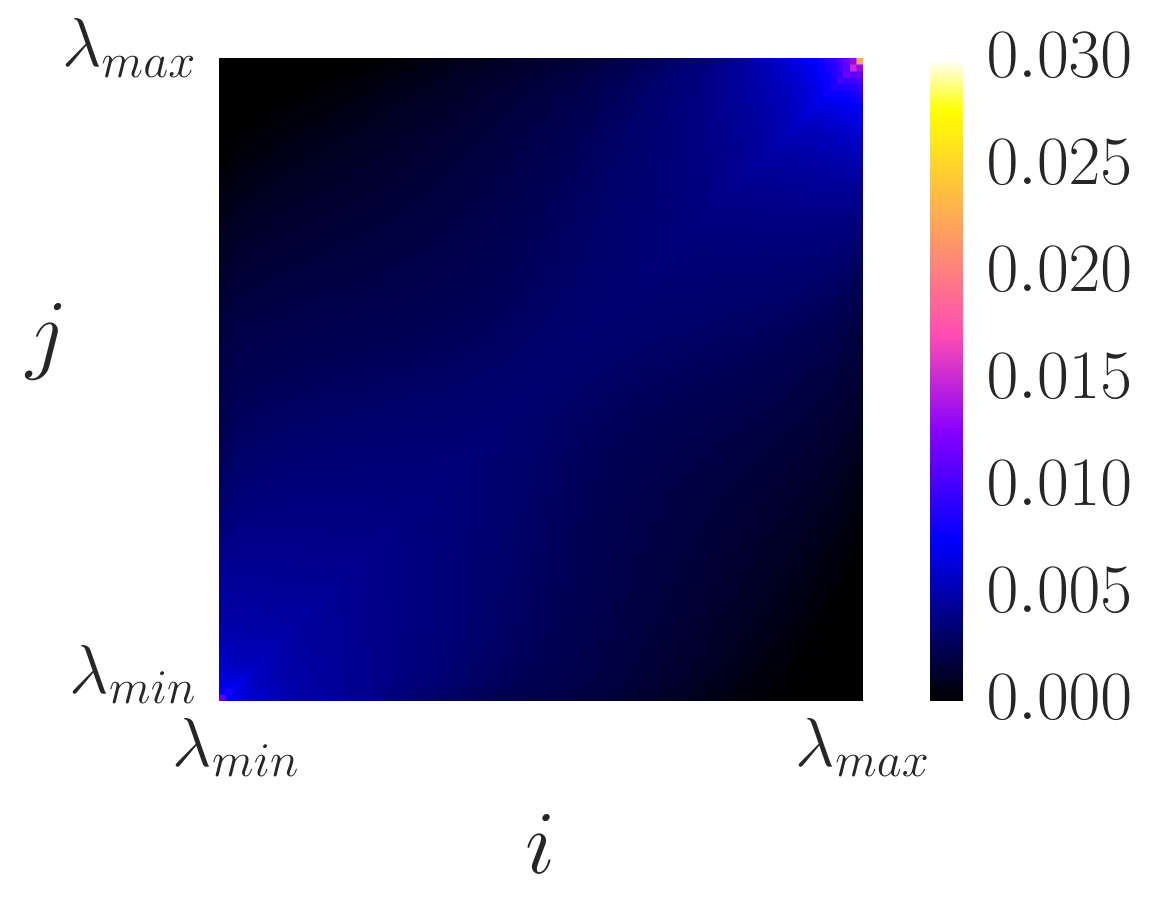}}
\caption{\label{fig:cov2211}This figure shows $\mathrm{Cov}(\lambda_i^2,\lambda_j^2)$ for type $(1,1)$, $N=5,10$ and $g_2=-2.0,-2.4,-3.5$. The axes correspond to the rows/columns of the correlation matrix, and the colour value of a given pixel indicates the value of the covariance between this pair.}
\end{figure}

\begin{figure}
\subfloat[][\label{fig:cov2220l5}$N=5$ $g_2=-2.50$]{\includegraphics[width=0.32\textwidth]{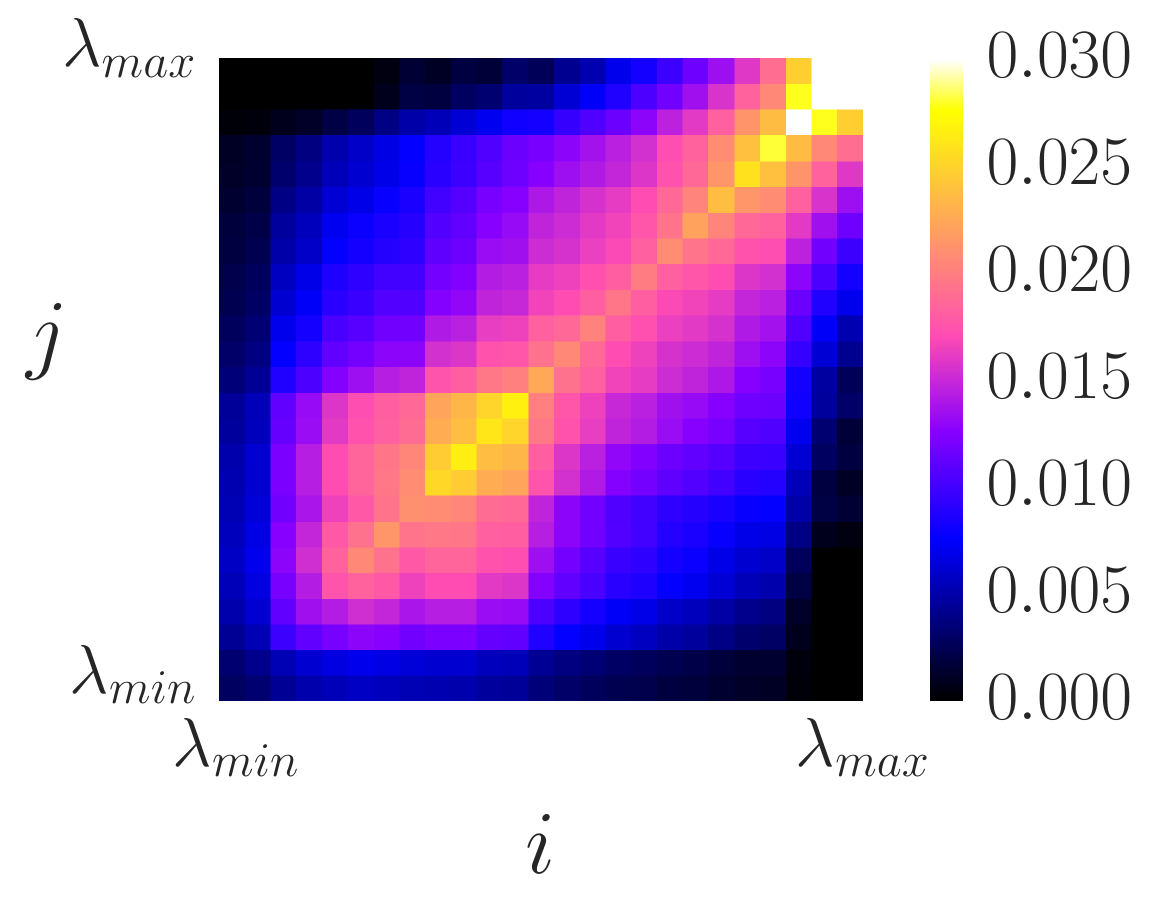}}
\subfloat[][\label{fig:cov2220gc5}$N=5$ $g_2=-2.80$]{\includegraphics[width=0.32\textwidth]{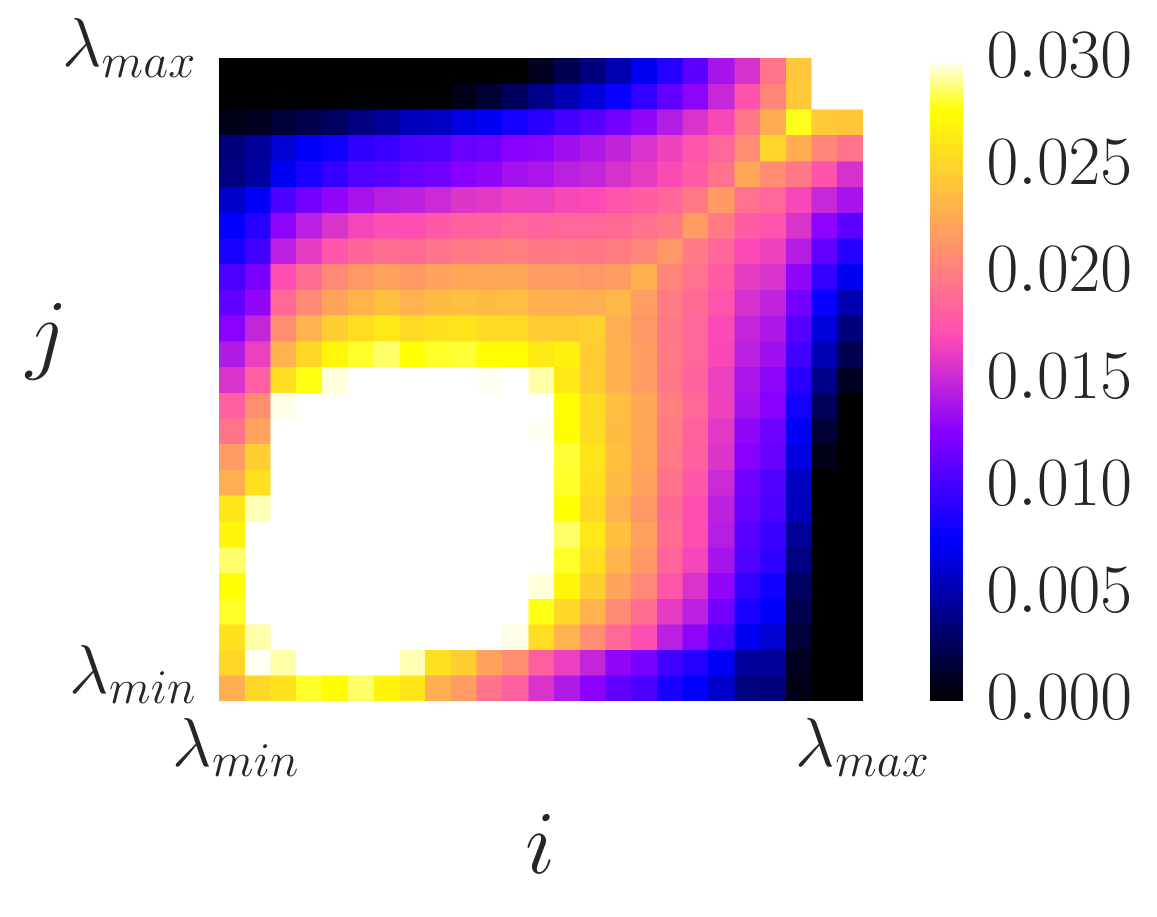}}
\subfloat[][\label{fig:cov2220h5}$N=5$ $g_2=-3.50$]{\includegraphics[width=0.32\textwidth]{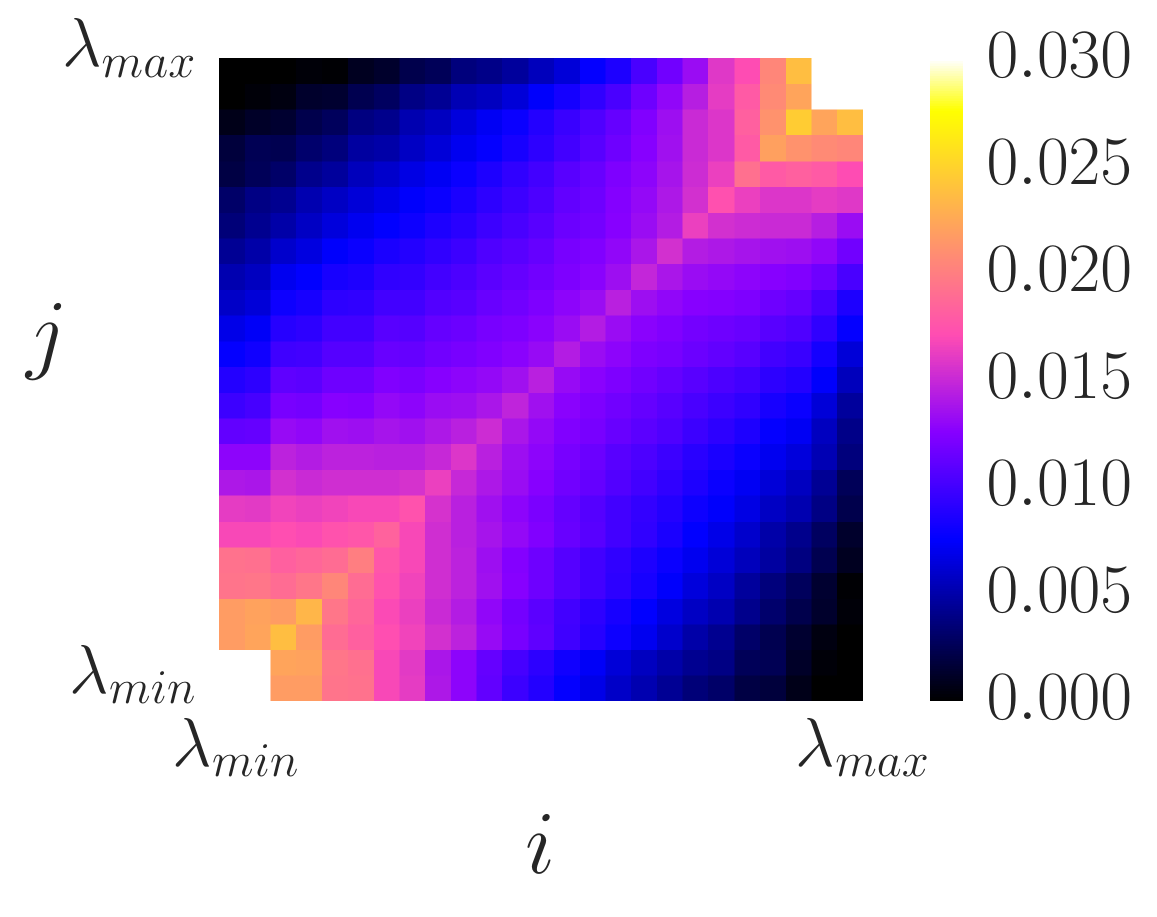}}

\subfloat[][\label{fig:cov2220l10}$N=10$ $g_2=-2.50$]{\includegraphics[width=0.32\textwidth]{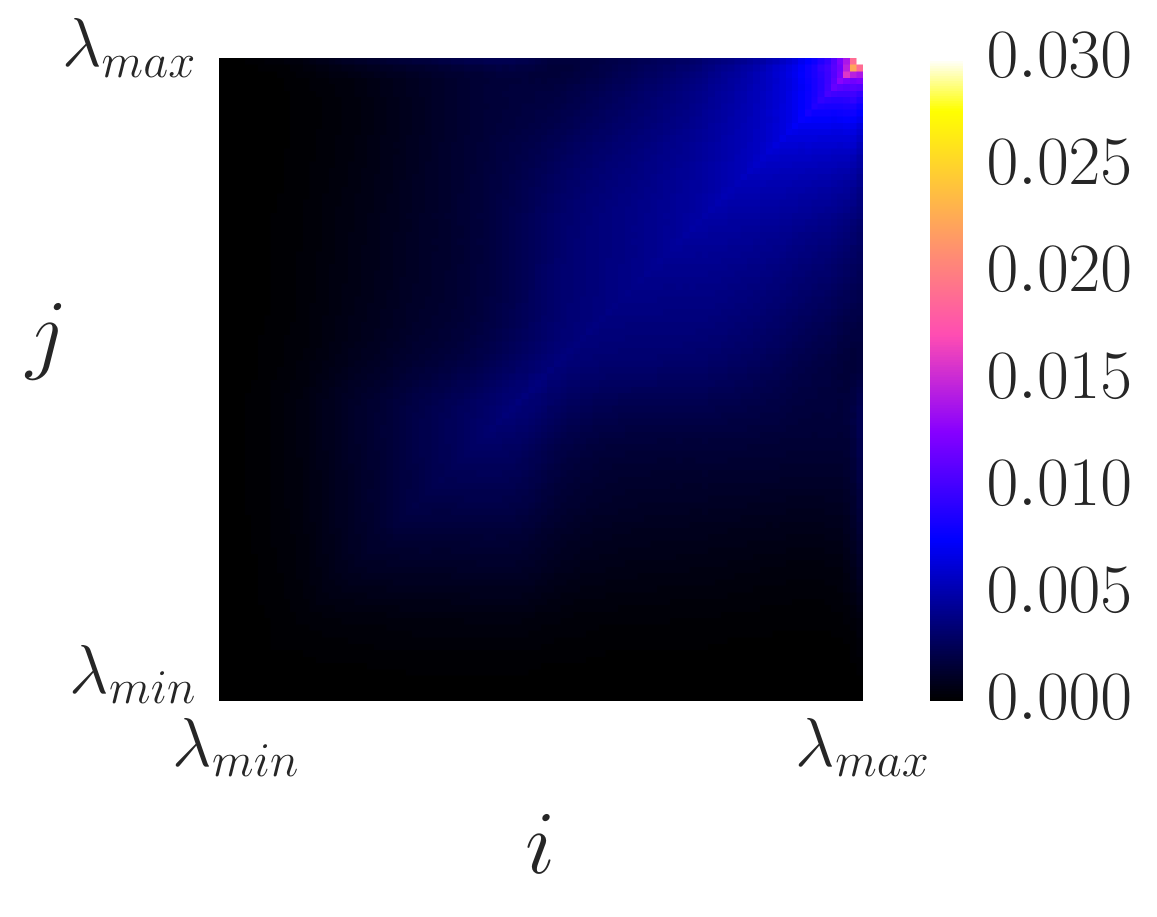}}
\subfloat[][\label{fig:cov2220gc10}$N=10$ $g_2=-2.80$]{\includegraphics[width=0.32\textwidth]{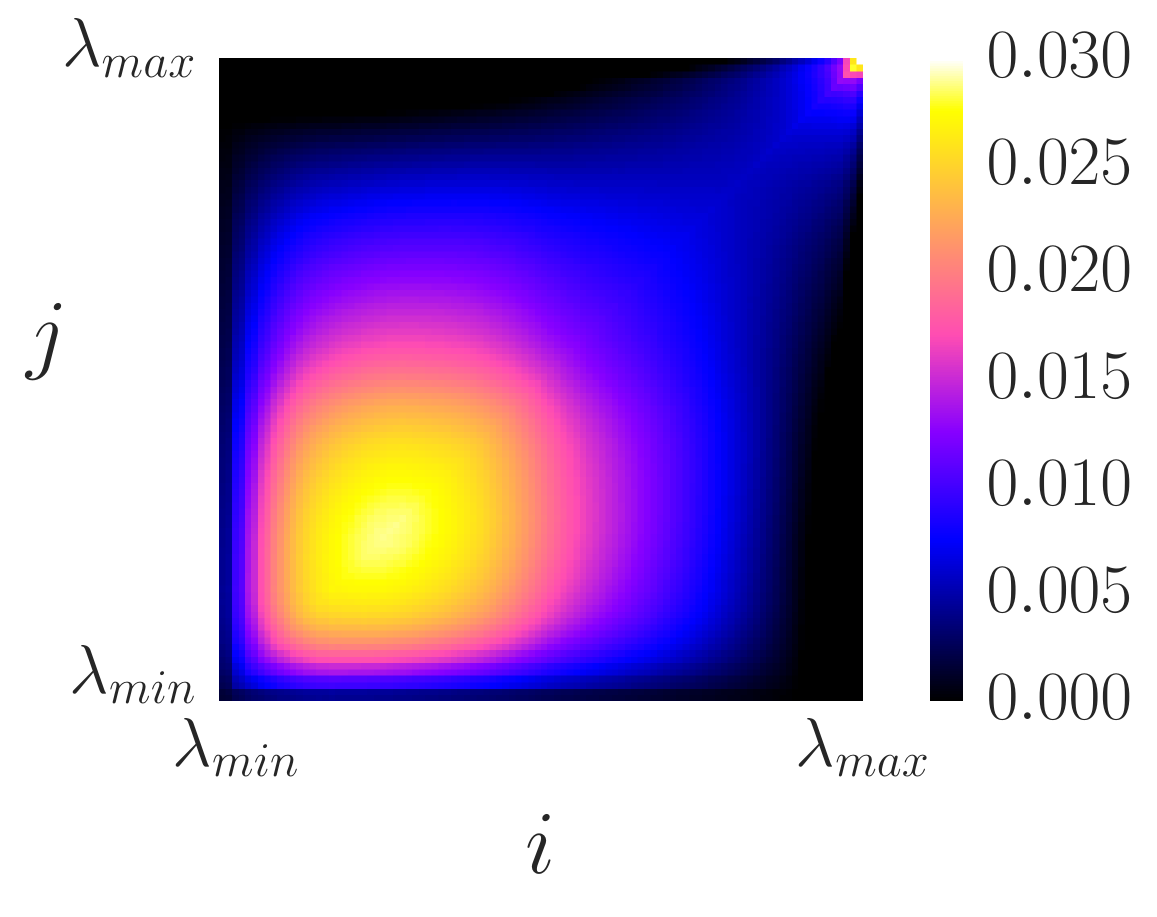}}
\subfloat[][\label{fig:cov2220h10}$N=10$ $g_2=-3.50$]{\includegraphics[width=0.32\textwidth]{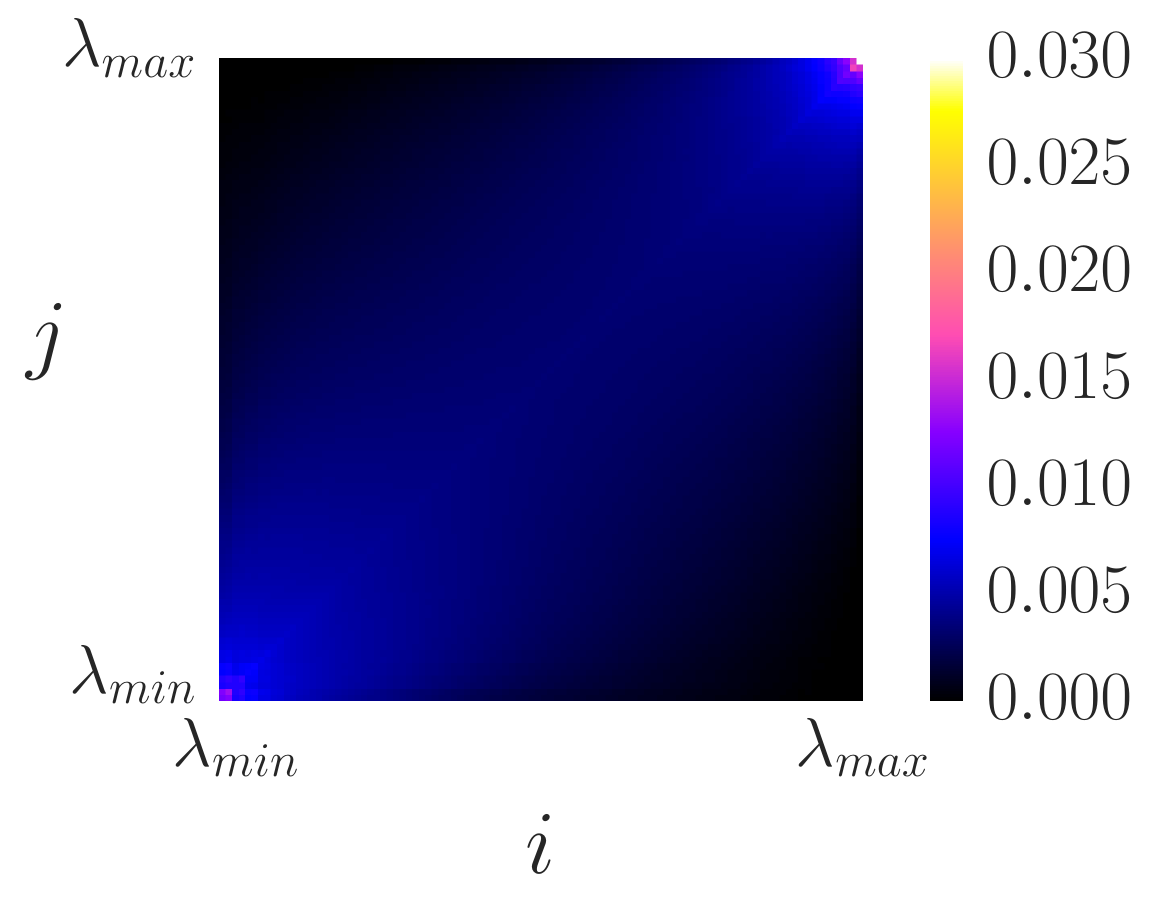}}
\caption{\label{fig:cov2220}This figure shows $\mathrm{Cov}(\lambda_i^2,\lambda_j^2)$ for $N=5,10$ and $g_2=-2.50,-2.80,-3.50$. The $x$ axis shows the $i$ label, while the $y$ axis shows the $j$ label, and the colour value of a given pixel indicates the value of the covariance between this pair.
}
\end{figure}

\begin{figure}
\subfloat[][\label{fig:cov2213l5}$N=5$ $g_2=-3.35$]{\includegraphics[width=0.32\textwidth]{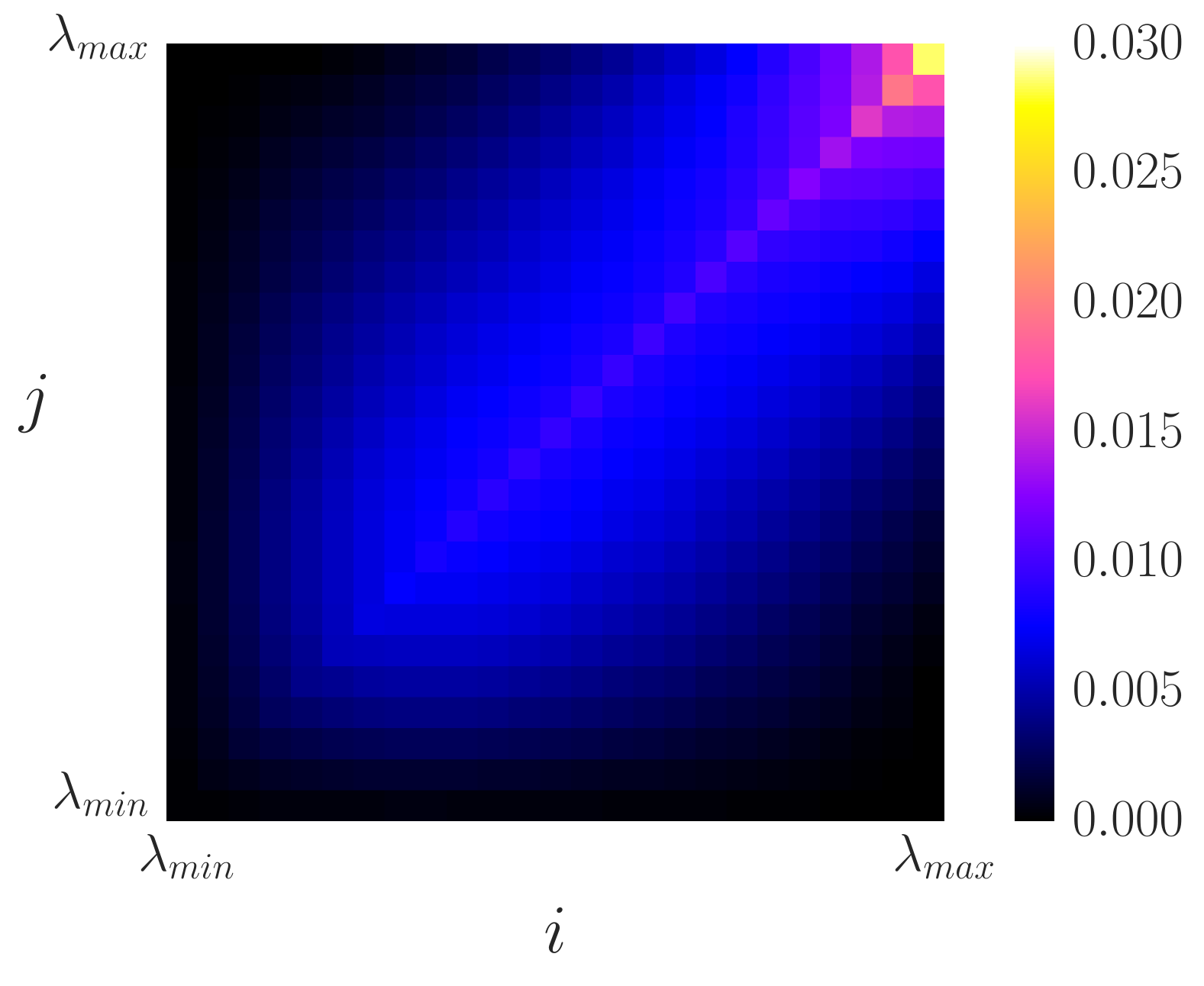}}
\subfloat[][\label{fig:cov2213gc5}$N=5$ $g_2=-3.70$]{\includegraphics[width=0.32\textwidth]{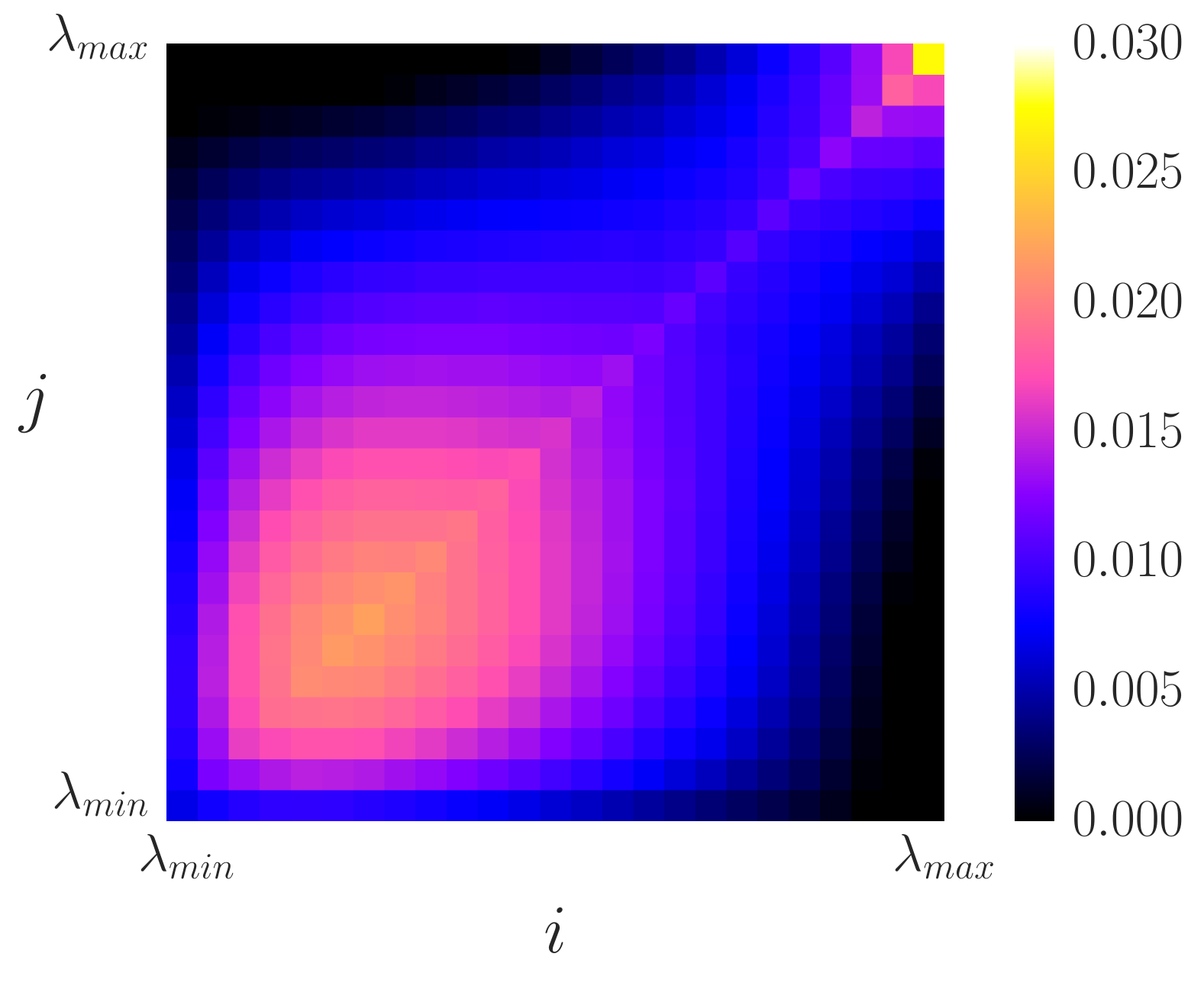}}
\subfloat[][\label{fig:cov2213h5}$N=5$ $g_2=-4.00$]{\includegraphics[width=0.32\textwidth]{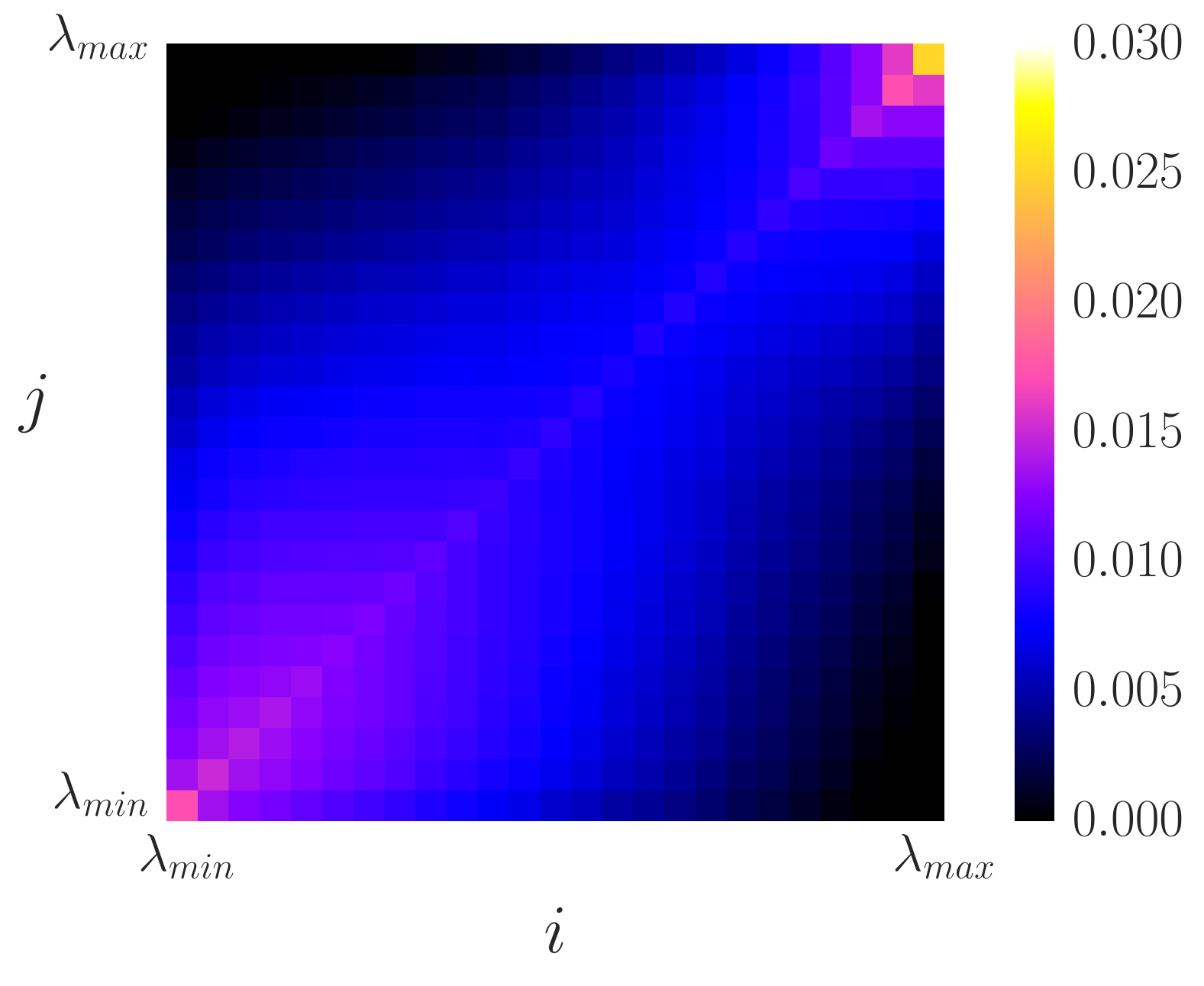}}

\subfloat[][\label{fig:cov2213l8}$N=8$ $g_2=-3.35$]{\includegraphics[width=0.32\textwidth]{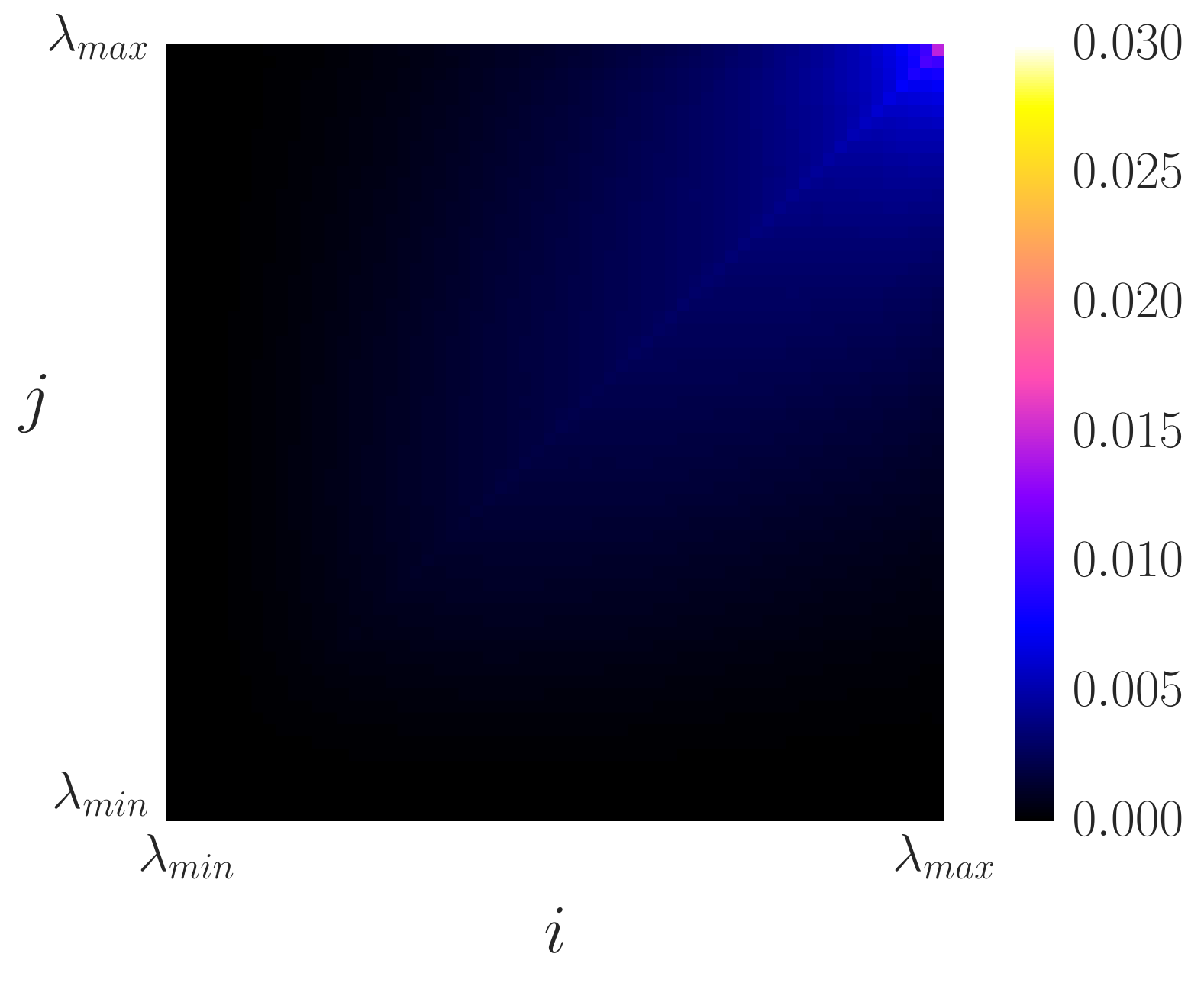}}
\subfloat[][\label{fig:cov2213gc8}$N=8$ $g_2=-3.70$]{\includegraphics[width=0.32\textwidth]{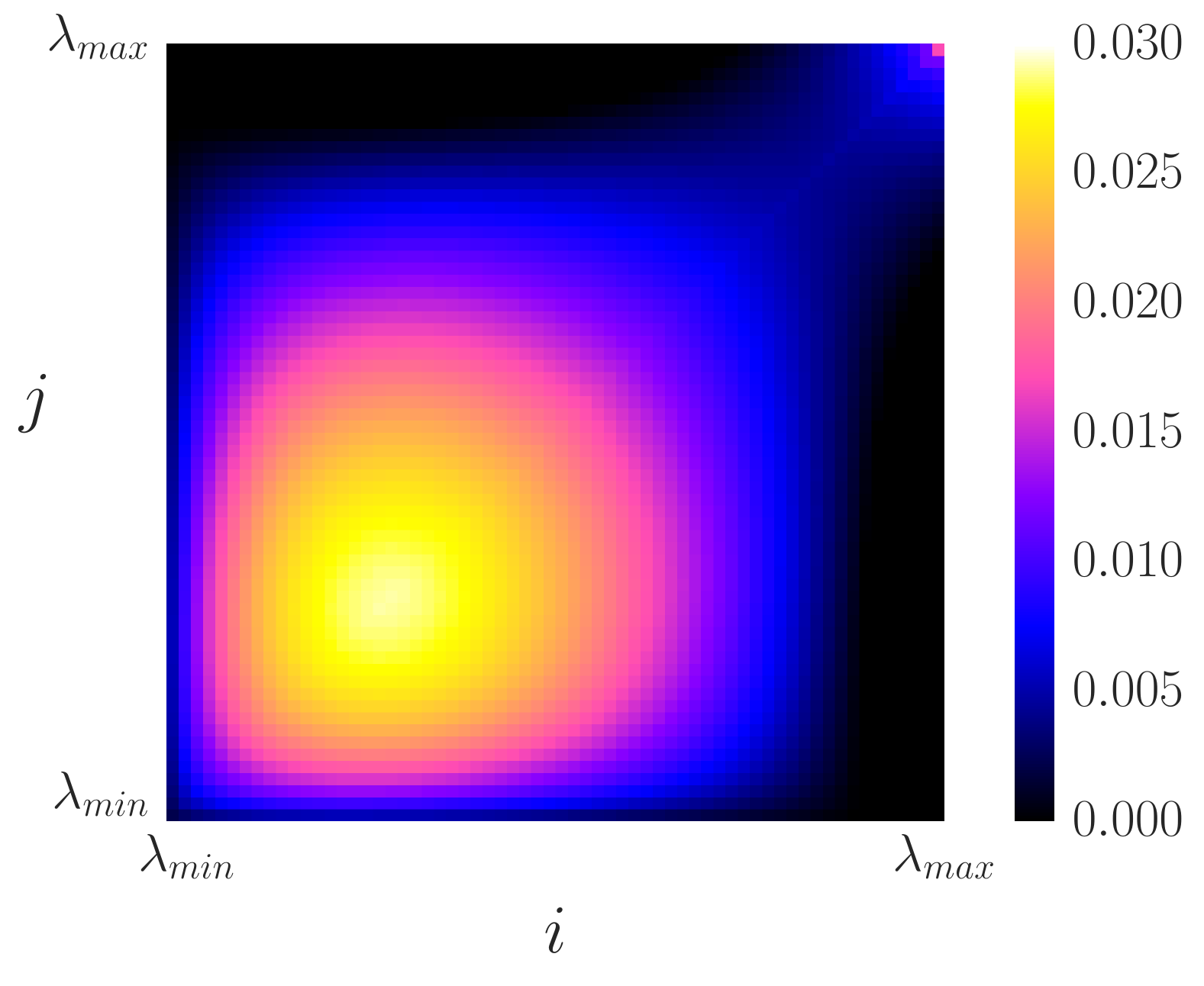}}
\subfloat[][\label{fig:cov2213h8}$N=8$ $g_2=-4.00$]{\includegraphics[width=0.32\textwidth]{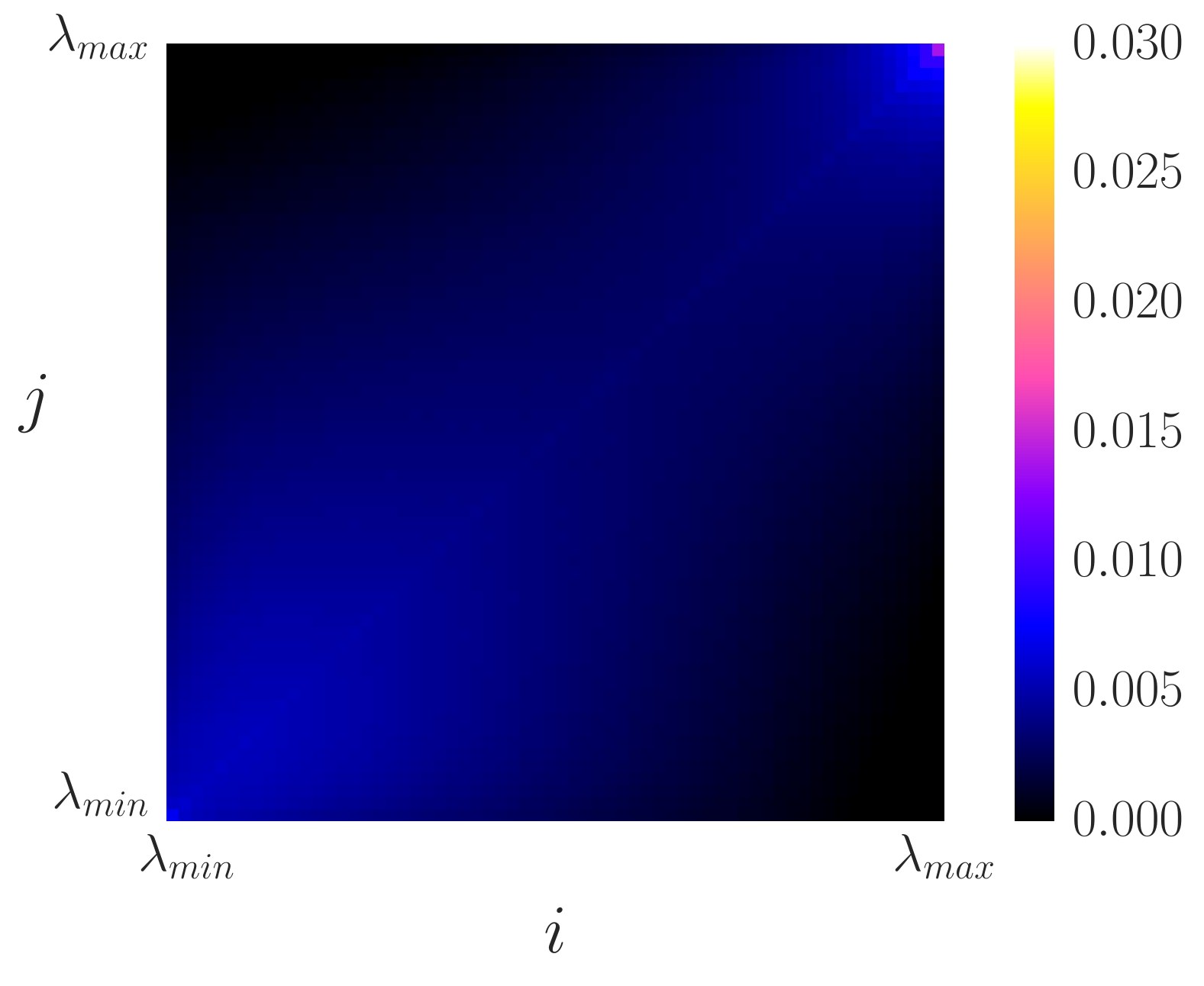}}
\caption{\label{fig:cov2213}This figure shows $\mathrm{Cov}(\lambda_i^2,\lambda_j^2)$ for type $(1,3)$, $N=5,8$ and $g_2=-3.35,-3.75,-4.0$. The $x$ axis shows the $i$ label, while the $y$ axis shows the $j$ label, and the colour value of a given pixel indicates the value of the covariance between this pair.}
\end{figure}

To understand this scaling we look at the plots of $\mathrm{Cov}(\lambda_i^2,\lambda_j^2)$ in Figures \ref{fig:cov2211}, \ref{fig:cov2220}, and \ref{fig:cov2213}.
We plot $\mathrm{Cov}(\lambda_i^2,\lambda_j^2)$ as a colour value map, and to calculate the covariances we only use geometries that are further away than the autocorrelation time of the Monte Carlo chain.
Since the geometries we examined here have a symmetry $\lambda_i=-\lambda_{n_d-i}$ the spectrum is symmetric on both the horizontal and vertical axis and we do not lose information by cutting the plots to only include the correlations between the positive eigenvalues, the axes are labelled $\lambda_{min}$ to $\lambda_{max}$, where we denote the lowest positive eigenvalue as $\lambda_{min}$ .

\begin{figure}
\subfloat[][$g_=-2.00$]{\includegraphics[width=0.32\textwidth]{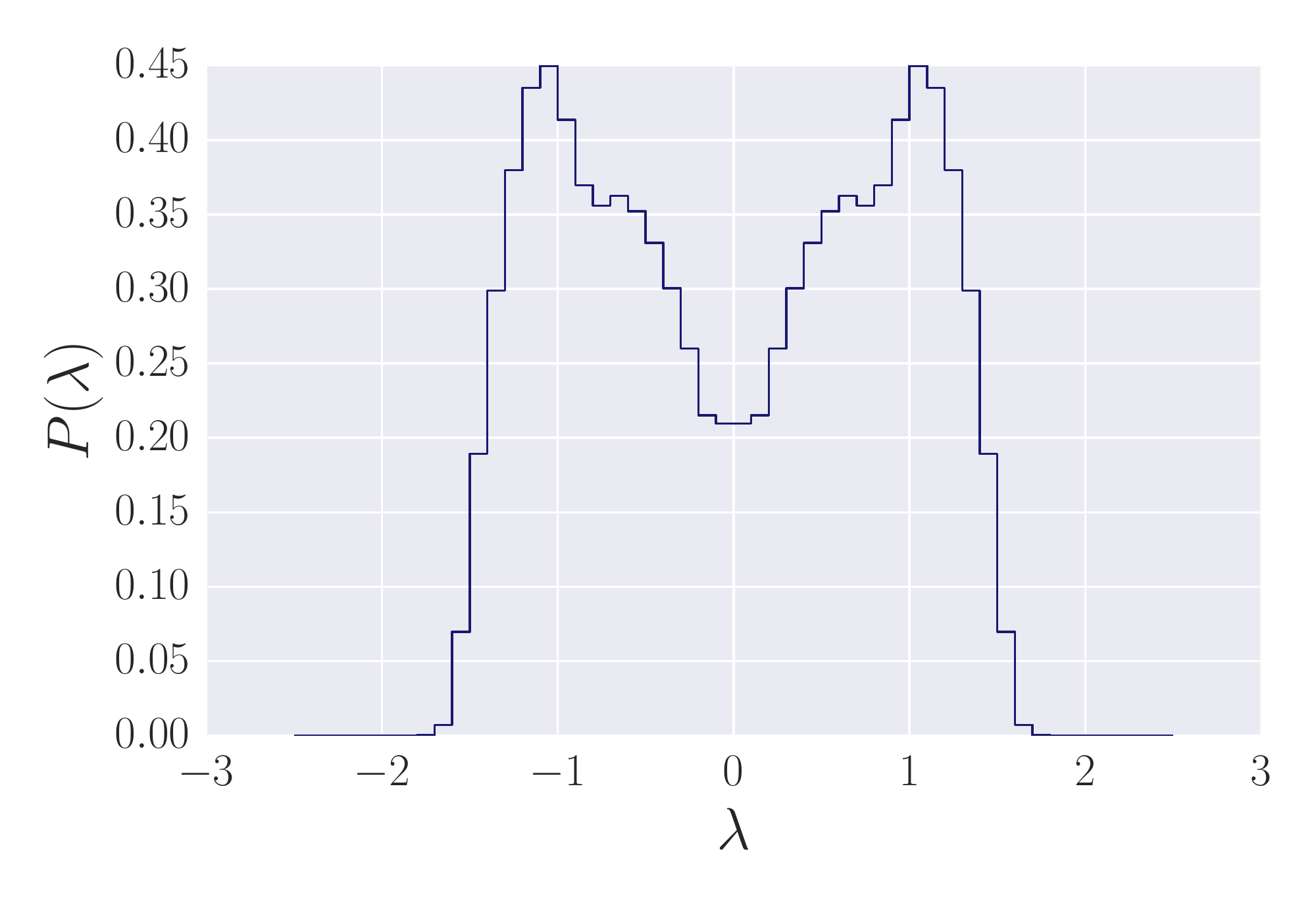}}
\subfloat[][$g_=-2.40$]{\includegraphics[width=0.32\textwidth]{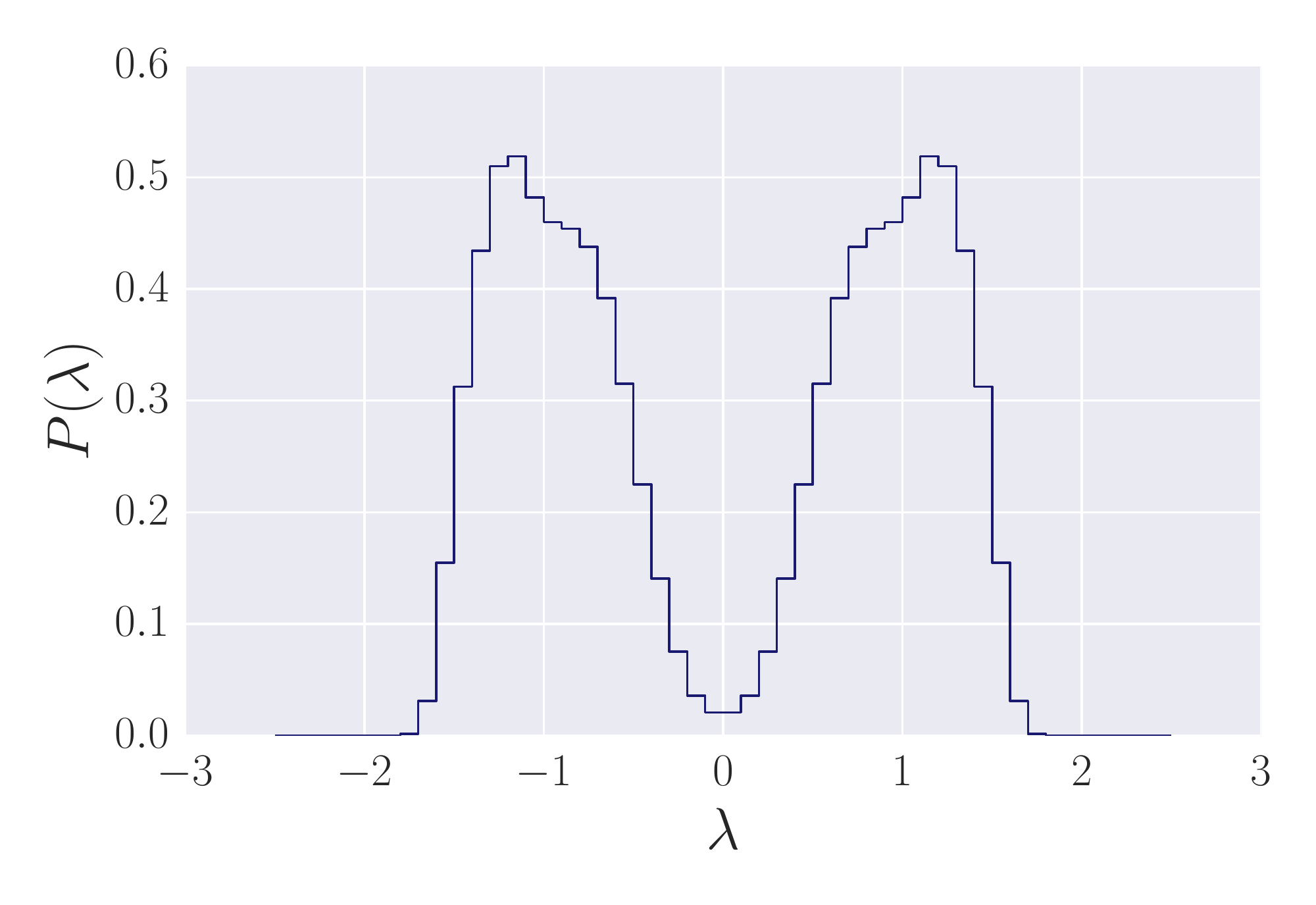}}
\subfloat[][$g_=-3.50$]{\includegraphics[width=0.32\textwidth]{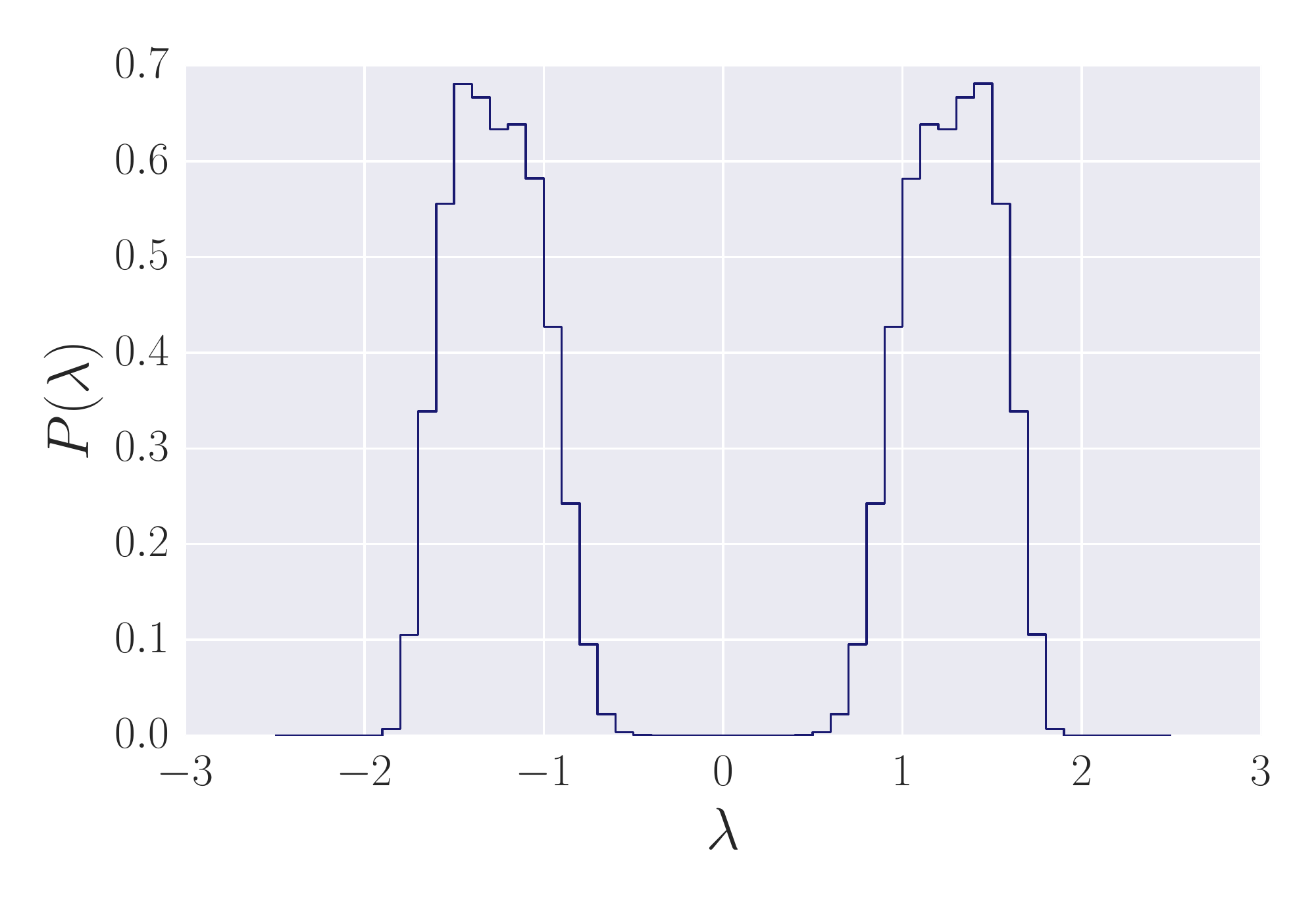}}

\caption{\label{fig:11EvHist}Histogram of the eigenvalues for the type $(1,1)$ case at $g_2=-2.0,-2.4,-3.5$, from left to right, to show the behaviour of the eigenvalues as the phase transition is crossed.}
\end{figure}

It becomes clear that there are two effects driving the scaling to be $N^2$ away from the phase transition, one is that the correlation is concentrated around the diagonals $i=j$ (and $i=n_d-j$ in the full plot), hence not all $N^4$ terms in the sum contribute to it.
The other effect, is that the values of $\mathrm{Cov}(\lambda_i^2,\lambda_j^2)$ become lower with increasing $N$.
For type $(1,1)$ both of these effects are at play even at the pseudo-critical point, which we can see very clearly in Figure \ref{fig:cov2211_gc10}.
There the correlations are much weaker than in the $N=5$ case above, indicating that the pseudo-critical point might wash out in the $N\to \infty$ limit.
This indicates that either the transition for type $(1,1)$ is more of a cross-over type, or that we have not found the right observables to explore it.
For more clarity we can look at the histogram of eigenvalues, (again only using uncorrelated samples), as we did in \cite{barrett_monte_2015}.
These are shown in figure \ref{fig:11EvHist} and show that the distribution of eigenvalues changes from being concentrated around $0$ to two peaks around $\mathrm{abs}(\lambda)=1.5$ approximately.
So clearly the behaviour of the geometry does change, it just seems to do so without an accompanying change in the correlation behaviour.

For type $(2,0)$ and $(1,3)$ the colour value map for the covariance at the phase transitions (Figures \ref{fig:cov2220gc5},\ref{fig:cov2220gc10},\ref{fig:cov2213gc5} and \ref{fig:cov2213gc8}) show strong correlations in square patches, hence indicating contributions of some, but not all off diagonal terms.
For type $(2,0)$  figure \ref{fig:cov2220gc10} shows a maximal correlation that is only very slightly lower than in figure \ref{fig:cov2220gc5}, while for type $(1,3)$ in figure \ref{fig:cov2213gc8} the maximal correlation value is stronger than in Figure \ref{fig:cov2213gc5}.
The plots for much lower or higher $g_2$ on the other hand (Figures \ref{fig:cov2220l5},\ref{fig:cov2220l10},\ref{fig:cov2220h5},\ref{fig:cov2220h10},\ref{fig:cov2213l5},\ref{fig:cov2213l8},\ref{fig:cov2213h5}, and \ref{fig:cov2213h8}) show a clear focus on the diagonal line, and in addition have entries of smaller value.
We thus expect the scaling at the phase transition to be different from $N^2$, but can not predict the exact scaling from the correlation.

\begin{figure}
\subfloat[][$(1,1)$]{\includegraphics[width=0.32\textwidth]{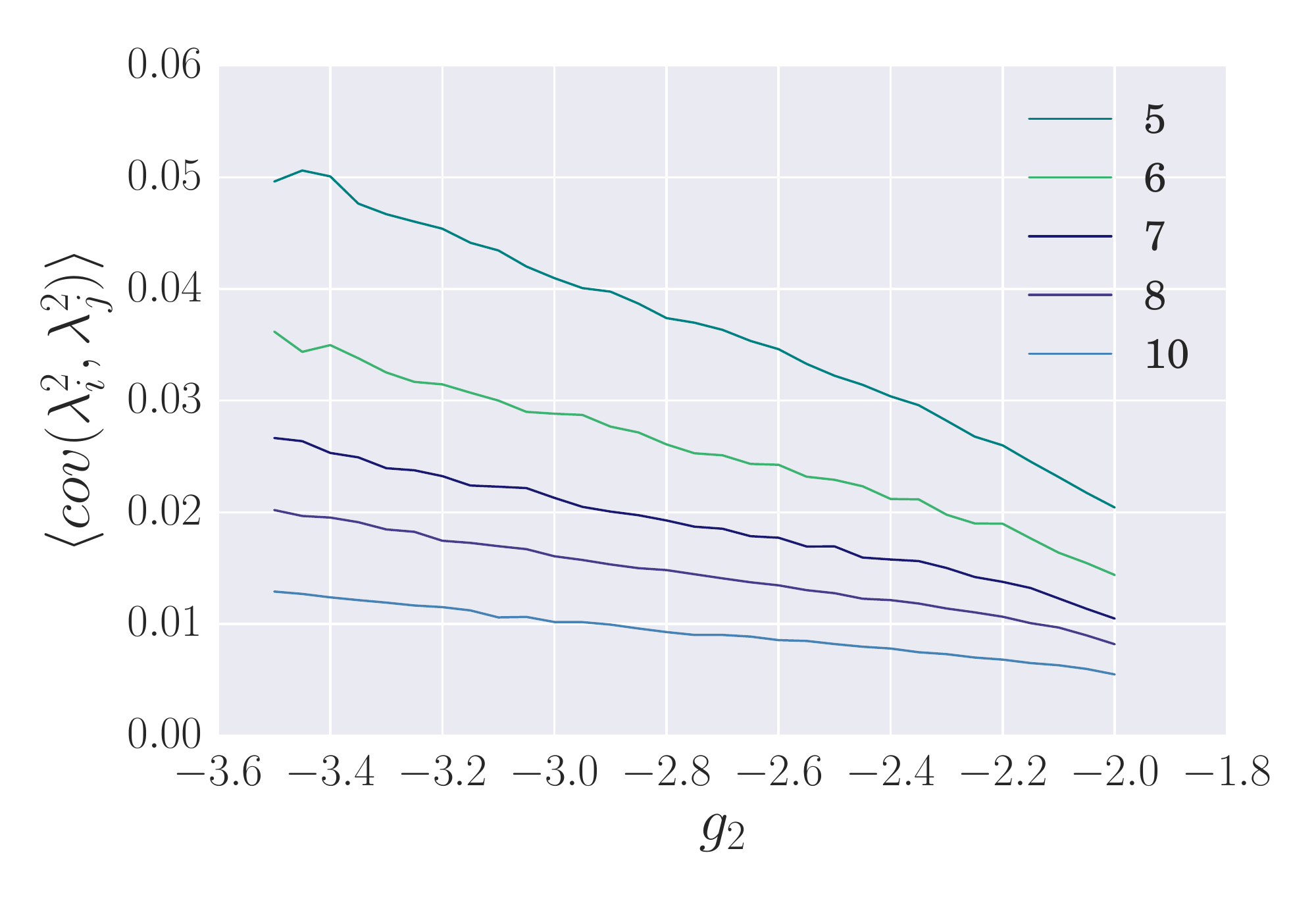}}
\subfloat[][$(2,0)$]{\includegraphics[width=0.32\textwidth]{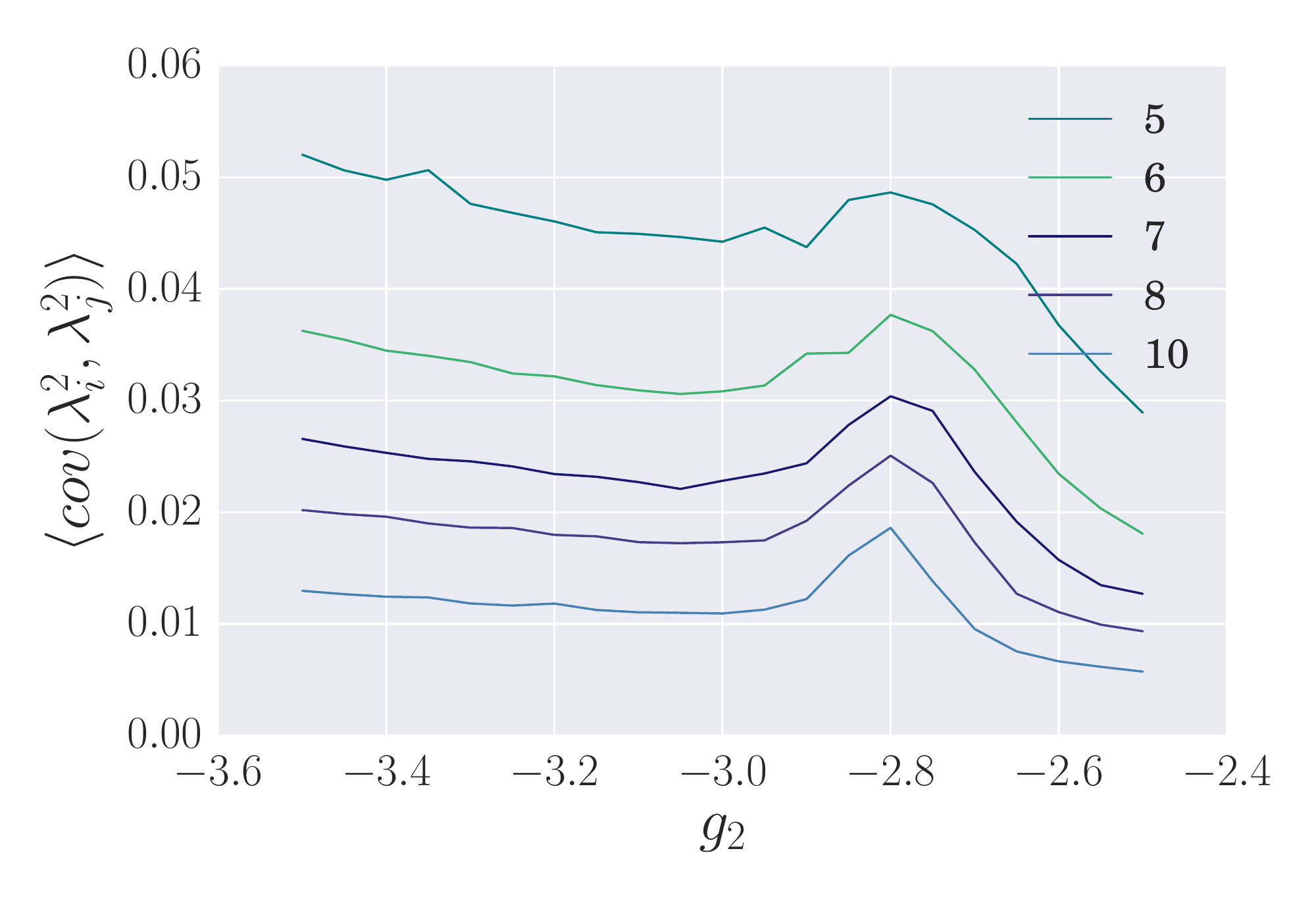}}
\subfloat[][$(1,3)$]{\includegraphics[width=0.32\textwidth]{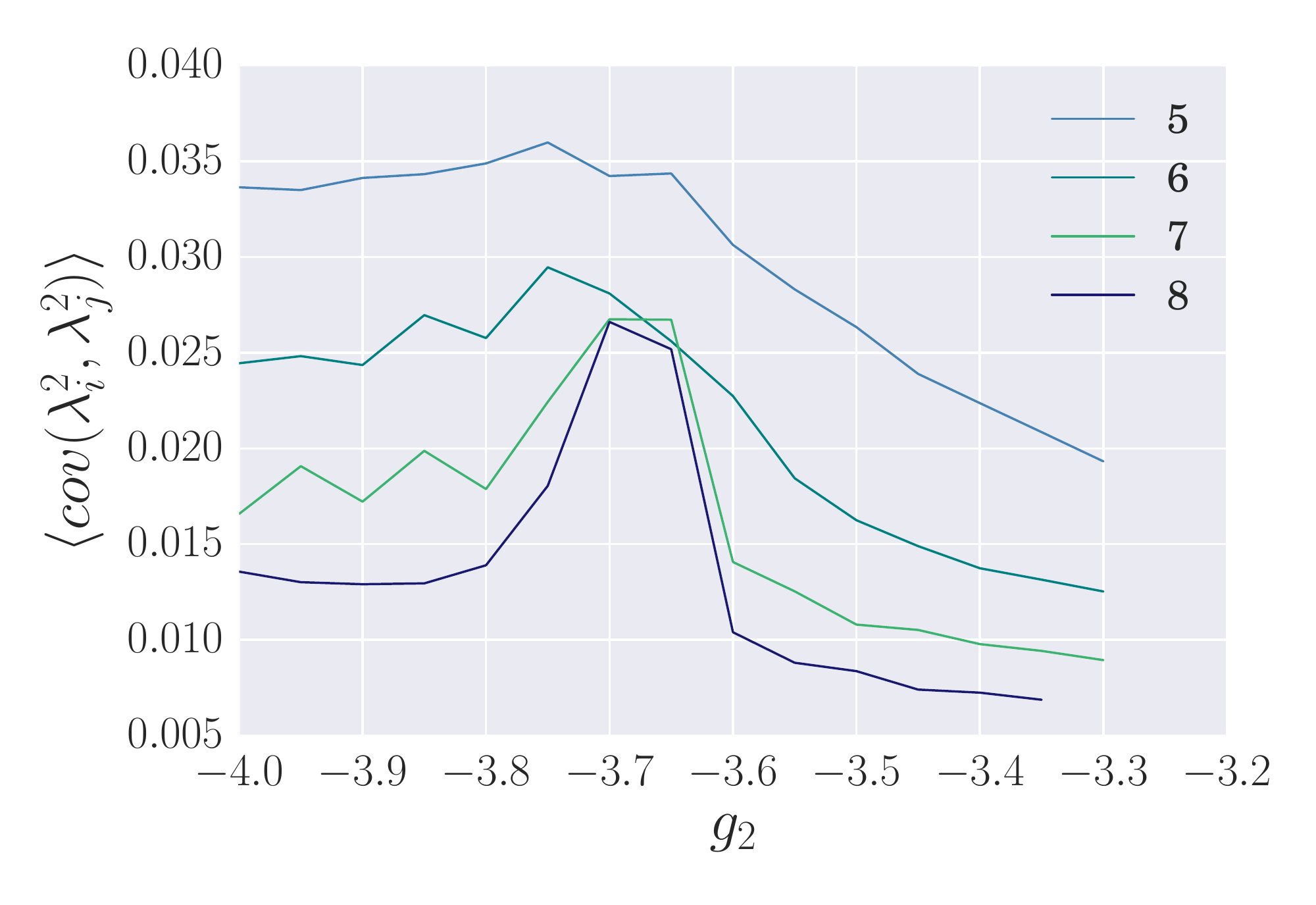}}
\caption{\label{fig:AvCov}The average value of $\mathrm{Cov}(\lambda_i^2,\lambda_j^2)$ plotted against $g_2$.}
\end{figure}

The colour value maps give a good impression of the general behaviour of the correlation, however it is hard to compare the overall magnitude between values of $g_2$ with them.
We thus plotted the average of $\mathrm{Cov}(\lambda_i^2,\lambda_j^2)$, which we will call the mean covariance, against $g_2$, to compare this behaviour.
In Figure \ref{fig:AvCov} we can see that the average covariance becomes weaker as $N$ rises.
We can explain this behaviour, if we assume that a $g_2$ dependent correlation length $\xi$ exists.
For small $N$ this correlation length covers a larger fraction of the overall geometry than at large $N$, hence the average correlation is stronger at small $N$.
We can make this slightly more precise by considering that our system volume grows like $N^2$, hence the linear extension of the system $L$ grows like $L \sim N^{\frac{2}{d}}$, assuming that all directions grow equally.
The behaviour at the pseudo-critical point is then dependent on the growth of this correlation length with $N$ as the pseudo-critical point is approached and on the dimension of the system.
In a system of higher dimension the linear extension will grow slower with $N$.
The three geometries we have examined seem to fall into different classes with regard to this behaviour.
For type $(1,1)$ the mean covariance does not change as the pseudo-critical point is approached, in fact the pseudo-critical point is not observable in this plot at all.
This indicates that the correlation length grows slower than $L$ even at the pseudo-critical point, hence washing out correlations.
In type $(2,0)$ on the other hand, the mean covariance does peak around the pseudo-critical point.
This peak becomes clearer with larger $N$, yet the maximal value of the mean covariance stays of roughly the same height over the average value for all $N$.
In terms of the correlation length this seems likely to indicate that the correlation length and the linear extension follow the same power law in $N$, hence preserving the feature and sharpening it up.
The most interesting case is the case $(1,3)$ in which the peak remains of almost constant height with increasing $N$.
This indicates that the correlation length $\xi$ grows faster than the linear extension $L$ of the system.

\begin{figure}
\subfloat[][$(1,1)$]{\includegraphics[width=0.49\textwidth]{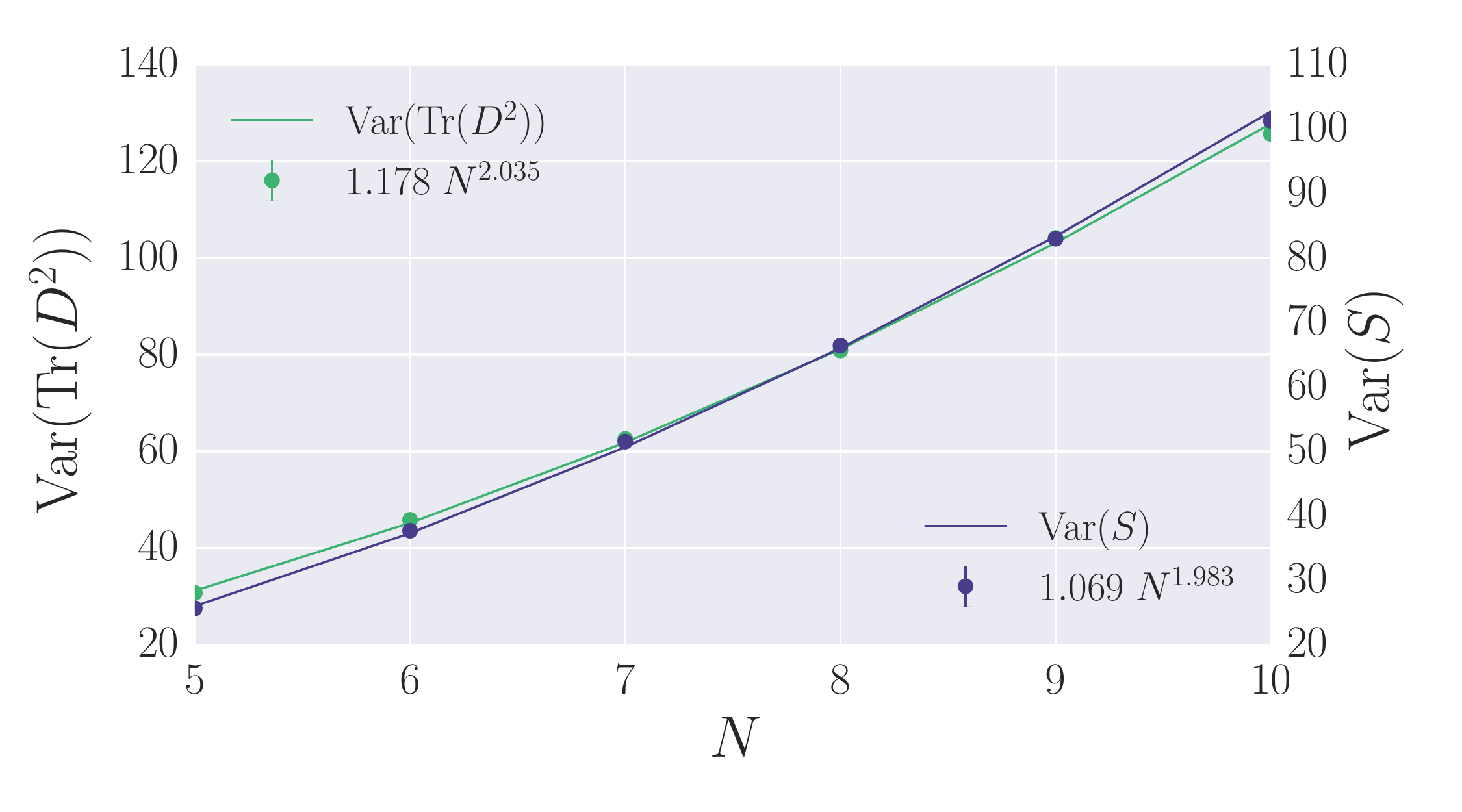}}
\subfloat[][$(2,0)$]{\includegraphics[width=0.49\textwidth]{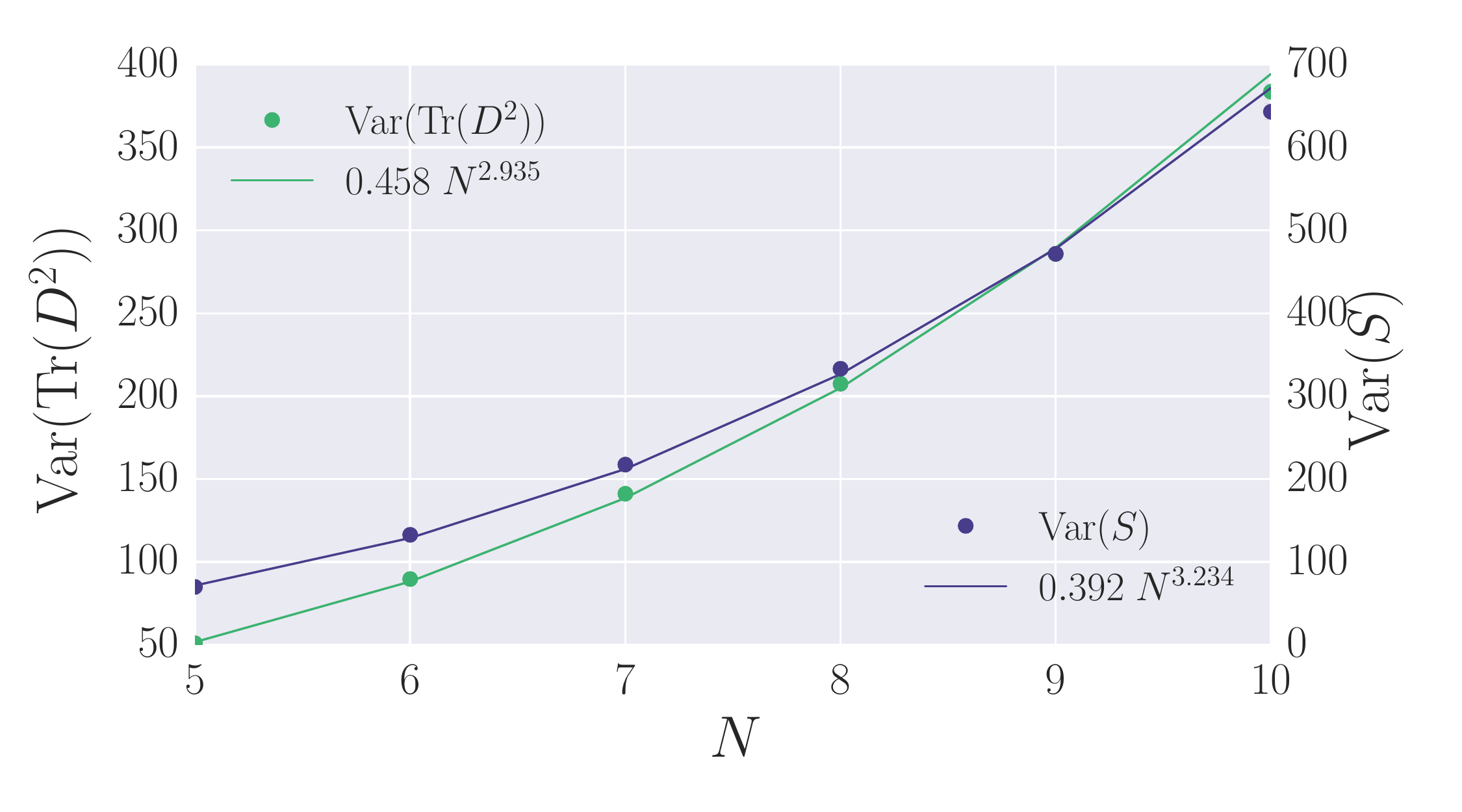}}

\subfloat[][$(1,3)$]{\includegraphics[width=0.6\textwidth]{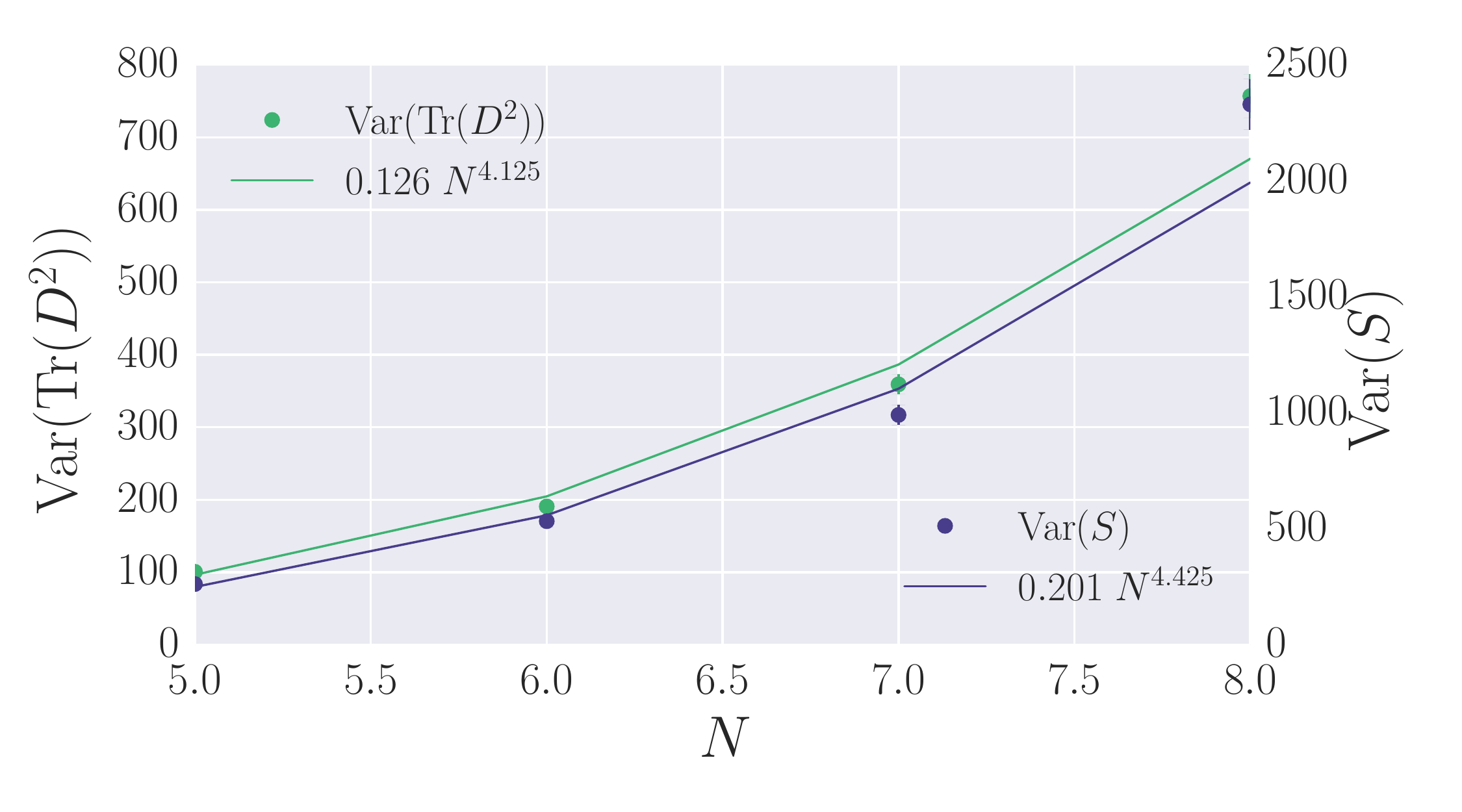}}
\caption{\label{fig:VarmaxN}Maximum of the variances plotted against $N$.}
\end{figure}

To better understand how exactly the variance scales with $N$ at the phase transition we can fit the maximal values of $\mathrm{Var}(\Tr{D^2})$ and $\mathrm{Var}(S)$ with a  polynomial $a N^b$.
These fits are shown in Figure \ref{fig:VarmaxN} and the values for $a$ and $b$ found in this manner are collected in Table \ref{tab:scalingN}.

Of particular interest in these fits is of course the parameter $b$, which shows how exactly the variance at the phase transition scales with $N$.
For type $(1,1)$ we find that $b$ from both fits is compatible with a scaling of $N^2$.
For type $(2,0)$ on the other hand the difference between the two values of $b$ is bigger than the statistical error on the fits.
However since there are likely additional systematic errors related to the sample size of our simulations and the resampling they are still compatible with each other and with the hypothesis that the variances scale like $N^3$.
For type $(1,3)$ we find that the $b$ fits are compatible with a scaling of the variance like $N^4$ at the phase transition.
We can try and make sense of these scalings by using a finite size scaling ansatz, in which $\chi$ stands for the variance, and $\Phi$ is some universal scaling function.
Normally the scaling ansatz includes the system size $L$, however for our system we do not have an immediate length parameter, instead we have a system volume proportional to $N^2$, hence
\begin{align}
\chi(\xi,N) = N^{2-2x} \Phi(\frac{N^{\frac{2}{d}}}{\xi})\;.
\end{align}
At the critical point the correlation length diverges and $\xi^{-1} \to 0$, hence the variance diverges with the power $N^{2-2x}$.
For our systems this would imply that $x=0,-0.5,-1$ for the types $(1,1),(2,0),(1,3)$ respectively.

\begin{table}
\caption{\label{tab:scalingN}Best fit to the scaling of $Max(\mathrm{Var}(\Tr{D^2})$ and $Max(\mathrm{Var}(S))$ with $a N^b$.}
\begin{tabularx}{\textwidth}{X X X X X X X X }
\toprule
type & & a & b& & a & b\\
\midrule
$(1,1)$& $\Tr{D^2}$&$1.178\pm  0.062$& $2.035\pm  0.027$ & $S$ &$1.069\pm  0.052$&$1.983\pm  0.025$\\
$(2,0)$& $\Tr{D^2}$& $0.458\pm  0.031$&$2.935\pm  0.035$ & $S$ & $0.392\pm  0.041$& $3.234\pm  0.054$ \\
$(1,3)$ & $\Tr{D^2}$&$0.126\pm  0.069$ &$4.125\pm  0.303$& $S$& $0.201\pm  0.146$ &$4.425\pm  0.394$\\
\bottomrule
\end{tabularx}
\end{table}

\subsection{A critical exponent?}
A quantity of particular interest around higher order phase transition are critical exponents.
These exponents characterise the universal behaviour of the system at the phase transition which is governed by conformal field theories.

\begin{figure}
\subfloat[][$(1,1)$]{\includegraphics[width=0.32\textwidth]{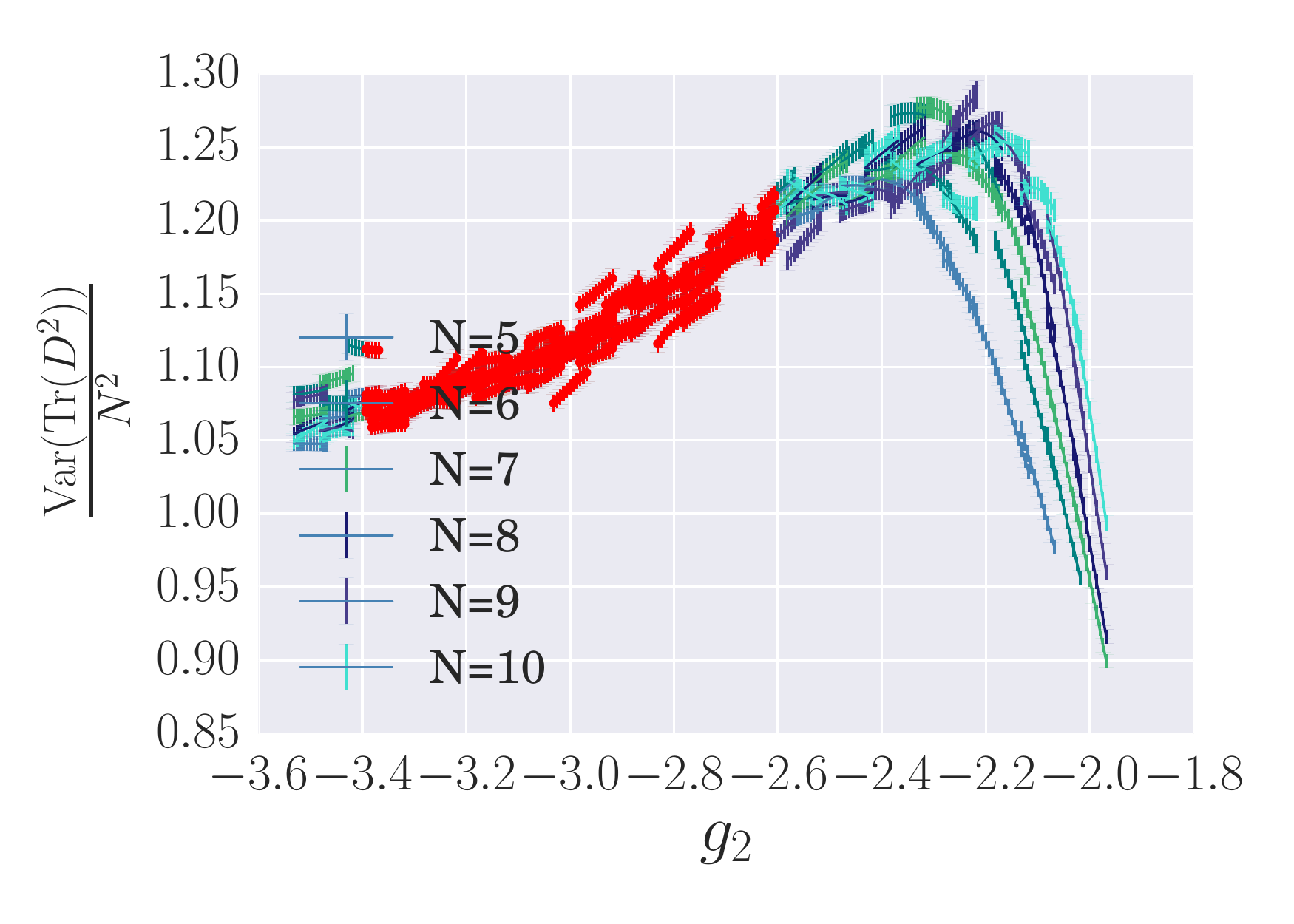}}
\subfloat[][$(2,0)$]{\includegraphics[width=0.32\textwidth]{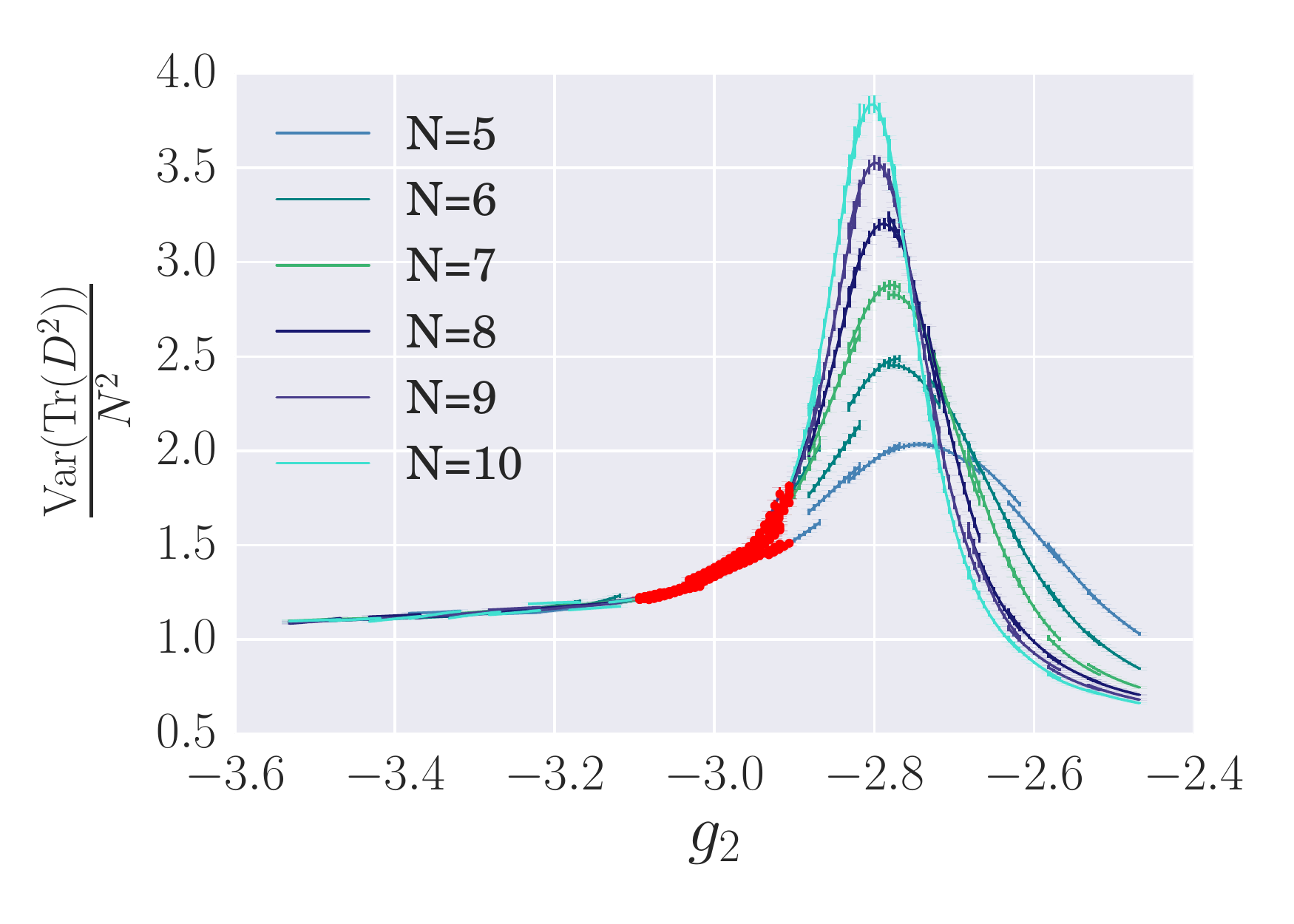}}
\subfloat[][$(1,3)$]{\includegraphics[width=0.32\textwidth]{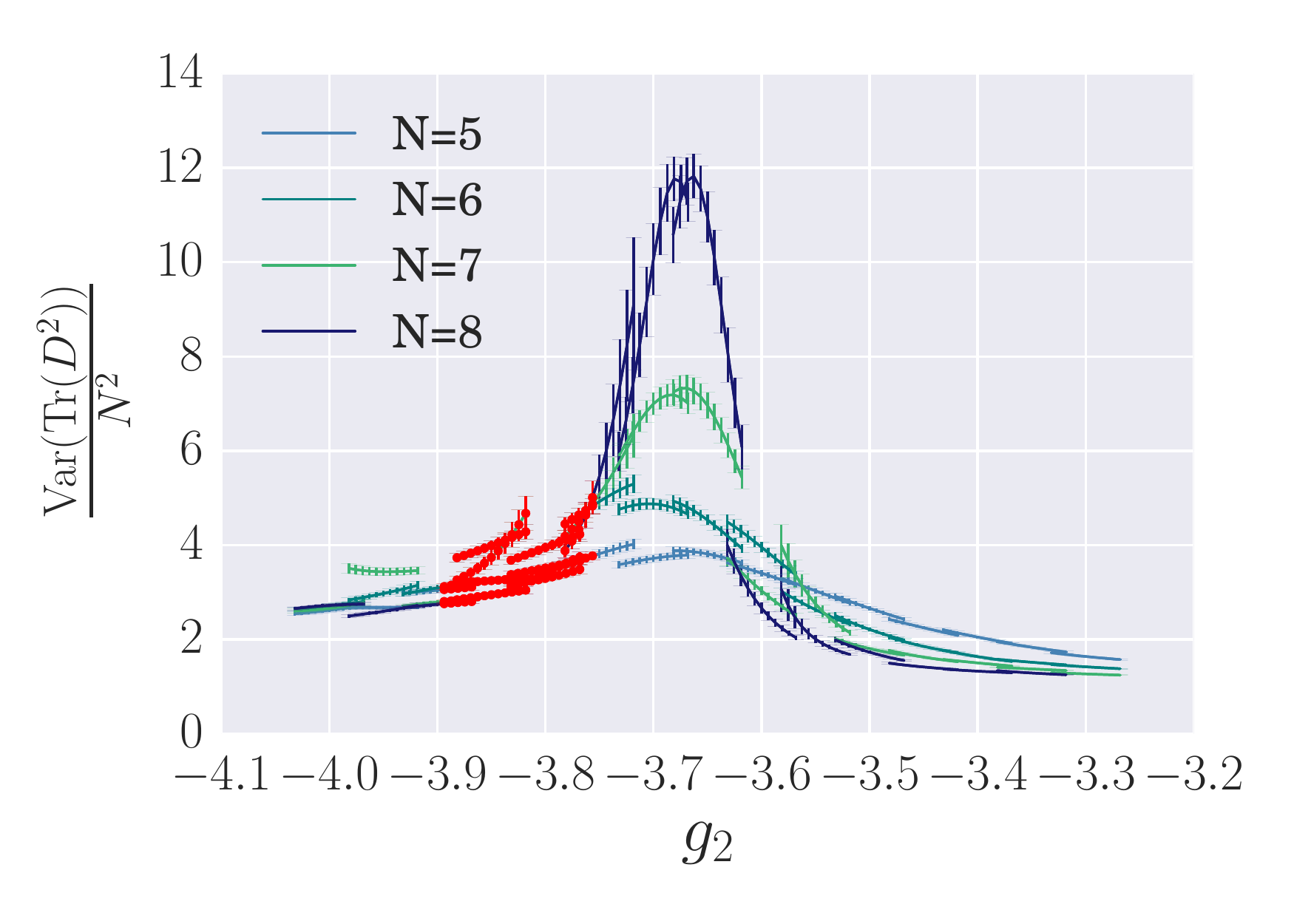}}
\caption{\label{fig:marked}Variance of $\Tr{D^2}$ with the region we fit for the critical exponent marked in red.}
\end{figure}

In our case we can determine the fall-off exponent on the left side of the peak.
In principle we expect the Variance to fall off according to
\begin{align}
\mathrm{Var}(S(g_2)) &\sim |g-g_c|^{\mu} & \mathrm{Var}(\Tr{D^2}) &\sim |g-g_c|^{\nu}
\end{align}
away from the peak.
This function is not obeyed exactly at the peak, due to finite size effects.
Our data only allows us to determine the exponent on the left side of the transition, for which we fit the data to the regions marked in red in Figure \ref{fig:marked}.
This region is chosen to avoid the finite size effects that become important close to the pseudo-critical point.

\begin{figure}
\subfloat[][$(1,1)$]{\includegraphics[width=0.32\textwidth]{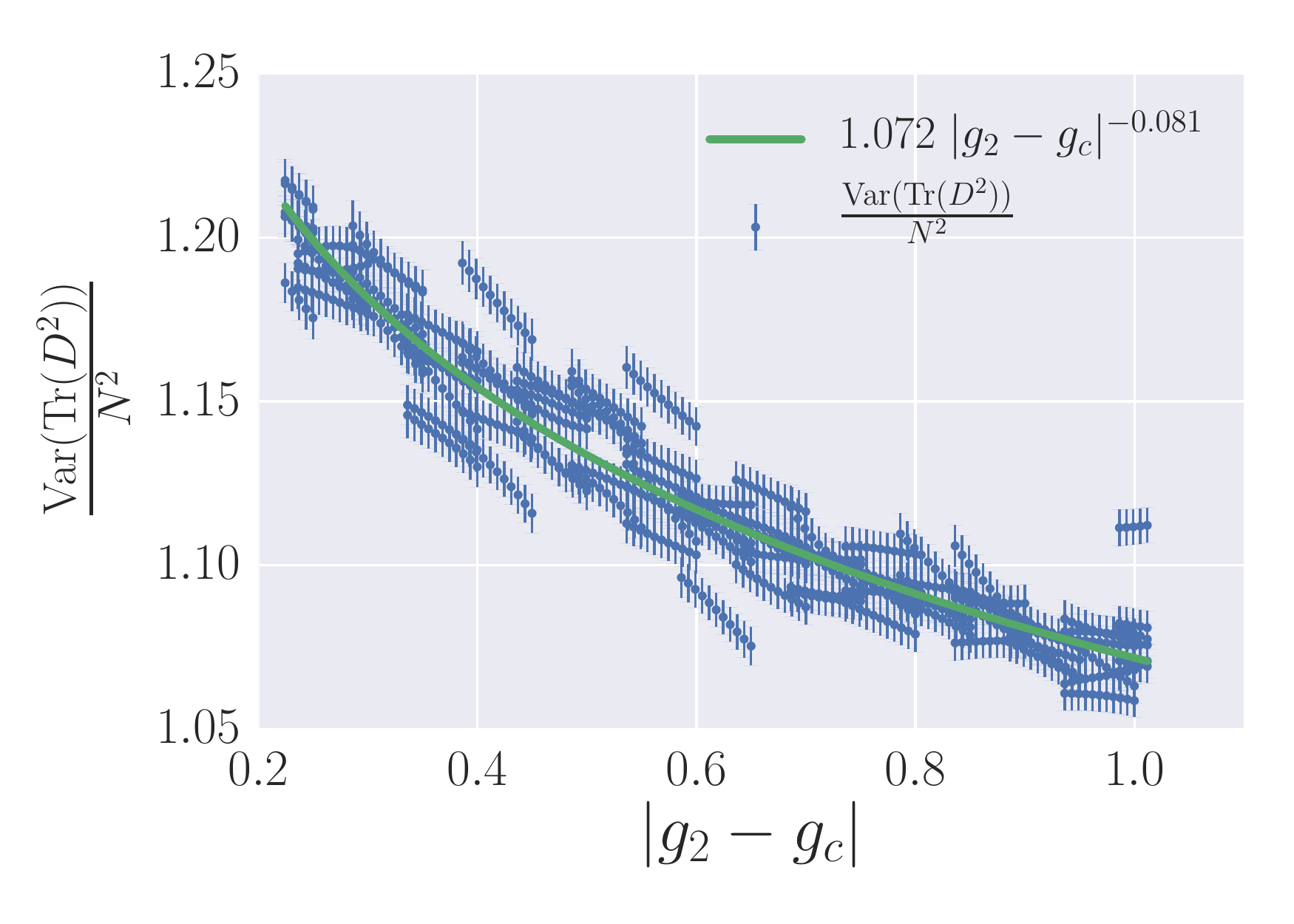}}
\subfloat[][$(2,0)$]{\includegraphics[width=0.32\textwidth]{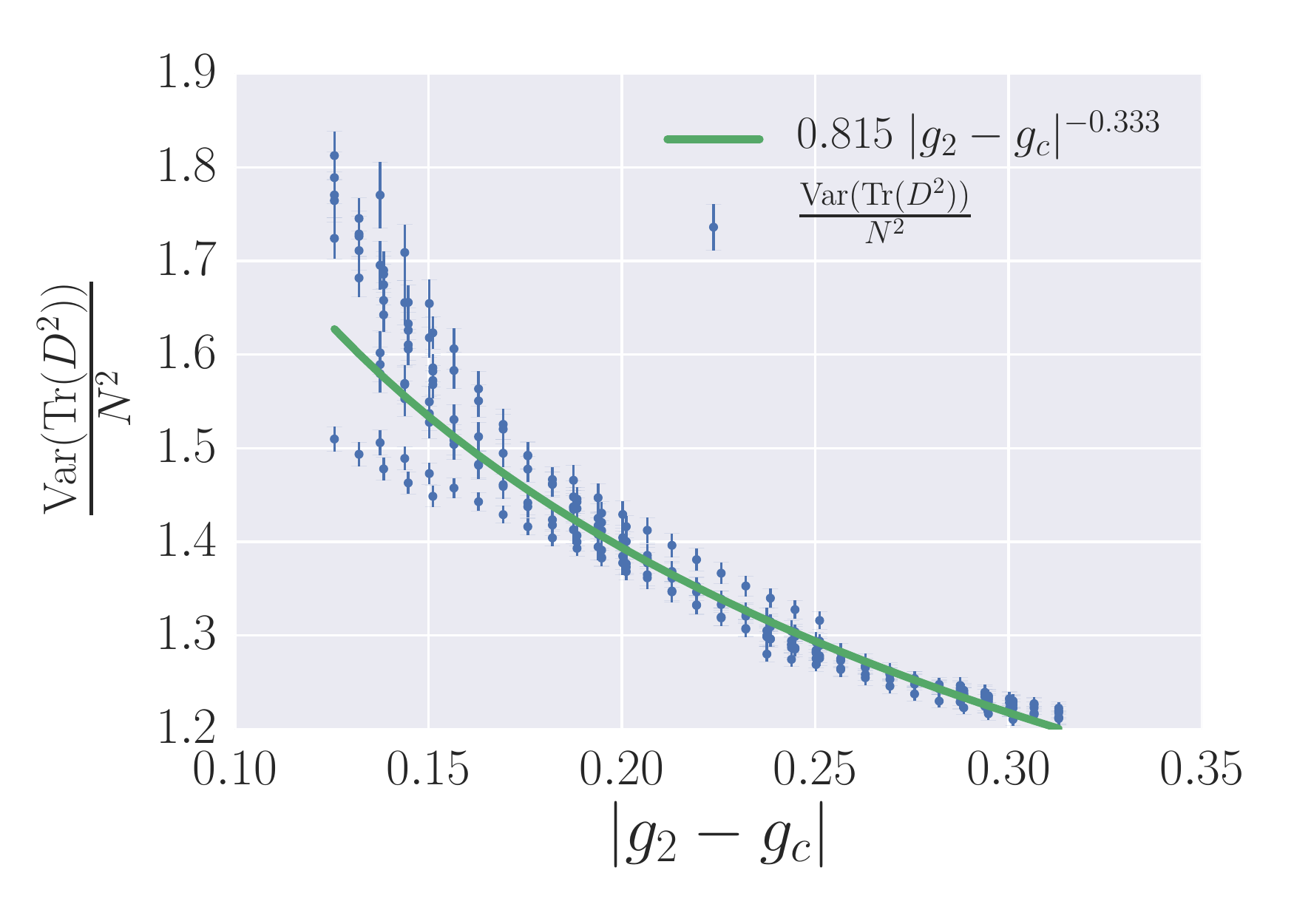}}
\subfloat[][$(1,3)$]{\includegraphics[width=0.32\textwidth]{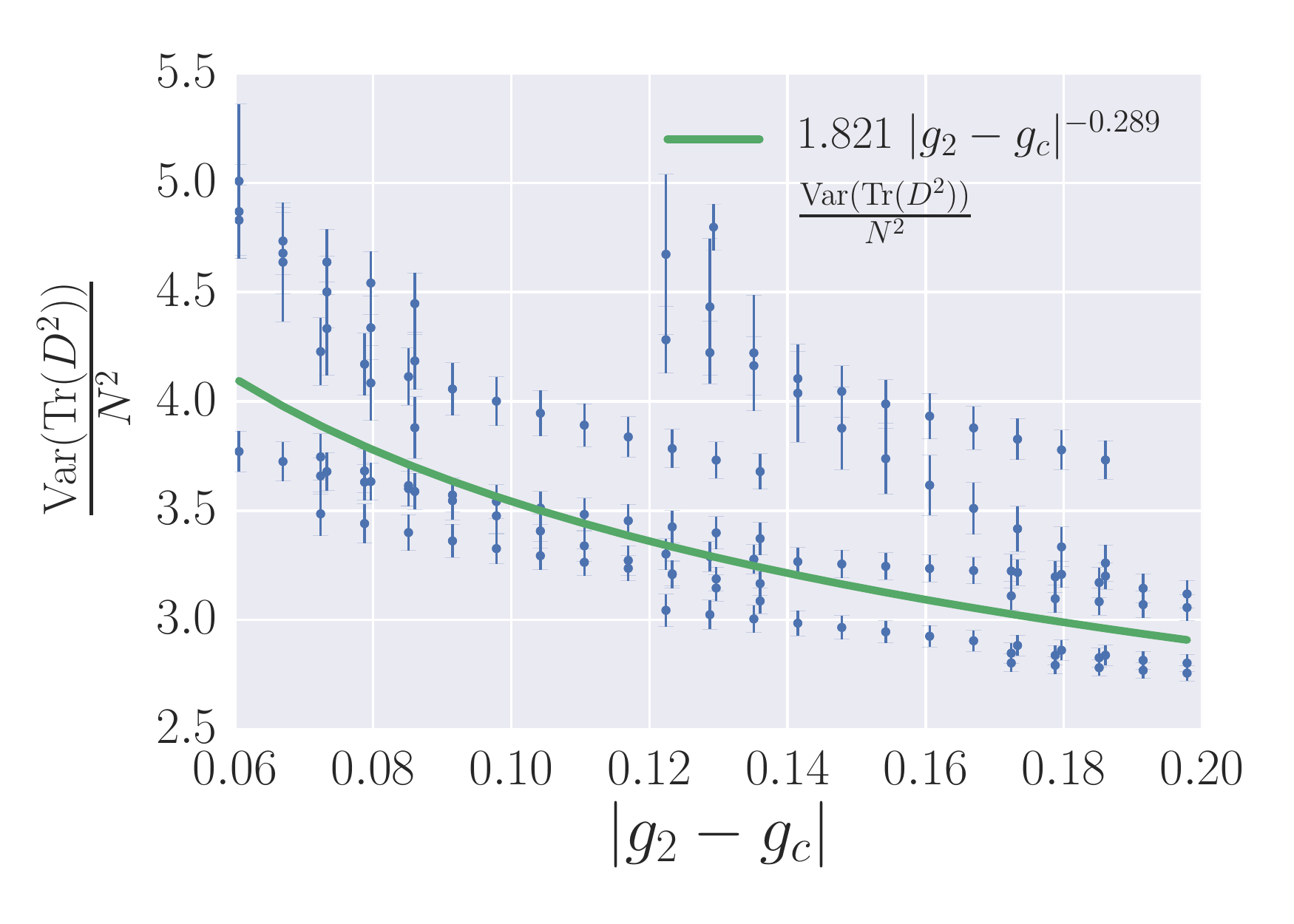}}
\caption{\label{fig:critexp}Fit of the collapsed values of $\mathrm{Var}(\Tr{D^2})$ against $|g_2-g_c|$ for the different geometries.}
\end{figure}

We fit to the same region for both $\mathrm{Var}(S)$ and $\mathrm{Var}(\Tr{D^2})$ for each type of geometry.
Since the data away from the phase transition obeys an $N^2$ scaling extremely well we can use this scaling to collapse the data and thus combine the data for different $N$, to improve statistics.
The resulting fits for $\mathrm{Var}(\Tr{D^2})$ are shown in Figure \ref{fig:critexp} and the best fit critical exponents are shown in Table \ref{tab:critexp}.
\begin{table}
\caption{\label{tab:critexp}The critical exponents $\mu$ and $\nu$ as determined in our fits.}
\begin{tabularx}{\textwidth}{X X X X}
\toprule
 & $(1,1)$ & $(2,0)$ & $(1,3)$\\
 \midrule
$\mu$ &$-0.130\pm  0.001$ &$-0.609\pm  0.011$ &$-0.457\pm  0.027$\\
$\nu$ & $-0.081\pm  0.001$ &$-0.333\pm  0.005$&$-0.289\pm  0.017$\\
\bottomrule
\end{tabularx}
\end{table}
Unfortunately the values we find do not immediately point towards any similarities with known theories, hence further study is necessary.

\section{\label{sec:conclusion}Conclusion}
We have established the location of the phase transition for all three geometries observed, and found evidence that the transitions are of second or higher order for type $(1,1)$ and $(2,0)$.
Unfortunately the data we were able to generate does not allow us to make any confident statement about type $(1,3)$, although there are other arguments, in particular the growth of correlation indicated by the covariance between eigenvalues, that indicate that it should also be of higher order.

The action and the $\Tr{D^2}$ term of all geometries scale as $N^2$, which is proportional to the number of eigenvalues of the Dirac operator.
Hence the system shows linear scaling with the volume, as one would expect.
Away from the phase transition the variances of these quantities also show this simple scaling, however at the phase transition this changes.
The system at the phase transition scales like $N^2$ for geometries of type $(1,1)$, like $N^3$ for geometries of type $(2,0)$ and like $N^4$ for geometries of type $(1,3)$.
These different scalings are interesting, and show that the random geometries of different types behave very differently.
This is particularly interesting for type $(1,1)$ and $(2,0)$ which looked extremely similar under the examinations conducted in \cite{barrett_monte_2015}.
The growth of the variances is driven by the strength of the correlation between the different eigenvalues of the system.
A stronger correlation between them makes the variance grow stronger than $N^2$.
Of particular interest for future investigations is the question how this growth relates to the dimension of the average geometries at the phase transition.
We start to address this question more fully by exploring dimension estimators in~\cite{barrett_druce_glaser}.
It is also an interesting speculation whether there exist classes of random geometries that show a growth stronger than $N^4$.
The number of terms that can contribute grows like $N^4$, so this might be an upper limit, on the other hand the values of the terms could grow with powers of $N$, hence allowing for unlimited growth of the variance.

As we saw in Figure \ref{fig:AvCov}, away from the phase transition the average covariance shrinks with rising $N$, which is expected since a finite range of correlation will correlate a smaller fraction of points in a larger system.
On the other hand for type $(1,3)$ and $(2,0)$ it clearly shows a peak that becomes clearer with $N$.
The growth of these peaks seems likely to be related to the correlation length of the system, and the question whether it diverges or not.
These observations are intriguing and further and more detailed study, in particularly including an investigation into possible definitions of a correlation length and a more in depth analysis of correlation in general is necessary.

In addition to the scaling with $N$ we were also able to determine the fall-off critical exponent of the variances on the left side.
These might be first pieces of a puzzle in trying to fit these complicated matrix models with a clear physical motivation into the general literature on random matrices.
In \cite{gnutzmann_universal_2004} a classification of random matrices by symmetry properties that are quite similar to those defined for a real non-commutative geometry is introduced, so a connection between these theories seems likely.
It would also be interesting to find a mean field model for our system, which would allow us to study the critical exponents in more detal and to explore if the model obeys hyperscalign laws.

\section{Acknowledgements}
I would like to thank John Barrett for introducing me to NCG and many long and fruitful discussions.
I would also like to thank Denjoe O'Connor, Sumati Surya and Des Johnson for discussions and encouragement concerning this work.
I am also grateful for access to the University of Nottingham High Performance Computing Facility.
During this work I was supported by funding from
the European Research Council under the European Union Seventh
Framework Programme (FP7/2007-2013) / ERC Grant Agreement n.306425 “Challenging General Relativity”
and also received funding from the People Programme (Marie Curie
Actions) of the European Union's Seventh Framework Programme FP7/2007-2013/ under REA grant
agreement n.706349 "Renormalisation Group methods for discrete Quantum Gravity"
The dissemination of this work has been supported by COST Action MP1405 ``Quantum structure of spacetime (QSPACE)".
\bibliography{ref}

\end{document}